\documentclass[useAMS,usenatbib,usegraphicx]{mn2e}

\usepackage{epsf,rotating}
\usepackage{setspace}
\usepackage{subfigure}

\newcommand{\lya}{{\rm Ly}\alpha}

\newcommand{\hkpc}{h^{-1}{\rm kpc}}
\newcommand{\hmpc}{h^{-1}{\rm Mpc}}
\newcommand{\mpc}{{\rm Mpc}}

\newcommand{\kms}{\;{\rm km}\,{\rm s}^{-1}}
\newcommand{\cms}{\;{\rm cm}^{-2}}
\newcommand{\cmc}{\;{\rm cm}^{-3}}

\newcommand{\Zsolar}{\;{\rm Z}_{\odot}}
\newcommand{\msolar}{\;{\rm M}_{\odot}}

\newcommand{\logrm}{{\rm log}}

\newcommand{\vw}{{v_{\rm wind}}}

\newcommand{\gad}{{\sc Gadget-2}}

\newcommand{\CII}{{\hbox{C\,{\sc ii}}}}
\newcommand{\CIII}{{\hbox{C\,{\sc iii}}}}
\newcommand{\CIV}{\hbox{C\,{\sc iv}}}

\newcommand{\SiII}{{\hbox{Si\,{\sc ii}}}}
\newcommand{\SiIII}{{\hbox{Si\,{\sc iii}}}}
\newcommand{\SiIV}{\hbox{Si\,{\sc iv}}}
\newcommand{\SiV}{\hbox{Si\,{\sc v}}}
\newcommand{\SiXII}{\hbox{Si\,{\sc xii}}}
\newcommand{\NV}{\hbox{N\,{\sc v}}}

\newcommand{\OV}{\hbox{O\,{\sc v}}}
\newcommand{\OVI}{\hbox{O\,{\sc vi}}}
\newcommand{\OVII}{\hbox{O\,{\sc vii}}}
\newcommand{\OVIII}{\hbox{O\,{\sc viii}}}

\newcommand{\HI}{{\hbox{H\,{\sc i}}}}

\newcommand{\NeVIII}{{\hbox{Ne\,{\sc viii}}}}
\newcommand{\MgII}{{\hbox{Mg\,{\sc ii}}}}
\newcommand{\MgX}{{\hbox{Mg\,{\sc x}}}}
\newcommand{\autovp}{{\sc AutoVP}}

\title[Metal-line absorption and physical conditions]{The intergalactic medium over the last 10 billion years II: Metal-line absorption and physical conditions}

\author[B. D. Oppenheimer et al.]{ 
\parbox[t]{\textwidth}{\vspace{-1cm}
Benjamin D. Oppenheimer$^1$, Romeel Dav\'e$^2$, Neal Katz$^3$, 
Juna A. Kollmeier$^4$,
David H. Weinberg$^5$}
\\\\$^1$ Leiden Observatory, Leiden University, PO Box 9513, 2300 RA Leiden, the Netherlands
\\$^2$ Astronomy Department, University of Arizona, Tucson, AZ 85721, USA
\\$^3$ Astronomy Department, University of Massachusetts, Amherst, MA 01003, USA
\\$^4$ Observatories of the Carnegie Institution of Washington, Pasadena, CA 91101, USA
\\$^5$ Astronomy Department, Ohio State University, Columbus, OH 43210, USA
}

\begin{document}

\pubyear{2011}

\maketitle

\label{firstpage}

\begin{abstract}
We investigate the metallicity evolution and metal content of the
intergalactic medium (IGM) and galactic halo gas from $z=2\rightarrow
0$ using 110-million particle cosmological hydrodynamic simulations.
We focus on the detectability and physical properties of UV resonance
metal-line absorbers observable with {\it Hubble's} Cosmic Origins
Spectrograph (COS).  We confirm that galactic superwind outflows are
required to enrich the IGM to observed levels down to $z=0$ using
three wind prescriptions contrasted to a no-wind simulation.  Our
favoured momentum-conserved wind prescription deposits metals closer
to galaxies owing to its moderate energy input, while the more
energetic constant wind model enriches the warm-hot IGM (WHIM) 6.4
times more.  Despite these significant differences, all wind models
produce metal-line statistics within a factor of two of existing
observations.  This is because $\OVI$, $\CIV$, $\SiIV$, and $\NeVIII$
absorbers primarily arise from $T<10^5$ K, photo-ionised gas that is
enriched to similar levels in the three feedback schemes.  $\OVI$
absorbers trace the diffuse phase with $\rho/\bar{\rho}\la 100$, which
is enriched to $\sim 1/50\Zsolar$ at $z=0$, although the absorbers
themselves usually exceed $0.3 \Zsolar$ and arise from inhomogeneously
distributed, un-mixed winds.  Turbulent broadening is required to
match the observed equivalent width and column density statistics for
$\OVI$.  $\CIV$ and $\SiIV$ absorbers trace primarily $T\sim 10^4$ K
gas inside haloes ($\rho/\bar{\rho}\ga 100$), although there appear to
be too many $\CIV$ absorbers relative to observations.  We predict COS
will observe a population of $\NeVIII$ photo-ionised absorbers tracing
$T<10^5$ K, $\rho/\bar{\rho}\sim 10$ gas with equivalent widths of
10-20 m\AA.  $\MgX$ and $\SiXII$ are rarely detected in COS S/N=30
simulated sight lines ($dn/dz\ll 1$), although simulated $\SiXII$
detections trace halo gas at $T=10^{6-7}$~K.  In general, the IGM is
enriched in an outside-in manner, where wind-blown metals released at
higher redshift reach lower overdensities, resulting in higher
ionisation species tracing lower-density, older metals.  At $z=0$, the
90\% of baryons outside of galaxies are enriched to
$\bar{Z}=0.096\Zsolar$, but the 65\% of unbound baryons in the IGM
have $\bar{Z}=0.018\Zsolar$ and contain only 4\% of all metals, a
large decline from 20\% at $z=2$, because metals from early winds
often re-accrete onto galaxies while later winds are less likely to
escape their haloes.  We emphasise that our results are sensitive to
how metal mixing is treated in the simulations, and argue that the
lack of mixing in our scheme may be the largest difference with other
similar publications.

\end{abstract}


\section{Introduction}  

The intergalactic medium (IGM) contains the majority of the Universe's
baryons.  The best available IGM tracer is the intervening $\lya$
absorption forest observed in quasar spectra \citep{rau98}.
Contributing a lesser but still significant amount of absorption are
intervening metal lines that trace the enrichment history of the IGM
\citep[e.g.][]{tyt87,cow95,bur96,son01,sch03,sim04,dan08,tri08}.
Since metal absorption is often found far from the sites of star
formation, analytic models and simulations strongly suggest that
metals must be transported out from galaxies into the IGM via
large-scale galactic outflows
\citep[e.g.][]{teg93,nat97,agu01a,spr03b,opp06,cen06b,kob07,som08,wie09b,tor10,cho11,tes11}.

The observation and interpretation of IGM metal-line absorbers
provides critical insights into the physics of galaxy formation.
First, galactic outflows that enrich the IGM concurrently regulate
the growth of galaxies by removing gas from star-forming regions.
Second, the census of IGM metals provides a unique and independent
constraint on the integrated amount of star formation in our Universe;
indeed, the metal census within galaxies falls significantly short
of the expectations from cosmic star formation, which is known
as the ``missing metals problem'' \citep[e.g.][]{fuk04,dav07}.
Finally, the amount and physical state of the metals in the IGM
alters how gas cools \citep[e.g.][]{sut93, gna07, wie09a}, thereby
impacting the efficiency by which galaxies receive gas from the IGM
\citep[e.g.][]{kat96, ker05, opp10}.  Hence studying IGM metal
absorbers provides unique insights into the cycle of baryons between
galaxies and their surrounding gas.

Despite the importance of IGM metal absorbers, the chasm between
observations and a theory of metal lines remains wide.  Many basic
questions concerning IGM chemical enrichment remain poorly understood.
Answering such questions is now becoming feasible owing to the
recent installation of the {\it Cosmic Origins Spectrograph} (COS)
on the Hubble Space Telescope (HST), whose central mission is to
improve our knowledge of the low-redshift IGM.  Key questions
include:

\noindent $\bullet$ {\it What is the metallicity of the IGM?}  At
redshifts of $z=2-3$ measurements range between $10^{-3.5}-10^{-2.0}
\Zsolar$ \citep{son96,dav98,sch03,sim04}, while by $z=0$ a value of
$0.1 \Zsolar$ is typically quoted \citep{cen01, fang01, che03}.  While
this suggests an increasing metallicity with time, the interpretation
is complicated by an evolution in how metals trace the IGM baryon
distribution.  Owing to cosmic expansion, a given column density $\HI$
system traces a higher overdensity at lower redshifts~\citep{dav99}; a
similar trend, modulated by evolving photo-ionisation rates, also
exists for metals.  One consequence of these effects is that summing
the column densities of metal lines to find an integrated ion density
($\Omega_{\rm ion}$) does not necessarily translate into a global
metallicity ($\Omega_{\rm metal}$).  \citet{opp06} argued that the
non-trivial ionisation correction from $\CIV$ to total carbon
abundance evolved with redshift, resulting in a relatively flat
$\Omega_{\CIV}$ that masks a significant increase in the metallicity
of the IGM.  These measurements have now been made down to $z\sim 0$
\citep[e.g.][]{dan08, coo10}, and COS should provide significantly
more insights into the relationship between metallicity and density
over most of cosmic time.

\noindent $\bullet$ {\it What temperature is the enriched IGM gas?}
High-ionisation lines commonly observed in the IGM such as $\CIV$
and $\OVI$ can arise both from photo-ionisation and collisional
ionisation.  The low densities and hard quasar-dominated metagalactic
flux can produce such high ionisation states, as can hotter gas in
the warm-hot intergalactic medium (WHIM) of gas shock-heated during
large-scale structure formation.  There is vigorous debate as to
whether observed high-ionisation lines trace collisionally-ionised
hot metals at $T>10^5\,$K \citep[e.g.][]{cen06b, dan08,
cen11, smi11, tep11} or low-density photo-ionised gas at $T<10^5\,$K
\citep[e.g.][hereafter OD09]{tho08b,tri08,opp09a}.  The increased sensitivity
of COS enables multiple ions to be detected in many systems, allowing
independent constraints on the nature of high-ionisation systems.

\noindent $\bullet$ {\it How homogeneously are metals distributed
in the IGM?} If metals are well-mixed throughout the IGM, metal
lines will cool more baryons, possibly leading to increased gas
accretion onto galaxies and hence an increase in star formation
\citep{wie09b}.  If metals are poorly mixed as observations suggest
\citep{sim06a, sch07}, then a census of $\OVI$ cannot straightforwardly
be converted to a census of baryons \citep[e.g.][]{tri00b, dan05}
since $\OVI$ only traces the baryons in over-enriched regions
\citep[OD09; ][]{tep11}.  The analysis of observed metal-line systems
often yields inconclusive answers, with complex absorbers suggesting
multi-phase gas components \citep[e.g.][]{pro04, sav05, dan06,
coo08, how09, nar09}.  The higher signal-to-noise provided by COS
will be critical for disentangling these effects.

A number of reasons exist for our less complete knowledge of metal
absorbers compared to that of the $\lya$ forest.  Observationally,
metal lines are weaker and, therefore, harder to detect.  Moreover,
many key ionisation states are buried within the $\lya$ forest or
occur at X-ray wavelengths where current sensitivities are limited
\citep[e.g.][]{che03, nic05, cen06b, fang10}.  Theoretically, IGM
enrichment depends sensitively on the transport mechanism out of
galaxies, which is poorly constrained, while the $\lya$ forest is less
affected \citep[Kollmeier et al. 2003; OD09; ][]{dav10, tes11}.
Furthermore, converting metal ions into metallicities typically
requires a detailed knowledge of the photo-ionisation rates at
energies poorly constrained by the data or, if collisionally ionised,
is highly sensitive to gas temperatures \citep[e.g.][]{hec02} and
hence detailed cooling rates \citep[e.g.][]{gna07}.  However, there
are many more metal transitions observable compared to hydrogen
transitions, providing an opportunity to synthesise a wider suite of
observations into a single coherent framework to determine the
physical state of IGM metals.  Hydrodynamic simulations that evolve
the metallicity and ionisation state of the IGM self-consistently
along with the enriching galaxy population offer a way to develop such
a framework within a modern hierarchical context.

In this paper, we employ state-of-the-art cosmological hydrodynamic
simulations to study the $z=0-2$ IGM as traced by high ionisation
metal lines.  Our strategy involves examining what can be learned by
observations targeting quasar absorption line spectra randomly
sampling the IGM, which represents a volume-weighted measure.  We
focus on high ionisation lines including $\OVI$, $\CIV$, $\SiIV$,
$\NeVIII$, $\MgX$, and $\SiXII$.  Low ionisation lines (e.g. $\CII$,
$\SiII$, $\MgII$) are not addressed as they trace higher density gas
where self-shielding renders the assumption of a uniform ionisation
background insufficient.  We briefly comment on $\NV$ and $\OV$
predictions.  This paper directly follows Dav\'e et al.\
(\citeyear{dav10}, Paper~I in this series, hereafter D10), which
discussed the $\lya$ forest and the physical conditions of baryons in
the IGM.  The next paper in this series will study correlations
between galaxies and absorbers (Kollmeier et al. in prep., Paper III).

While we will show predictions for observable quantities with COS, we
do not intend these to be used for direct comparisons with data.  We
are more interested in showing how plausible variations in the physics
manifest themselves in forefront observations, particularly in terms
of answering some of the questions outlined above.  For instance, we
will discuss how integrated column densities trace the underlying IGM
metallicity, how various ions observable with COS trace the
temperature and density of the enriched IGM baryons, and how absorber
alignment statistics can constrain the homogeneity of metals in the
IGM.  We will do all this for several prescriptions of galactic
outflows, to better understand how COS observations can constrain this
important physical process of galaxy formation.

A description of our cosmological hydrodynamic simulations and our
artificial spectra follows in \S2.  \S3 concentrates on the evolving
physical conditions of metals, subdivided by phase between
$z=2\rightarrow 0$.  We simulate COS metal-line absorber observations
in \S4 and relate them to their physical conditions in \S5,
demonstrating that only a fraction of IGM metals are probed via UV
resonance absorption lines.  We spend \S6 exploring physically
plausible model variations both to demonstrate the inherent
uncertainties in interpreting metal-line observations and to distill
what can be learned using relatively straight-forward statistics in a
sample size similar to the one that COS will observe.  \S7 discusses
the successes and failures of our simulations and compares them to
simulations by other groups.  We conclude in \S8.  For readers who
want a briefer overview of the results, we suggest first reading the
conclusions section and following the sections listed after each of
the main findings for more detail.

\section{Methods and Simulations}

We utilise our modified version \citep{opp08} of the \gad~N-body +
Smoothed Particle Hydrodynamic (SPH) code \citep{spr05} to evolve a
series of cosmological simulations to $z=0$.  The details of the
simulations can be found in \S2 of D10; they have also been used and
described in \citet{opp10}.  Briefly, our cosmology agrees with the
WMAP-7 constraints \citep{jar11} and uses the following values:
$\Omega_{\rm M}=0.28$, $\Omega_{\rm \Lambda}=0.72$, $h\equiv H_0/(100
\kms \mpc^{-1})=0.7$, a primordial power spectrum index $n=0.96$, an
amplitude of the mass fluctuations scaled to $\sigma_8=0.82$, and
$\Omega_{\rm b}=0.046$.

\subsection{Wind Implementations}

We briefly recap the implementations of galactic winds in these
simulations; for a more extensive description see \citet{opp08}.
The basic technical implementation is the one described in
\citet{spr03a} and incorporated into \gad.  In a given timestep, a
particle in a star-forming region has $p_{ej}= \eta\times{\rm
SFR}/m_{\rm p}$ of being ejected, where $\eta$ is the mass-loading
factor, SFR the star-formation rate, and $m_{\rm p}$ the particle
mass.  The ejection velocity, $\vw$, is oriented according to ${\bf v}
\times \nabla\Phi$, where ${\bf v}$ is the particle velocity and $\nabla\Phi$
is the gradient of the gravitational potential, resulting in winds
propagating perpendicular to galactic disks.  To allow the particle
to escape the ISM of its host, hydrodynamic forces are turned off
until one of two criteria are met: either $1.95\times 10^{10} \kms/
\vw$ years have passed or (more often) the particle flows out to a
density of less than 10\% of the star formation critical density.

In our constant wind (cw) model, we employ an ejection velocity of
$\vw= 680 \kms$ and a mass loading factor of $\eta= 2$ for all
galaxies.  The slow wind model has $\vw= 340 \kms$ and $\eta=2$.  Both
of these wind models are similar to \citet{spr03a}, except for their
assumed velocity; they use $\vw=484 \kms$.  The momentum-conserved
wind model, inspired by the \citet{mur05} models of radiation
pressure-driven winds, has $\vw = 3 \sigma \sqrt{f_{\rm L}-1}$ where
$f_{\rm L}$ is the luminosity in units of the Eddington 
luminosity required to expel gas from a galaxy potential, and $\eta =
\sigma_0/\sigma$ where $\sigma_0= 150 \kms$ and $\sigma$ is the
galaxy's internal velocity dispersion calculated by a group finder run
during the evolution of the simulation \citep[see ][]{opp08}.

The metallicity of an ejected wind particle is the metallicity of
the SPH particle at the time of ejection.  SPH particles are
continually self-enriched while they are eligible for star formation.
In this scheme, both ejected winds and newly formed stars have the
metallicity of the SPH particle at the time of the ejection or
formation event; wind particles are essentially ejected ISM SPH
particles.  Note that only particles above the critical star-formation
density threshold are eligible to form stars or to become launched
as a wind.  Cooling, which is never suppressed, invariably allows
the launched winds to cool and maintain a temperature of $\sim 10^4$
K while decoupled.  Once the hydrodynamic forces are restored, which
always occurs deep inside haloes given the two criteria listed
above, the SPH particle interacts normally.  In practice, re-coupled
wind particles show a bimodal behaviour: (i) they shock heat to
temperatures above the peak cooling efficiencies for metal-enriched
gas ($T\sim 10^5-10^6$ K), or (ii) they shock heat to a lower
temperature where metal-line cooling is efficient and then rapidly
cool to $T\sim 10^4$ K.  The most important determining factor
appears to be $\vw$, which we will show when we consider IGM
enrichment patterns among the different wind models in \S\ref{sec:metphys}.

\subsection{Simulations}

Our naming convention for our simulations continues the precedent we
established in our recent papers: r[boxsize]n[number of particles per
  side][wind model], where the initial letter ``r'' indicates the
particular choice of cosmology above.  The main simulations that we
explore are evolved in a cubic periodic volume that is $48 \hmpc$ on a
side: r48n384nw (nw: no winds), r48n384cw (cw: constant winds),
r48n384sw (sw: slow winds), and r48n384vzw (vzw: momentum-conserved
winds).  These simulations were also used in \citet{opp10}, D10, and
\citet{dav11a, dav11b}.  The main simulations are listed in Table
\ref{table:sims}.  The gas and dark matter particle masses are
$3.56\times 10^7 M_\odot$ and $1.81\times 10^8 M_\odot$, respectively.

\begin{table*}
\caption{Simulations}
\begin{tabular}{lcccccccc}
\hline
Name$^{a}$ &
$L^{b}$ &
$N_{\rm side}^{c}$ &
$\vw^{d}$ &
$\eta^{e}$ & 
$E_{\rm wind}/E_{\rm SN}^{f}$ &
$Z$ Cooling$^{g}$ &
$f_{\tau}^{h}$ &
$\Omega_{*}/\Omega_{b}(z=0)^{i}$
\\
\hline
\multicolumn {8}{c}{}\\
r48n384nw     & 48 & 384 & --              & --                  & 0.0  & CIE & 0.9  & 0.213 \\
r48n384cw     & 48 & 384 & 680             & 2                   & 0.96 & CIE & 0.9  & 0.045 \\
r48n384sw     & 48 & 384 & 340             & 2                   & 0.24 & CIE & 0.8  & 0.124 \\
r48n384vzw    & 48 & 384 & $\propto\sigma$ & $\propto\sigma^{-1}$ & 0.45-0.56 & CIE & 0.67 & 0.097 \\
r16n128cw     & 16 & 128 & 680             & 2                   & 0.96 & CIE & 0.9  & 0.040 \\
r16n128cw-PI  & 16 & 128 & 680             & 2                   & 0.96 & PI  & 0.9  & 0.037 \\
r16n128vzw    & 16 & 128 & $\propto\sigma$ & $\propto\sigma^{-1}$ & 0.41-0.54 & CIE & 0.67 & 0.096 \\
r16n128vzw-PI & 16 & 128 & $\propto\sigma$ & $\propto\sigma^{-1}$ & 0.41-0.56 & PI  & 0.67 & 0.088 \\
\hline
\end{tabular}
\\
\parbox{25cm}{
$^a$Suffix: nw- no-wind, cw- constant wind, sw- slow wind, vzw- momentum-conserved winds.\\
$^b$Box length of cubic volume, in comoving $\hmpc$.\\
$^c$Number of dark matter and SPH particles per box side.\\
$^d$Wind velocity in $\kms$.\\
$^e$Mass loading, where $\dot{M}_{\rm wind}=\eta\dot{M}_{*}$.\\
$^f$Wind energy divided by total SNe energy assuming $E_{\rm SN}=9.7\times 10^{48} {\rm erg}/\msolar$.  Range of values at $z=$0,1,2,3 for vzw.\\
$^g$Metal-line cooling.  CIE- Collisional Ionising Equilibrium cooling rates, PI- Photo-ionisation-dependent cooling rates.\\
$^h$Factor $\lya$ optical depth is scaled in determining ionisation background normalisation.\\
$^i$Fraction of baryons in stars at $z=0$.\\
}
\label{table:sims}
\end{table*}

Four other simulations listed in Table \ref{table:sims} are evolved
in a $16 \hmpc$ box with $2\times128^3$ particles, the same resolution
as the main simulations.  We use these to test the effects of
metal-line cooling with photo-ionisation on metal-line observations,
because \citet{tep11} and \citet{smi11} demonstrate that this could
significantly alter observations of $\OVI$ versus using collisionally
ionised metal-line cooling rates.  Despite their smaller volume,
we will demonstrate in \S\ref{sec:Zcool} that these simulations'
results remain converged compared to the 48 $\hmpc$ simulations,
likely because absorbers are mainly associated with sub-$M^*$
galaxies that are sufficiently sampled in both size volumes.  We
add the suffix PI to those simulation names evolved using the
photo-ionised cooling rates of \citet{wie09a}.

\subsection{Spectra}\label{sec:spectra}

Spectra are generated as described in \S2.3 of D10.  Briefly, 70
continuous lines of sight are created from $z=2\rightarrow 0$ for
each model using our spectral generation code {\tt specexbin}
\citep{opp06}.  We generate mock spectra using only the strongest
transition for an ion (e.g. the 1031.93\AA~line for $\OVI$, the
1548.19\AA~line for $\CIV$) and convolve the spectra with the
COS line spread function (LSF).  We add Gaussian random noise with
a signal to noise ratio S/N=30 per 6 $\kms$ pixel to simulate the
highest quality data obtainable using COS.  We employ the continuum
provided directly by the simulation, and do not attempt to mimic
the observational procedure of fitting a continuum, since at low
redshifts the sparseness of absorption features usually results in
a robust and reliable continuum fit.

We fit absorbers in the spectra using {\sc AutoVP}~\citep{dav97},
obtaining a column density $N$, line width $b$, and redshift $z$
for each absorber.  We consider each metal transition separately
and do not combine them into a single spectrum with all the
transitions.  This assumes that, in observations, each transition
can be reliably identified and fit without contamination from other
lines, which is typically true at low redshifts.

\subsection{Ionisation Background}\label{sec:ionbkgd}

\begin{figure}
\includegraphics[scale=0.81]{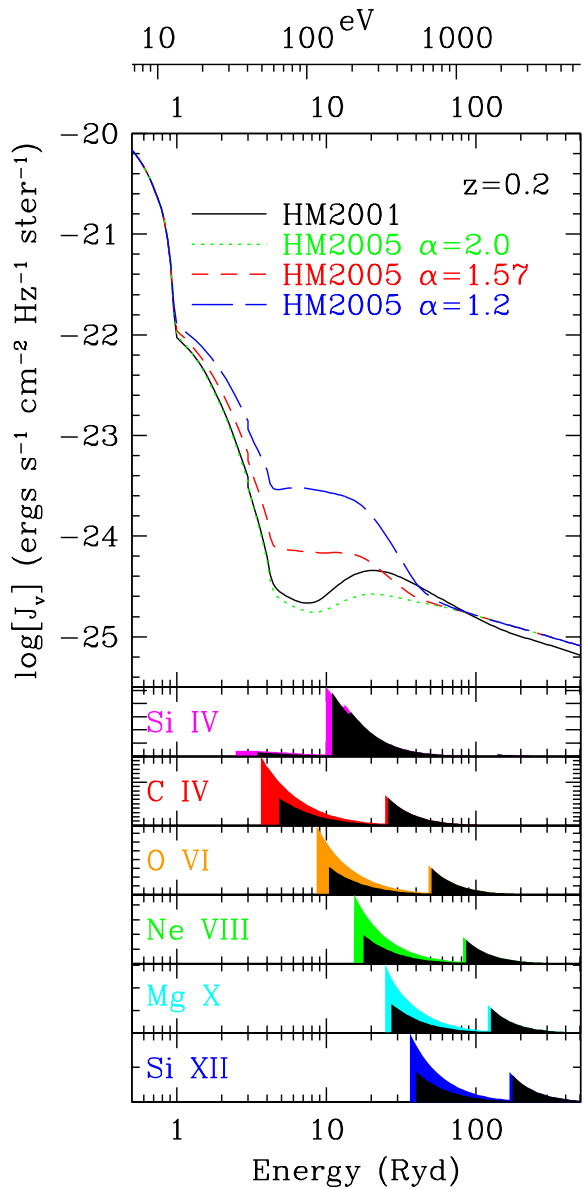}
\caption[]{The ionisation background spectra explored in this paper.
  HM2001 is the \citet{haa01} spectrum multiplied by 1.5 to match the
  observed $\lya$ flux decrement (D10).  Three other backgrounds
  generated using CUBA \citep{haa01} with varying slopes for quasar
  spectra above the Lyman limit are also displayed (HM2005).  Energies
  are listed in units of Rydbergs as well as electron volts (eV, top
  x-axis).  Below are photo-ionisation cross-sections for various
  ions.  Coloured regions show the magnitude of the cross-section from
  the ionisation state below to the specified ionisation state, while
  black regions show the cross-section from the specified state to the
  state above.  For these panels, levels of $10^{-18} {\rm cm}^2$ are
  indicated by large ticks on the $y$-axis and $10^{-19} {\rm cm}^2$
  are indicated by small ticks, with the $y$-axis starting at zero in
  all cases.}
\label{fig:ionbkgd}
\end{figure}

The metagalactic flux amplitude is adjusted to match the observed
evolution of the $\lya$ forest mean flux decrement.  We multiply the
$\lya$ optical depth by the same values of $f_\tau$ that we calculated
in D10 and presented in Table~\ref{table:sims} of that paper.  In
practice, we renormalise our CLOUDY \citep[][version 08.00]{fer98}
ionisation tables so that the magnitude of the ionisation background
is multiplied by $1/f_\tau$, which effectively shifts the ionisation
fractions of all ions by $1/f_\tau$ in density under the assumption of
an optically thin gas.  The vzw normalised \citet[][HM2001]{haa01}
$z=0.2$ background (i.e. $J_{\nu}$ multiplied by $1.5$) is shown in
Figure \ref{fig:ionbkgd} in black.  Three other backgrounds are also
displayed, which will be described in \S\ref{sec:ionbkgdvar} when we
explore the effects of variations in the ionisation background.

We also display the photo-ionisation cross-sections of various ions
used in CLOUDY \citep{ver95,ver96} on a linear scale in the bottom
panels of Figure \ref{fig:ionbkgd}.  The coloured regions show the
ionisation cross-section of the next lowest ionisation state to the
specified ionisation state, while the black shading indicates the
cross-section from that state to the next highest state.  For example,
the magenta shading indicates the cross-section for $\SiIII$ to
$\SiIV$, and the black shading indicates the cross-section for $\SiIV$
to $\SiV$.  With the exception of $\SiIV$, all cases shown are
lithium-like ions and display a trend of increasing ionisation energy
and declining cross-section with higher atomic number.  Metal ion
cross-sections are complex, which is exemplified by the $\SiIII$ and
$\SiIV$ cross-sections peaking above 10 Ryd even though $\SiIII$ has
the lowest ionisation energy (2.5 Ryd) of all the species we consider.
The multiple offset wedges, two in the case of lithium-like ions,
correspond to different subshells in the ground state.

\subsection{Chemodynamical Model} \label{sec:chemo}

We follow the production of four heavy elements (C, O, Si, \& Fe) from
three sources: Type II Supernovae (SNe), Type Ia SNe, and AGB stars as
explained in \citet{opp08}.  The details of the enrichment scheme are
in \S2.1 of D10.  Cosmic oxygen and silicon production is dominated by
Type II SNe, while carbon production is augmented significantly by AGB
stars but still primarily arises from SNe \citep{opp08}.  We do not
follow neon and magnesium directly when generating $\NeVIII$ and
$\MgX$ statistics but instead use oxygen as their proxy, which is a
reasonable approximation given that all three elements arise primarily
from Type II SNe.  The resulting abundances are [Ne/O]=0.16 and
[Mg/O]=-0.21 for \citet{chi04} SNe yields when using \citet{asp05}
solar abundances, which we scale to throughout this paper whenever
quoting metallicities.  We assume the same Type II SNe origin for
nitrogen by using [N/O]=-0.40, but note that this does not include
secondary nitrogen production in AGB stars, which is likely
significant.  The solar metal mass fraction by weight, $\Zsolar$, is
0.0122 and the mass fractions of carbon, nitrogen, oxygen, neon,
magnesium, and silicon are 0.00218, 0.00062, 0.00541, 0.00103,
0.00061, and 0.00067 respectively.  We also do not consider depletion
of these elements onto dust grains, although recent results suggest
that there may be dust in the IGM~\citep{men10,zu11} that contains
as much as 50\% of the mass of intergalactic metals.

\section{Metal Physical Conditions} \label{sec:metphys}

\subsection{Density-Temperature Phase Space} \label{sec:rhotcond}

\begin{figure*}
\includegraphics[scale=1.0]{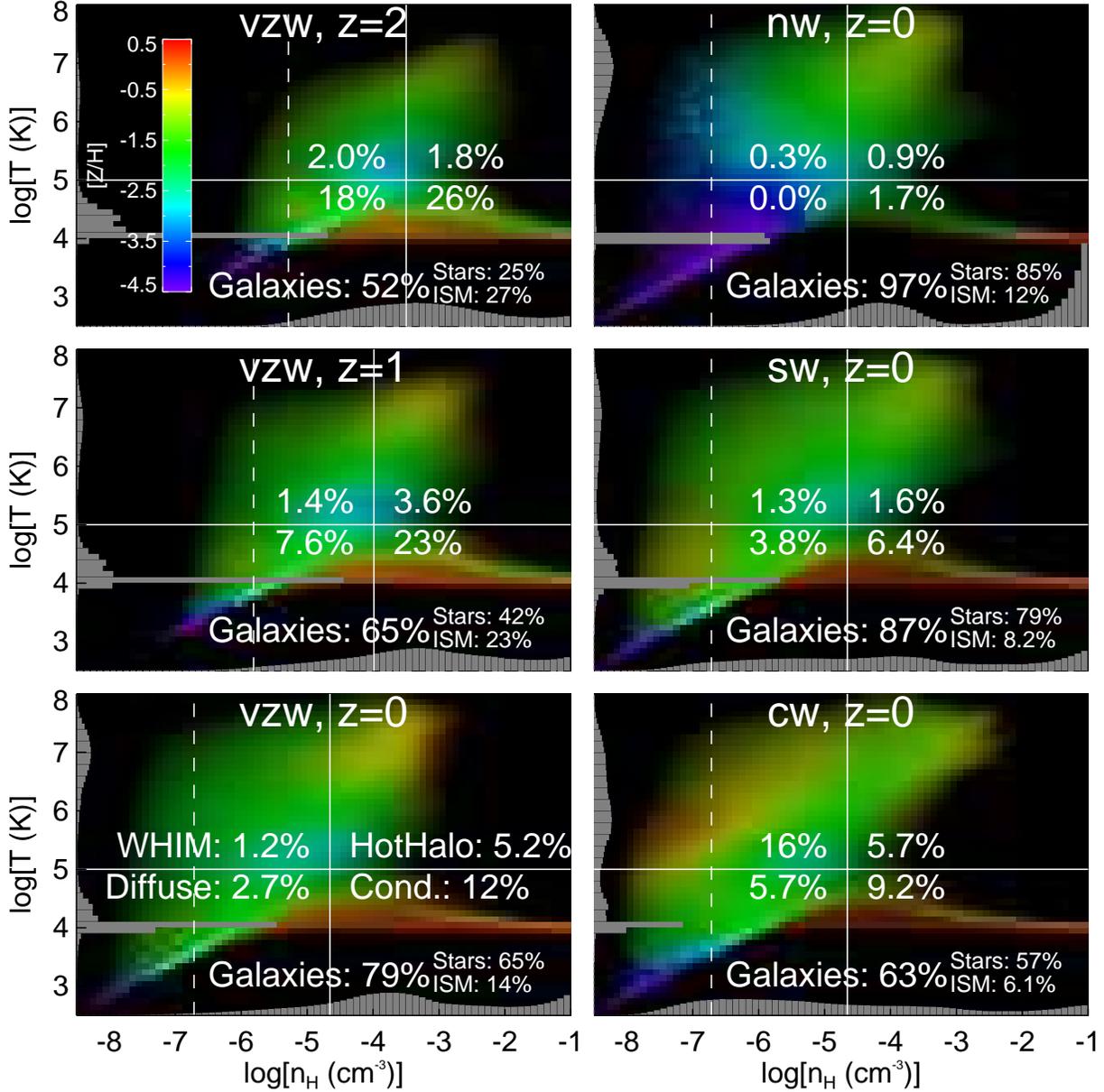}
\caption[]{Density-temperature phase space diagrams showing the
  distribution of metals divided into four phases by the solid lines:
  horizontal divides ``hot'' and ``cold'' phases, and vertical divides
  halo and IGM phases.  Colour indicates metallicity, and brightness
  corresponds to the baryon fraction.  Percentages indicate the
  fraction of global metals in each phase (bottom right- condensed,
  bottom left- diffuse, top left- WHIM, top right- hot haloes, plus
  the percentage in galaxies subdivided into stars and ISM is also
  listed). Grey histograms show the gaseous metal fraction as a
  function of density ($x$-axis) and temperature ($y$-axis) including
  only metals outside galaxies.  The vertical dashed line corresponds
  to the mean overdensity of the Universe.  The left panels show the
  evolution of the momentum-conserved wind (vzw) model at $z=2,1,0$
  and the right panels show the three other wind models at $z=0$ (nw-
  no winds, sw- slow winds, \& cw- constant winds).  }
\label{fig:Zphasespaces}
\end{figure*}

The density-temperature phase space diagrams of metals residing in
the IGM are displayed in the left panels of Figure \ref{fig:Zphasespaces}
for our preferred vzw model at $z=2$, 1, and 0.  The diagrams are
coloured by metallicity while the brightness indicates the fraction
of baryons.  On the right side are depicted the $z=0$ phase space
diagrams for the other three wind models.  Histograms along the x-
and y-axes depict the relative fraction of metal mass at a given
density and temperature, respectively.

The metals populate the cosmic phase diagram differently than the baryons
(see e.g. D10) with the density and temperature dependence of
metallicity linked to the type of outflows that put them there, which
we will discuss in \S\ref{sec:windphys}.  Here we focus on generic
trends seen among all the wind models, contrasting them with the no-wind
case.

Following D10, we divide the gas component into four phases, 
demarcated by the solid lines in the figures:

\begin{itemize}
\item Diffuse ($T<T_{\rm th}, \delta<\delta_{\rm th}$),
\item WHIM ($T>T_{\rm th}, \delta<\delta_{\rm th}$),
\item Hot halo ($T>T_{\rm th}, \delta>\delta_{\rm th}$),
\item Condensed ($T<T_{\rm th}, \delta>\delta_{\rm th}$),
\end{itemize}

\noindent where $T_{\rm th}=10^5$ K represents the division between
the ``cool'' and ``hot'' phases, and $\delta_{\rm th}$ represents the
division between bound (halo) gas and unbound (IGM) gas.  These
definitions are designed to provide a physically-motivated demarcation
between cosmic baryon phases.  $T_{\rm th}=10^5$ K is near the peak in
the helium and metal cooling curve that causes a minimum in the
temperature distribution of baryons, while the density threshold
separates truly intergalactic baryons, i.e. those not associated with
gravitationally collapsed structures, from baryons that lie
predominantly within the virial radii of galaxy halos.  Following
\citet{kit96}, we set
\begin{equation}\label{eqn:deltath}
\delta_{\rm th} = 6 \pi^2 (1+0.4093(1/f_\Omega-1)^{0.9052})-1, \\
\end{equation}
where
\begin{equation}
f_\Omega = \frac{\Omega_m (1+z)^3}{\Omega_m (1+z)^3+(1-\Omega_m-\Omega_\Lambda)(
1+z)^2+\Omega_\Lambda}.
\end{equation}
$\delta_{\rm th}$ evolves from a value of $\approx 60$ at high
redshift to $\approx 120$ by $z=0$.  As in D10, we adopt Equation
\ref{eqn:deltath} to be 1/3rd the \citet{kit96} mean enclosed density,
which corresponds to the density at $r_{200}$ for objects just
undergoing collapse.  The condensed phase does {\it not} include
star-forming ISM gas, which has $n_{\rm H}\ge 0.13 \cmc$ and lies to
the right of the high-density edge of each diagram.  The dashed
vertical line indicates the mean overdensity of the Universe
($\delta\equiv\rho/\bar{\rho}-1=0$).  The percentages listed provide a
fractional accounting of the total metals in all gas and star
particles associated with each phase; we also show the percentage of
metals in galaxies in the lower right corner, and additionally divide
this phase into stars and and star-forming (ISM) gas.

Examining the evolutionary trends in the vzw model, we see that the
fraction of metals in the diffuse IGM drops rapidly with time, from
18\% to 3\% from $z=2\rightarrow 0$.  In contrast, the fraction in
the WHIM phase is quite small ($\sim 1-2\%$) and evolves downwards
more slowly.  Recall from D10 that the mass fraction of the WHIM
in this model increases from $10\rightarrow 24\%$ during this time,
while the fraction in the diffuse gas drops from $73\rightarrow
41\%$.  Hence overall, the fraction of metals in the IGM (i.e.
diffuse+WHIM) drops, a trend that is accentuated in the diffuse
phase by a drop in the baryonic mass fraction in this phase, while
for the WHIM the trend is countered by an increasing mass fraction.
The decrease in the IGM metal fraction is balanced by an increase
in the metal fraction in galaxies, from half at $z\sim2$ to almost
$80\%$ at $z=0$.  Hence, from $z=2\rightarrow 0$, fractionally the
metals move from low-density regions to higher density regions.
We emphasise, however, that galaxies continue to enrich the IGM up to
the present day and that the metallicity of the IGM does not actually
decline; only the {\it fraction} of metals in the IGM declines.
We return to this point in \S\ref{sec:outsidein}.

By $z=0$, we see that all the wind models have most of their metals
within galaxies.  In the no-wind case, only 3\% of all metals are
outside of galaxies, while in the cw case it is 37\%.  The
momentum-conserved and slow wind models are intermediate, with
10--20\% of their metals outside of galaxies.  Still, these $z=0$
numbers are not large; in all the wind models more metals are outside
of galaxies at high-$z$.  This means that the solution to the
``missing metals'' problem~\citep{bou07,dav07} at early epochs depends
more on the accounting of metals outside of galaxies than today.

Examining the temperature histograms along the $y$-axis (including
only metals outside galaxies), we see that the gas-phase cosmic metals
in all the models at all epochs have a maximum at $T\sim 10^4$~K.
This is an artificial temperature floor, since we do not include
radiative cooling below $10^4$ K; however, this also is nearly the
equilibrium temperature where cooling equals photo-heating.  We will
show in \S\ref{sec:Zcool} that when we examine models including
metal-line cooling below $10^4$ K from \citet{wie09a}, metals cluster
around $\sim 10^4$ K, although with more variable temperatures
extending above and below owing to different equilibrium temperatures
dependent on density and metallicity.  

The central result is that most cosmic metals are not in hot gas,
either in the WHIM or hot halo gas.  At $z=0$, the IGM metallicity
distribution for all wind models has a bimodal temperature
distribution, with cool metals mostly at $T\sim 10^4$ K and hot metals
mainly residing at intracluster medium (ICM) temperatures ($T\ga
10^6$~K).  Few metals reside in between $10^5$~K and $10^6$~K (except
in the cw case) because metal-line cooling is very efficient there,
causing metal enriched gas to rapidly transition through these
temperatures (OD09).  In \S\ref{sec:mixing}, we state that the
bimodality is stronger in our simulations compared to simulations that
explicitly smooth metals, because metals confined to individual SPH
particles cool more rapidly over this intermediate temperature range.

The density histograms along the bottom of Figure \ref{fig:Zphasespaces}
for the vzw model (left panels) show a peak in metal fraction at
$n_{\rm H}\sim 10^{-3}-10^{-4} \cmc$ at all redshifts.  This
corresponds to an increasingly higher overdensity towards lower
redshifts, reflecting the result from \citet{dav07} that the peak
overdensity of metals increases with time.  This is qualitatively
true in the other wind models as well (not shown), although the
actual distribution of metals in overdensity depends strongly on
the particular wind  model as we discuss in \S\ref{sec:windphys}.

\begin{figure}
\includegraphics[scale=0.65]{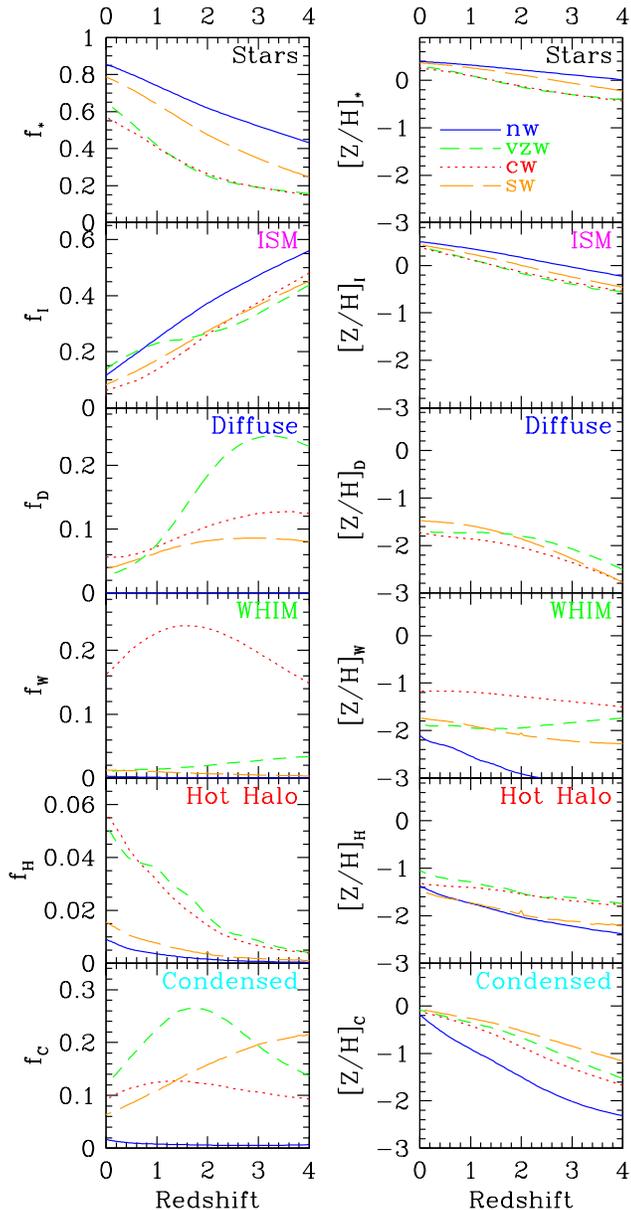}
\caption[]{The evolution of metal mass fractions (left) and
  mass-weighted mean metallicity (right) of the four wind models from
  $z=4\rightarrow 0$ subdivided by phase (1st-stars, 2nd-ISM,
  3rd-diffuse, 4th-WHIM, 5th-hot halo, \& 6th-condensed).  Four wind
  models are shown: solid blue- no-wind, dashed green-
  momentum-conserved winds, dotted red- constant winds, and
  long-dashed orange- slow winds. }
\label{fig:Zevol}
\end{figure}

Figure \ref{fig:Zevol} shows the mass fraction (left panels) and
mass-weighted mean metallicity (right panels) of six baryonic phases
(stars-top, ISM-2nd, diffuse-3rd, WHIM-4th, hot halo-5th, \&
condensed-bottom panels) for our four wind models, from
$z=4\rightarrow 0$.  At early epochs, star-forming gas holds the
majority of metals.  As time passes, metals become progressively
locked up into stars, so that by $z=0$ stars contain the majority of
cosmic metals.  Outflows suppress the metal content in stars and the
ISM somewhat but do not alter the overall trends.  The mean stellar
and ISM metallicities increase slowly from $z=4\rightarrow 0$, but
even at early epochs it is already close to solar, and by today it
exceeds solar.  This is consistent with the early, rapid enrichment in
galaxies typical in these models~\citep[e.g.][]{dav06,fin10} and is
relevant for the IGM because it shows that the material expelled from
galaxies has near-solar metallicities at all epochs.  The metallicity
of the IGM depends then on how far the metals are driven, and how they
mix with the unenriched ambient gas.

The metal fraction of the IGM phases (diffuse+WHIM) generally drops
from $z\sim 3\rightarrow 0$ in the wind models.  This reflects the
increasing difficulty that metals have in escaping from the denser
regions where they are formed to less dense regions, as depicted
in Figure~\ref{fig:Zphasespaces}.  While the fraction of metals in
the IGM phases is generally not large, it is still sufficient to
reproduce the observations of metal-line absorbers across cosmic
time~\citep[e.g.][]{opp06,opp09b}.

For the WHIM and hot halo phases, the $\delta_{\rm th}$ threshold that
divides these two phases, first introduced in D10, makes a stark
difference in the WHIM metal content compared to previous WHIM
definitions without such a density threshold.  The WHIM, which
contains 24\% of $z=0$ vzw baryons, holds only 1.2\% of the $z=0$
metals and remains steady at [Z/H]$\sim -1.8$ since $z=4$.  The WHIM
forms through shock-heating owing to the gravitational collapse of
large-scale structures \citep{cen99, dav99, dav01a, cen06a} while
metal-enriching galactic outflows contribute fractionally less to the
WHIM at late times \citep{opp08, cen11}.  The metal content of the hot
halo phase steadily increases, and by $z=0$ it holds more metals than
the combined IGM phases (cf. 5.2\% vs. 3.9\% for vzw).  The $y$-axis
histograms for the vzw model clearly show the growing importance of
hot halo metals from $z=2\rightarrow 0$.  The $z=0$ panel indicates
that most of these metals are at $T\ga 10^{6}$ K and hence are in the
intragroup or intracluster medium (ICM).  In all the wind models, the
hot halo metallicity reaches 0.1~$\Zsolar$ at $z=0$, but it also shows
a density gradient such that the portion that would emit most strongly
in X-rays has [Z/H]$\sim -0.5$ (see Figure~\ref{fig:Zphasespaces}).
This is comparable to the measured metallicity of the ICM
\citep[e.g.][]{fuka98,pete03} and reconfirms that these simulations
can generally reproduce the observed ICM metallicities \citep{dav08b}.

More broadly, the fraction of metals in the IGM increases at very
early epochs as winds enrich diffuse cosmic gas~\citep{opp09b}, but
then at later epochs the metals do not escape their haloes so easily
or else re-accrete back into haloes.  Hence, the metal fraction in the
diffuse phase peaks at $z\sim 3-4$ in these wind models
(Figure~\ref{fig:Zevol}).  Meanwhile, the metallicity in all the
phases increases, but slowly; the most rapid increase is in the
condensed phase gas, since this halo gas continues to be enriched even
to late times by outflows.  These trends, while generally true, are in
certain aspects quite sensitive to the wind model, as we will discuss
next.

\subsection{Dependence on the Galactic Wind Model}\label{sec:windphys}

The value of studying enrichment patterns lies in their high
sensitivity to galactic winds, upon which the process of galaxy
formation heavily depends \citep[e.g.][]{sch10, opp10}.  We now
contrast our vzw simulation to the three other wind models.  The
no-wind model produces nearly twice as many stars and metals as any
wind model, and is incapable of enriching the diffuse IGM appreciably
at any redshift, leaving it mostly pristine even by $z=0$
(Figure~\ref{fig:Zphasespaces}, upper right).  The necessity to enrich
the diffuse IGM to levels of at least [Z/H] $= -3$ at $z\ge2$
\citep[e.g.][]{sch03,sim04} is the main empirical requirement for
large-scale galactic outflows.  The diffuse phase metallicity of the
nw model remains below the y-axis in Figure~\ref{fig:Zevol}, only
reaching [Z/H] $= -6.1$ at $z=4$ and $-3.9$ by $z=0$.  This model
does, however, achieve a hot halo metallicity comparable to our other
feedback models, indicating that the enrichment of group and cluster
haloes at least in part owes to a mechanism other than galactic
outflows, such as dynamical stripping.  \citet{dav08b} showed that
their no-wind model reached observationally reasonable [Fe/H]
abundances via delayed Type Ia SNe from intragroup/intracluster stars,
but [O/H] abundances were too low compared to observations and the
baryonic fraction in stars was too high.  Overall, the no-wind model
strongly fails to match basic observations of both galaxies
\citep[e.g.][]{dav11a} and the IGM.

The constant wind (cw) model has a high wind speed of $\vw=680 \kms$
emanating from every galaxy, which pushes shock-heated metals into
voids.  This leads to a WHIM metal mass fraction more than 10 times
higher than any other model (Figure~\ref{fig:Zevol}, middle panel).
These metals remain hot, because adiabatic cooling from Hubble
expansion is not sufficient to return this gas to the photo-ionised
IGM equation of state \citep[EoS; ][]{hui97, sch00} within a Hubble
time.  The excess heating is also reflected in the phase space
diagram for this wind model (Figure~\ref{fig:Zphasespaces}, lower
right), which has a high-metallicity peak in underdense regions at
$T\sim 10^6$~K.  The fast winds also increase the WHIM baryon
percentage from 23 to 33\% (compared to the nw model), whereas the
vzw and sw models increase the WHIM baryonic fraction by one tenth
as much, a 1\% increase (D10).  As \citet{opp06} show, a similar
model with somewhat lower wind speeds ($\vw=484 \kms$) produces
$\CIV$ absorbers at $z\sim 3$ that are are too wide compared to the
observations, indicating that there is too much IGM heating at this
early epoch.

\begin{figure}
\includegraphics[scale=0.7]{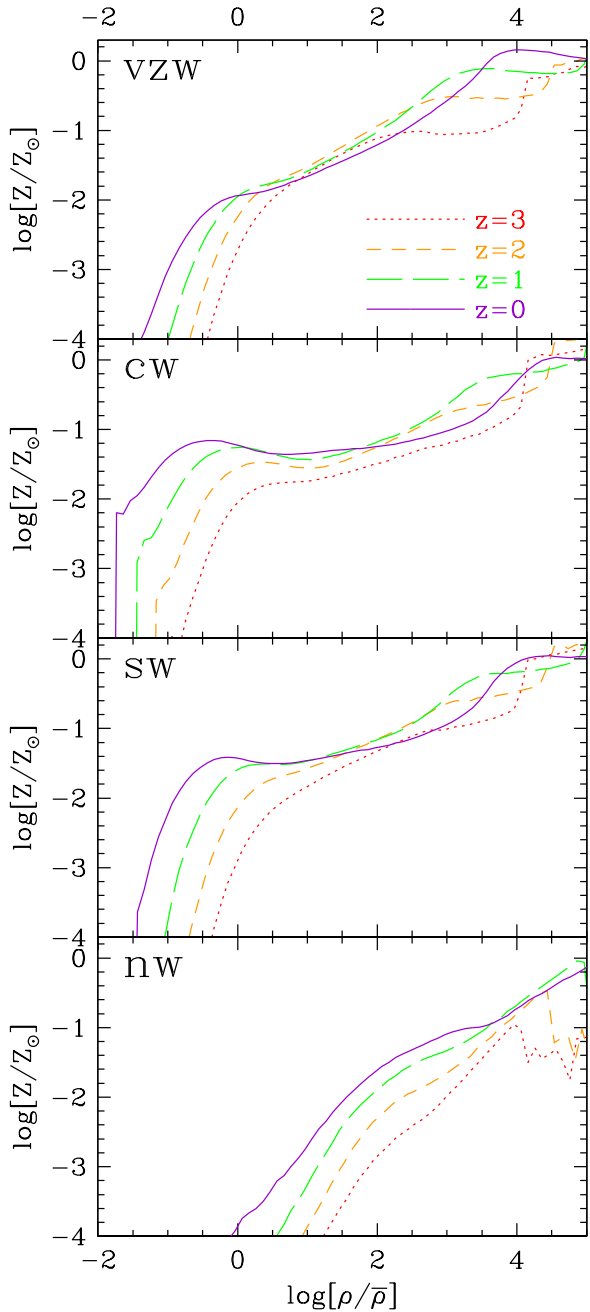}
\caption[]{Mass-weighted mean metallicity as a function of overdensity
  from $z=3\rightarrow 0$ for the four models: momentum-conserved
  winds (vzw), no winds (nw), constant winds (cw), and slow winds
  (sw).  }
\label{fig:rhoZ}
\end{figure}

The slow wind (sw) model with $\vw=340 \kms$ does not overly heat
the early IGM, and in some sense represents a hybrid of the vzw
model (similar average wind velocities) and the cw model (same mass
loading).  Like cw, this model creates a concentration of shock-heated
metals ($T\sim10^{4.5}$ K) in voids, but it is at lower temperatures
and is much less prominent owing to the sw model's lower wind speeds.
In this model low-mass galaxies eject winds at high velocities
relative to their escape velocity and high-mass galaxies, typically
in dense environments, eject winds at less than the escape velocity.
Hence, the sw model cannot populate the ICM with metals and,
therefore, the metal phase space diagram in this regime mimics that
of the no-wind model.

In the vzw model, winds from small galaxies have lower velocities.
Hence they deposit metals at higher densities where metal-line
cooling is efficient and can return these metals to $T\sim 10^4$
K.  In hot gas, the vzw model produces a strong metallicity-density
gradient by $z=0$ owing to its high outflow velocities from massive
galaxies.  Winds with $>1000 \kms$ speeds emanating from these
galaxies rarely escape the halo owing to the intervening halo gas, but
they do shock heat to temperatures high enough that the cooling times
are long, and hence the metals do not re-accrete onto the galaxies.

Figure \ref{fig:rhoZ} shows the mean metallicity-density relation
in our four wind models from $z=3\rightarrow 0$.  The no-wind model
fails to provide widespread IGM enrichment, but even dynamical
processes can deposit a smattering of metals at the cosmic mean
density by $z\sim 0$.  All the wind models show increased metallicities
at all the plotted overdensities relative to the no-wind case.  The
constant wind model results in a curious upturn in the mean metallicity
at void densities ($\rho<\bar\rho$), which is a generic feature of
strong feedback models \citep[e.g.][ their Figure 24]{cen11}.
Unfortunately this is very difficult to probe observationally using
UV resonance lines owing to the low densities, as we will discuss
in \S\ref{sec:absphyscond}.  The slow wind model has a less pronounced
upturn, and it shows a less widespread metal distribution particularly
at early epochs owing to its lower wind speed.  The vzw model shows
an interesting trend with overdensity, and the metallicity in filaments
($1\la \rho/\bar\rho\la 100$) is nearly time-invariant.  Continual
enrichment from galaxies pumps metals into these regions, but this
is approximately balanced by gravitational effects that draw the
metals back closer to galaxies and by the accretion of more pristine
gas.  At higher overdensities, the mean metallicity increases with
time, while at sub-mean densities, metals slowly diffuse deeper
into voids \citep{opp06}.  The differences between the models offer
an opportunity to probe outflows if the metallicity-density relation
and its evolution can be reliably measured, which is a central goal
of IGM enrichment observations.

Overall, the evolution of metal content by phase shows much more
dependence on feedback than our corresponding study of the abundance
of baryons by phase (D10).  While all the wind models significantly
suppress the global star formation rate and hence metal production,
the distribution of metals in the IGM shows clear differences.  The
velocity of the outflow and its motion through the ambient gas
determines to what temperature the gas is shock heated and to what
overdensity it reaches.  Gas in sufficiently dense regions cools
quickly, particularly since it is enriched.  Hence, the outflow
parameters critically determine the enrichment history of baryons
outside of galaxies, providing an opportunity to constrain such
parameters based on observations of IGM metal lines.

\subsection{Outside-In IGM Enrichment}\label{sec:outsidein}

A generic feature of our IGM enrichment models is that they have a
very early epoch where metals are distributed outwards into the
IGM, but over most of cosmic time the newly produced metals became
confined to ever decreasing regions around the galaxies that produced
them.  We call this ``outside-in" IGM enrichment, and it has
consequences not just for how metals enrich the IGM, but for how wind
material re-accretes onto galaxies \citep{opp10}.

We have already seen trends exemplifying outside-in enrichment
above.  The metal mass fraction in the IGM decreases steadily since
$z\sim 3$, and the peak overdensity of metals increases with time.
\citet{opp08} showed that the median distance reached by vzw outflows
(with a wide dispersion) is $\sim 100$~physical kpc, implying
widespread enrichment (in a comoving sense) early on and more
confined enrichment at later epochs.  The evolution of the vzw
metallicity-density relationship in Figure \ref{fig:rhoZ} (top
panel) shows that metals are in place at IGM overdensities at $z=3$
and that the metallicity is mostly invariant and actually declines
slightly at $\rho/\bar{\rho}\sim 2-10^3$ from $z=2\rightarrow 0$.
Newly produced metals are mostly confined to their own haloes and
fall back onto their parent galaxies primarily from these overdensities;
however winds continue to enrich IGM densities until $z=0$, especially
the winds from lower mass galaxies, often even reaching void
densities.  Additionally, some enriched WHIM material also falls
into galaxy haloes and eventually onto the central galaxy.

\begin{figure}
\includegraphics[scale=0.67]{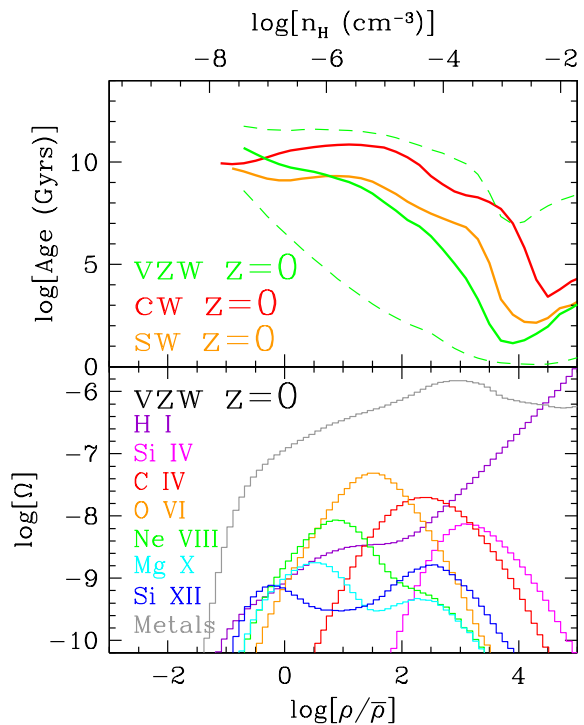}
\caption[]{The top panel shows median ages of wind-blown metals at
  $z=0$ as a function of density (solid lines) with the 16th and 84th
  percentage cuts (dashed lines, vzw only) for the momentum-conserved
  winds (vzw, green), constant winds (cw, red), and slow winds (sw,
  orange).  Lower overdensities are enriched earlier in all wind
  models in what we term ``outside-in'' enrichment.  The bottom panel
  shows cosmic ion densities of vzw metal species as a function of
  density, as well as, $\HI$ (purple lines) and total metal density
  (grey lines).  The two panels can be compared to show that in
  general higher ionisation species generally trace lower
  overdensities, which have older metals.}
\label{fig:rhoageomega}
\end{figure}

Another view of outside-in enrichment is provided by the age of
$z=0$ metals as a function of overdensity, shown in Figure
\ref{fig:rhoageomega}, top panel.  We define the metal age as the
time since a wind particle has most recently been launched, which
is a good indicator of the age for metals in the IGM as these baryons
are almost entirely enriched by winds.  The median age of wind-blown
metals is shown by the solid lines, and the 16 and 84\% cuts are
indicated by the dashed lines for the vzw model.

Figure \ref{fig:rhoageomega} shows directly that lower overdensities
are enriched earlier, particularly in the vzw model.  For this model,
the typical age of metals in voids is $\sim 10$~Gyr, reaching a
minimum of $\sim 1$~Gyr at overdensities of a few thousand.  The
constant-$\eta$ models show less dependence on age with overdensity in
the IGM, though at higher overdensities the metal ages are again
significantly younger.  The evolution of the metallicity-density
relation of the cw model (Figure \ref{fig:rhoZ}) shows a significantly
increasing metallicity within voids with time.  Most metals were
already in place at higher redshift in gas that continues to expand
into voids by $z=0$, but galaxies continue to enrich void
overdensities below $z=1$.  The same trends are true for the sw model,
but they are shifted to higher overdensities and younger ages owing to
its weaker winds.  While the median age trends with overdensity are
statistically significant and indicative of the history of metal
enrichment, the larger scatter in all models at most densities
prohibits a direct mapping of an overdensity to an age for a given
wind model.  

\citet{wie10} find a similar age-density anti-correlation using
their definition of mean enrichment redshift.  Our definition of
age is a lower limit compared to theirs, because we measure the
most recent wind launch as opposed to the redshift-weighted enrichment
age.  Both methods will find similar ages for wind particles that
rapidly enrich and launch once, as is typical for metals that escape
galaxies at high-$z$, but we will find significantly younger ages
for winds that recycle back onto galaxies and have multiple wind
launches and enrichment epochs.

To relate the IGM metal age to observables, we plot in the lower panel
of Figure \ref{fig:rhoageomega} histograms of the $z=0$ vzw cosmic ion
densities (in 0.1 dex bins of overdensity) for the ions that we follow
and for $\HI$.  In general, higher ionisation metal species trace
older IGM metals occupying lower overdensities in the vzw model.  For
example, observed $\NeVIII$ and $\OVI$ absorbers are likely to be
tracing fairly old ($\ga 5$~Gyr) metals, while $\CIV$ and $\SiIV$ more
often trace younger metals, although with a large age scatter.  Every
metal species shows a peak corresponding to the overdensity where the
species is primarily photo-ionised.  A second peak at higher
overdensities for $\SiXII$ and $\MgX$ corresponds to collisionally
ionised metals inside haloes.  $\NeVIII$ also shows an extended
shoulder toward halo densities corresponding to collisionally ionised
metals.  This figure foreshadows our analysis in
\S\ref{sec:absphyscond} showing that species including $\CIV$, $\OVI$,
and $\NeVIII$ are primarily photo-ionised in our simulations.
Interestingly, these three species exceed the cosmic density of $\HI$
at a range of IGM densities.

The idea that the IGM is enriched in an outside-in manner has a long
history \citep{teg93,nat97,mad01,sca02}.  In general, these models
postulated that the IGM underwent widespread early enrichment
typically during the reionisation epoch from some putative powerful
source such as Population~III stars, when halo potential wells were
still small.  Our model, while similar in character, is fundamentally
different in that it does not invoke very early enrichment from
mysterious sources.  Instead, the outside-in pattern arises from the
dynamics of galactic outflows from the epoch of early galaxies ($z\ga
6$) until today.  The source of IGM enrichment is the observed
population of high-$z$ star-forming galaxies that our simulations
naturally reproduce~\citep{dav06,bou07}.  Energetically moderate
outflows from these galaxies enrich a small, but rapidly growing
fraction of the IGM, in agreement with observed constraints from
high-$z$ metal-lines \citep[e.g.][]{rya09}.  These early outflows end
up enriching a greater comoving volume than lower redshift outflows,
which travel similar physical distances in our model \citep{opp08}.
In contrast, \citet{opp07} showed that enriching the IGM entirely
prior to $z\sim 6$ would violate observations of the $\CIV$ enrichment
at those epochs.

\section{Metal-Line Observables}

We present metal-line predictions for COS observations in this section
along with comparisons to other pre-COS data.  We consider statistics
assuming a S/N=30 per $6 \kms$ pixel in our simulated spectra,
convolved with the COS line spread function (LSF) assuming the FUV
G130M LSF at 1450\AA, just as we did in D10.  We model the LSF as a
central peak with a $1\sigma$ Gaussian width of $\approx 17\kms$ and
substantial non-Gaussian wings, keeping this fixed for all wavelengths
to avoid introducing artificial evolution from a varying LSF.  

We obtain line statistics by fitting Voigt profiles to individual
components using \autovp~\citep{dav97}.  Each component has a fitted
rest equivalent width ($W$), a column density ($N$), and linewidths
($b$-parameters).  An absorption system is defined as all components
that can be grouped together if they lie within 100 $\kms$ of another
component.  Systems are more easily comparable with observations,
because data quality does not always allow the easy and consistent
identification of components.  For example, it can be easier to divide
a system into more components if the S/N and resolution are higher.
We sum up system column densities to calculate cosmic ion densities
($\Omega$).  We discuss the equivalent width distribution of systems
versus components for $\OVI$, where observations for both have been
quantified.  We present results in this section for our four main
simulations, plus a vzw model with turbulent broadening added after
the completion of the simulation (OD09), motivated by the observed
$\OVI$ linewidths.

\subsection{Equivalent Width Distributions}

\begin{figure}
  \subfigure{\setlength{\epsfxsize}{0.45\textwidth}\epsfbox{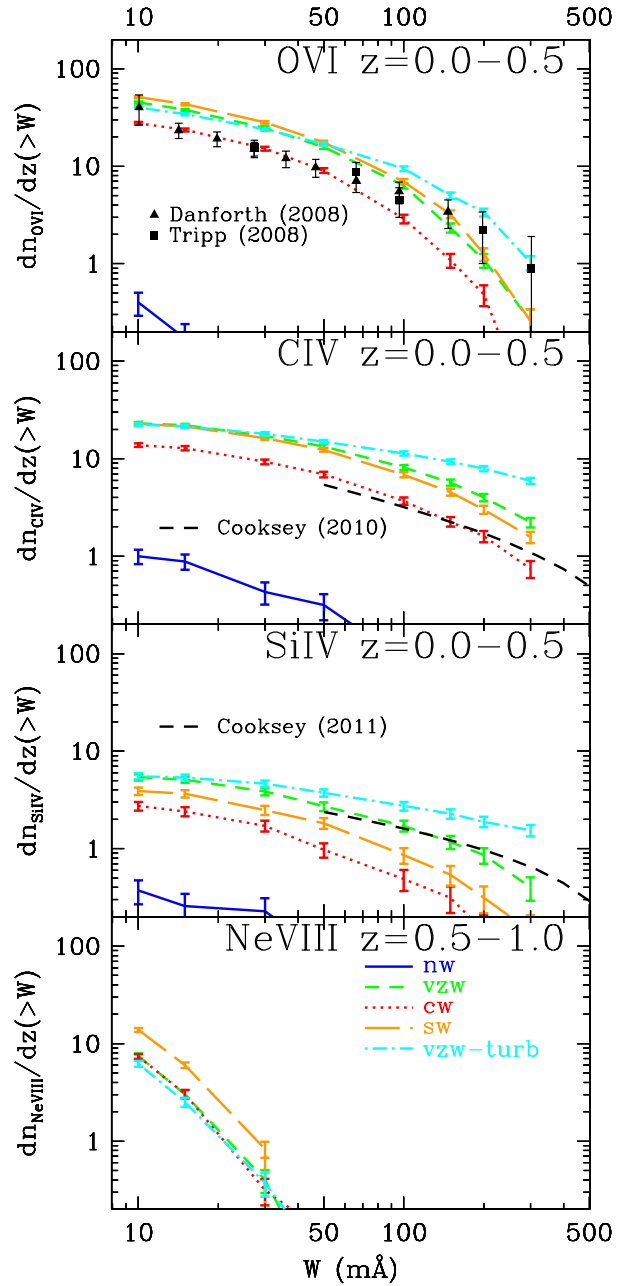}}
\caption{Cumulative equivalent width distributions of $\OVI$ (top),
  $\CIV$ (second), $\SiIV$ (third), and $\NeVIII$ (bottom) for our
  four wind models (solid blue- no-wind, dashed green-
  momentum-conserved winds, dotted red- constant winds, and
  long-dashed orange- slow winds, plus the vzw model with turbulent
  broadening (dot-dashed cyan).  The line types correspond to the same
  models throughout the paper.  Observations for $\OVI$ \citep{dan08,
  tri08} and $\CIV$ \& $\SiIV$ \citep{coo10, coo11} are also plotted.
  For the first three panels, we count all lines in our continuous
  synthetic spectra out to $z=0.5$ and divide by $\Delta z = 0.5$,
  while the fourth panel uses the same procedure for $z=0.5-1$.  Error
  bars correspond to 1-$\sigma$ Poisson errors in the models and
  observations.}
\label{fig:EW}
\end{figure}

Cumulative equivalent width distributions (EWDs), presented in Figure
\ref{fig:EW}, show the frequency of systems greater than or equal to
the $W$ indicated on the $x$-axis.  They offer a simple and direct
statistic to test theoretical models.  We show EWDs of four metal
lines: \OVI~1032\AA, \CIV~1548\AA, \SiIV~1394\AA~ (all between
$z=0-0.5$), and \NeVIII~770\AA~ (at $z=0.5-1.0$). We show predictions
for our four wind models and a fifth model (vzw-turb in cyan) of
turbulent $b$-parameters ($b_{\rm turb}$) added after the completion
of the simulation using {\tt specexbin} \citep{opp09a}.

The EWDs with no winds (blue solid lines) exhibit far fewer absorbers
than models with winds, and significantly underestimate the
observations of $\OVI$, $\CIV$, and $\SiIV$.  In general, the cw model
(red dotted lines) produce fewer absorbers than the other wind models
because this wind model (i) produces less metals overall owing to its
greater suppression of star formation and (ii) injects more of its
metals into the low-density WHIM where they are unobservable using
these transitions.  Compared to vzw, the sw model creates fewer $\CIV$
and $\SiIV$ absorbers, which we will show in \S\ref{sec:absphyscond}
trace the condensed phase within haloes that are enriched to higher
levels by vzw winds.  At the same time there exist more weak $\OVI$
and $\NeVIII$ lines in the sw model, because these arise from the
diffuse phase that is enriched to higher levels via sw winds (cf. vzw
and sw $z=0$ metallicity-density relations in Figure \ref{fig:rhoZ}).

We discuss turbulence as a model variation in \S\ref{sec:turb}, but we
present it here because this is part of our favoured model for $\OVI$
observations.  We also show the effect on $\CIV$, $\SiIV$, and
$\NeVIII$, although we argue later that this model may not be
applicable to absorbers tracing higher densities.  As in OD09, adding
$b_{\rm turb}<60\kms$ (see Equations \ref{eqn:bturb1} \&
\ref{eqn:bturb2} in \S\ref{sec:turb}) increases the number of
absorbers above 100 m\AA~, allowing us to fit the observed $\OVI$
EWDs, and it also blends some weaker components into stronger ones,
thus reducing their numbers (cf. vzw and vzw-turb model in the top
panel of Figure \ref{fig:EW}).  $\CIV$ and $\SiIV$ also show a
dramatic increase in the number of $\ge 100$ m\AA~components if we
apply this same density-dependent model.

The simulations presented here produce statistically identical $\OVI$
results to OD09 for the vzw model.  However, the vzw model appears to
over-produce the observed $\CIV$, while the cw model provides a better
fit to this data \citep{coo10}\footnote{Kathy Cooksey kindly provided
redshift frequency distribution fits for $\CIV$ and $\SiIV$, which can
be found at http://www.ucolick.org/$\sim$xavier/HSTCIV/ and
http://www.ucolick.org/$\sim$xavier/HSTSiIV/}. We predict that COS
will make many $\NeVIII$ detections, especially between 10-20 m\AA.
In general, the lithium-like ions are harder to detect for higher
atomic numbers corresponding to higher ionisation species, because
both their oscillator strengths and wavelengths decline, leading to
smaller equivalent widths for a given column density.

\subsection{Column Density Distributions}

Column densities provide a more direct accounting of the amount of
metals present than equivalent widths, but they are more difficult to
obtain from observations as COS often does not resolve individual
metal-line absorbers.  Nevertheless, \autovp~fits column densities $N$
and linewidths $b$ for every component.  Given that we have applied
COS instrumental characteristics (albeit with a perhaps optimistic S/N
ratio), these predictions should be representative of the level of
information obtainable from (the very best) COS spectra.  We emphasise
that these are distributions of inferred column densities obtained by
Voigt-profile fitting components and summing them together into
systems, and not necessarily distributions of true column densities of
physically identified, unblended components.

\begin{figure*}
  \subfigure{\setlength{\epsfxsize}{0.45\textwidth}\epsfbox{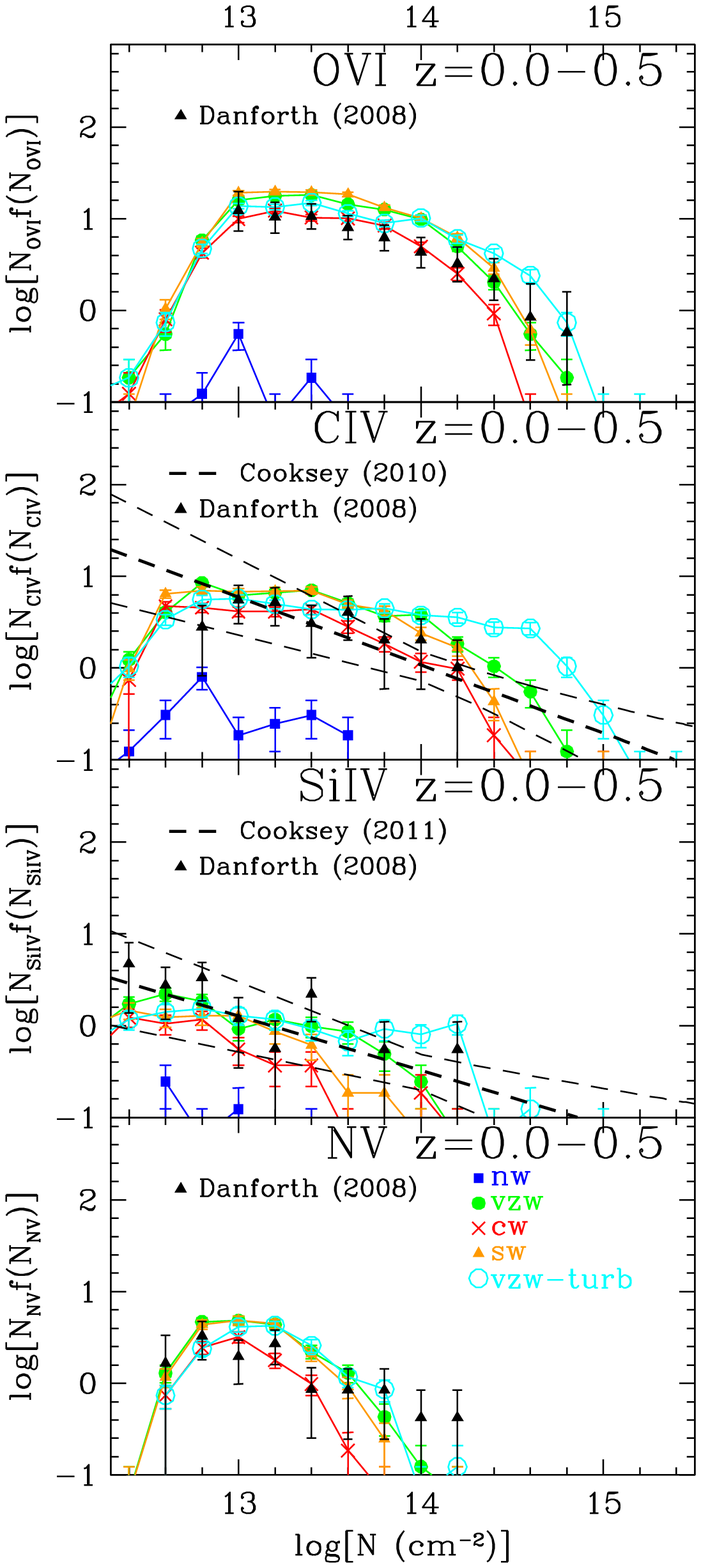}}
  \subfigure{\setlength{\epsfxsize}{0.45\textwidth}\epsfbox{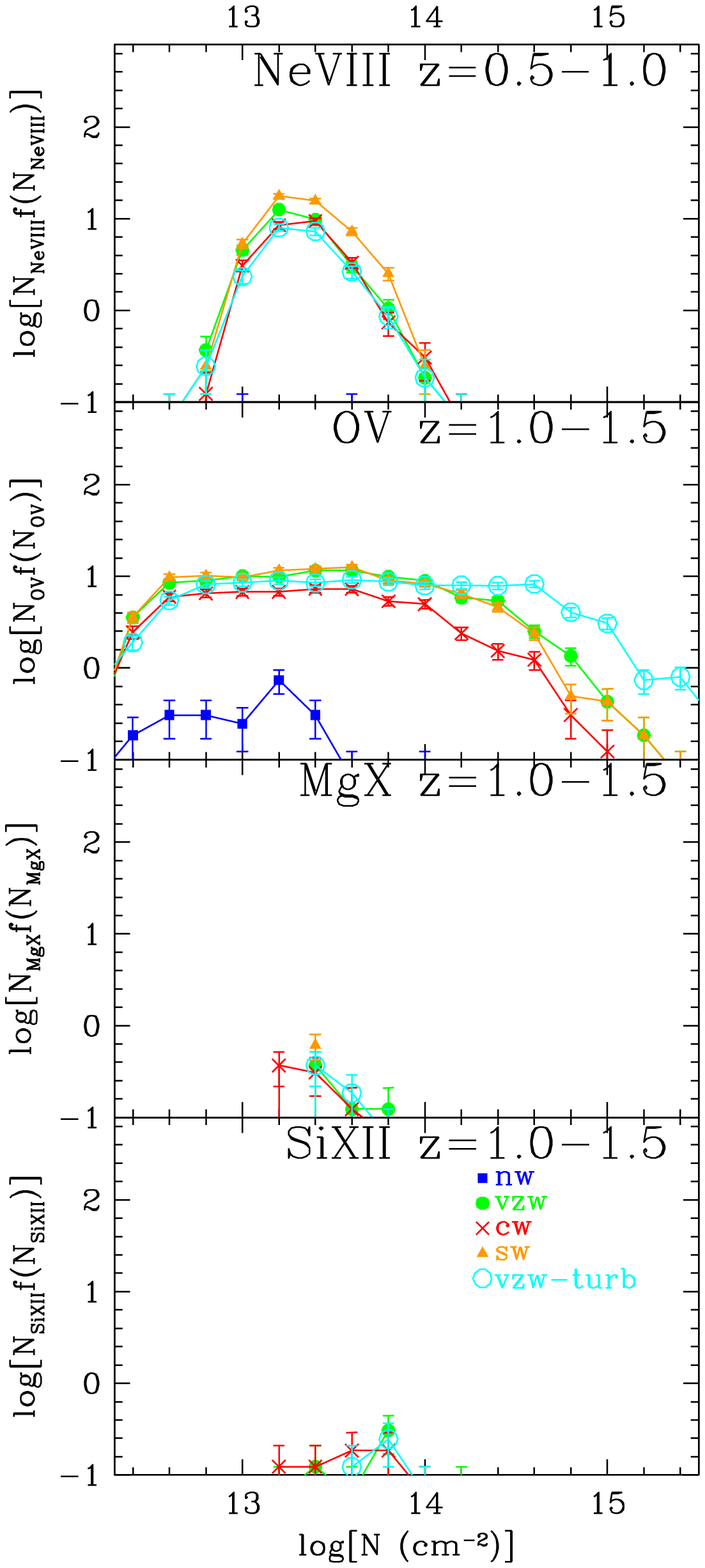}}
\caption{Column density distributions for our mock S/N=30 COS
  observations for six species: $\OVI$ 1032\AA, $\CIV$ 1548\AA,
  $\SiIV$ 1394\AA, $\NV$ 1239\AA, $\NeVIII$ 770\AA, $\OV$ 630\AA,
  $\MgX$ 609\AA, and $\SiXII$ 499\AA.  The four wind models and the
  turbulently broadened vzw model are generated from a path length of
  $\Delta z=35$ over a redshift range observable by the COS NUV
  channel.  Data from \citet{dan08} (triangles) and \citet{coo10,
  coo11} (dashed lines, thin lines encompass the 1-$\sigma$ error
  range) are also shown. Our simulations do not include AGB production
  of $\NV$, so these predictions should be taken as lower bounds.
  Error bars are 1-$\sigma$ Poisson errors.}
\label{fig:CDD}
\end{figure*}

We display the differential column density distributions (CDDs) of
systems in Figure \ref{fig:CDD}.  As in D10, we plot $f(N_{\rm ion})
\equiv d^2n/dN_{\rm ion} dz$ multiplied by $N_{\rm ion}$ to improve
the readability and to provide a dimensionless number that indicates
the relative number of lines per unit redshift for a given $N_{\rm
  ion}$.  Again our four basic wind models are plotted, along with
turbulent broadening added to the vzw model.  We show eight species
over $\Delta z=0.5$ for a redshift range that falls within the COS
far-UV (FUV) channel: $\OVI$, $\CIV$, $\SiIV$, \& $\NV$ all at
$z=0-0.5$, $\NeVIII$ at $z=0.5-1.0$, and $\OV$, $\MgX$, \& $\SiXII$ at
$z=1.0-1.5$.

The qualitative differences between the wind models and the data for
the column density distributions are similar to those obtained for the
equivalent widths as we discussed in the previous section.  Without
winds, very few IGM metal lines are present, and the predictions fall
well short of available observations.  The $\OVI$ CDD is consistent
with the data \citep{dan08}\footnote{Charles Danforth kindly provided
slightly updated versions of their data for $\OVI$, $\CIV$, $\SiIV$,
and $\NV$.} for the vzw and vzw-turb models, but the same models
produce too many $\CIV$ absorbers, especially above $10^{14} \cms$,
compared to the observations \citep{coo10}.  This disagreement is
statistically significant, and further data from COS should provide
much better constraints on weak $\CIV$ absorbers.  $\SiIV$ predicts
more strong lines and less weak lines than those observed
\citep{coo11}, though the discrepancy is not as large as with $\CIV$,
and the $z=0-1.2$ redshift range of this data set is larger and
centred at much higher redshift than in our simulations ($\langle z
\rangle \sim 0.9$ vs. 0.25).  Despite the different redshifts for
$\SiIV$, our simulated $\SiIV$ CDDs remain statistically identical
with $\langle z \rangle = 0.75$ and $1.25$ over the column density
range of detected absorbers (not shown).

We show $\NV$ CDDs but caution that we do not follow the unique
chemodynamic origin of nitrogen from secondary enrichment sources (AGB
stars).  Hence, the Type II SNe origin assumed here for nitrogen
likely represents a lower limit, as nitrogen is copiously produced in
AGB ejecta \citep[e.g.][]{mar01}.  Our models underestimate the
\citet{dan08} statistics for higher $\NV$ columns, which appears to
support this interpretation.

For $\NeVIII$ (upper right panel of Figure \ref{fig:CDD}), pushing
below $10^{13.8} \cms$ is rewarding because the CDD rises sharply.
At column densities $\ga 10^{14} \cms$, the line counts are predicted
to be well under one per unit redshift, but at $\sim 10^{13} \cms$
there should be $5-10$ per unit redshift.  Hence high S/N is essential
to detect significant numbers of these absorbers.  Note that our
estimates are probably optimistic because we assume detection based
on only the stronger component, but in actuality detection of the
weaker 780\AA~component is usually required for a robust line
identification.

We predict $\OV$ should be generously abundant in the spectra of $z
\ga 1$ quasars probed by COS over a wide range of column densities.
We show the 630 \AA~$\OV$ statistics, because it is likely
representative of a large number of lower ionisation species with
far-UV transitions that should contribute significantly to intervening
absorption in higher redshift quasars probed by COS.  $\OV$ should be
among the strongest owing to a large oscillator strength and oxygen
being the most abundant metal.

$\MgX$ and $\SiXII$ absorbers are very rare in our sight lines and
definitely require winds to be present at all.  We find only 2 $\MgX$
and 6 $\SiXII$ absorbers over $\Delta z=35$ with $W\ge 10$ m\AA~in the
vzw model.  We argue in \S\ref{sec:absphase} that we may not be
resolving the conductive interfaces from which very high-ionisation
lines such as $\NeVIII$ and $\SiXII$ may arise, and hence that these
absorbers may be more common than our current models predict.

Looking at the vzw-turb model, in theory the total column density
should be conserved (unlike $W$) when adding turbulence in the way
that we do using {\tt specexbin}.  But as one can see, in practice the
column densities are increased when turbulence is added.  This occurs
because for strong saturated lines, which preferentially receive the
largest turbulent contributions, \autovp~tends to fit a higher column
density given only a single line with no doublet information.  Also,
weaker lines may become blended or harder to detect.  Hence, there is
a noticeable increase in the predicted line frequencies especially for
$\CIV$ at $10^{14.0} \cms$ and $\SiIV$ at $10^{13.8} \cms$, which
correspond to the column densities where these transitions fall off
the linear curve of growth and begin to become saturated without
turbulent broadening.  COS spectra will be able to use the two doublet
transitions to perform a curve of growth analysis and obtain more
accurate $N$s for saturated components.  Including such information in
\autovp~may mitigate the increase seen at high-$N$; we leave such an
analysis for a more careful comparison to real data.

\subsection{Cosmic Ion Densities}

Integrating over a CDD or summing the column densities of lines
provides a global view of the amount of cosmic metal absorption.
To obtain the cosmic density of an ion, we sum up all the absorber
column densities between $10^{12-15} \cms$ using

\begin{equation} \label{equ:omega}
\Omega({\rm ion}, z) = {H_0 m_{\rm metal} \over c \rho_{crit}} {\Sigma N({\rm ion},z) \over \Delta X(z)},
\end{equation}

\noindent where $H_0$ is the Hubble constant, $c$ is the speed of
light, $\rho_{crit}$ is the critical density of the Universe, and
$m_{\rm species}$ is the atomic weight of the given species.  $\Delta
X(z)$ is the pathlength in a $\Lambda$CDM Universe where
$X(z)=\frac{2}{3\Omega_{\rm M}}\sqrt{\Omega_{\rm M} (1+z)^3+
\Omega_{\Lambda}}$.  We evaluate Equation \ref{equ:omega} from the
Voigt-profile fitted column densities, just as is done for observations.
We choose the column density range both for historical
reasons~\citep{son01}, and because it quantifies lower-column metal
lines that are expected to arise in the IGM as opposed to those
within the galactic ISM.

\begin{figure}
  \subfigure{\setlength{\epsfxsize}{0.45\textwidth}\epsfbox{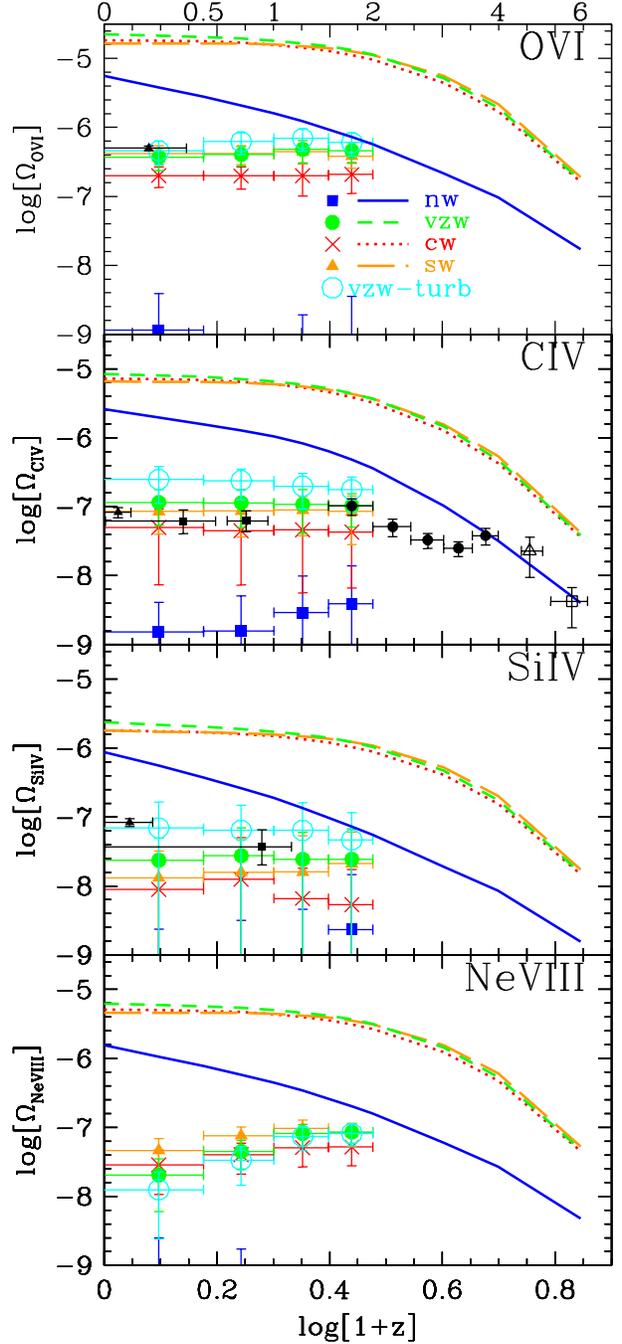}}
  \caption[]{Cosmic ion density evolution of $\OVI$ (top), $\CIV$
    (second), $\SiIV$ (third), and $\NeVIII$ (bottom) for the same
    models described in Figures \ref{fig:EW} and \ref{fig:CDD}.  Large
    symbols show simulated $\Omega_{\rm ion}$s calculated from S/N=30
    COS spectra, with horizontal error bars indicating $\Delta z=0.5$
    bins and vertical error bars indicating the variance among $\Delta
    z = 0.5$ sight lines.  Continuous lines show the evolution of
    $\Omega_{\rm metal}$ calculated by summing up all the mass outside
    galaxies of oxygen (top panel), carbon (second), silicon (third),
    and neon (bottom).  Data are from \citet{dan08} (solid triangles),
    \citet{coo10,coo11} (solid squares), \citet{dod10} (solid
    circles), \citet{pet03} (open triangle), and \citet{rya09} (open
    square) and are all scaled to a $\Omega_{\rm M}=0.28$,
    $\Omega_{\rm \Lambda}=0.72$, $H_{0} = 70 \kms$ cosmology.
    Redshifts are listed on the top axis.  }
\label{fig:omega}
\end{figure}

We bin the $\Omega_{\rm ion}$s in bins of $\Delta z=0.5$ and plot
the points with error bars in Figure \ref{fig:omega} for $\OVI$,
$\CIV$, $\SiIV$, and $\NeVIII$.  For simplicity, we use the same
COS FUV instrument parameters for the entire $z=0-2$ range even
though most redshifts fall outside of the COS wavelength range.  We
show the same four wind models as before, and also the vzw model
with turbulent broadening.  We also plot $\Omega_{\rm metal}$
computed by summing up the total mass density of all the SPH particles
outside of galaxies ($n_{\rm H}<0.1 \cmc$) for the appropriate metal
species (continuous lines).

$\Omega_{\rm ion}$s usually do not change by much if we adjust the
column density bounds from $10^{12}$ to $10^{13} \cms$ or from
$10^{15} \cms$ to any higher value, because the vast majority of an
ion's cosmic density resides in $10^{13-15} \cmc$ components in S/N=30
spectra.  The most dramatic exception is $\CIV$ at $z\ga 1$, which can
increase by a factor of more than two when including lines above
$N_{\CIV}>10^{15} \cms$.

Models with outflows produce roughly constant $\Omega_{\OVI}$ and
$\Omega_{\CIV}$ from $z=2\rightarrow 0$, and agree with observations
where available.  While $\OVI$ agrees with the data in all three
observables thus far explored, $\CIV$ only agrees for $\Omega_{\CIV}$
because too many weak and too few strong absorbers exist compared to
the current data \citep{coo10}.  In general, the non-evolution in
these species reflects the non-evolution in the total mass densities
of metals outside of galaxies ($\Omega_{\rm metal}$s, lines in Figure
\ref{fig:omega}), but the ion densities are shifted lower by a factor of
$\sim30-100$.  The majority of the Universe's metals are produced
after $z=2$, but this is offset by more metals remaining nearer to galaxies
and being reincorporated into stars, i.e.  ``outside-in" enrichment.
Note that the ratio $\Omega_{\rm metal}/\Omega_{\rm ion}$ cannot be
treated as a global ionisation correction for an ion (e.g. C/$\CIV$ or
O/$\OVI$), because we will show in \S\ref{sec:omegacorr} that
$\Omega_{\rm ion}$ calculated using Equation \ref{equ:omega} does not
recover the total cosmic density of that ion.  

In contrast to the other ions shown, $\Omega_{\NeVIII}$ shows a steady
decline from $z=1.5\rightarrow 0$ for the vzw model.  The photo-ionised
$\NeVIII$ component is within the detection limits of COS at higher
redshift, and becomes harder to detect at low redshifts at wavelengths only 
accessible using the {\it Space Telescope Imaging Spectrograph} (STIS)
and the {\it Far UV Spectroscopic Explorer} (FUSE).  Observing $\NeVIII$
at wavelengths between 1100-1700\AA~with COS could be ideal to trace the
peak in $\Omega_{\NeVIII}$.

The turbulent vzw model predicts higher $\Omega$s than the vzw model
even though the total column density theoretically should be
preserved.  As explained previously, the broader profiles with
turbulent broadening are fitted more accurately by \autovp~than
unbroadened thin saturated profiles.  This is especially true if
the cosmic density of an ion is dominated by a few strong absorbers.

In summary, outflows are generically required to enrich the IGM as
observed, and there is a mild sensitivity in observations of $W$, $N$,
and $\Omega_{\rm ion}$ to the particular form of galactic outflows
employed.  While momentum-conserved winds plus turbulence best
matches the $\OVI$ data (as it was in part designed to do), it
overproduces $\CIV$ and $\SiIV$ absorption.  $\NeVIII$ shows a
particularly steep column density distribution owing to a
preponderance of weak photo-ionised lines, motivating deep
observations to detect this interesting IGM tracer.  The cosmic metal
density is generally constant from $z\sim 2\rightarrow 0$, in contrast
to its marked rise from $z\sim 6\rightarrow 2$~\citep{opp06,opp08},
and this is broadly reflected in the constancy of $\Omega_{\rm ion}$
for most ions over this redshift range.

\subsection{Component Linewidths} \label{sec:linewidths}

\begin{figure}
  \subfigure{\setlength{\epsfxsize}{0.44\textwidth}\epsfbox{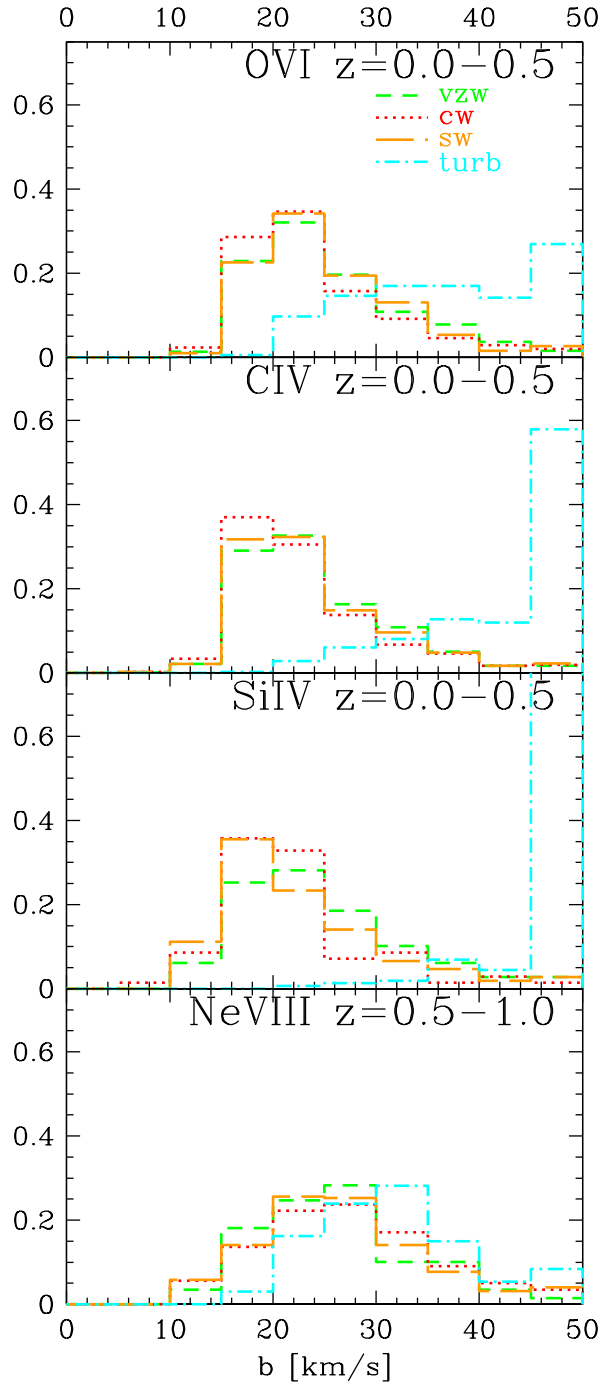}}
  \caption[]{Simulated COS linewidth histograms for $\OVI$ (top),
    $\CIV$ (second), and $\SiIV$ (third) between $z=0-0.5$ for all
    components with $W\ge 30$ m\AA~and $\NeVIII$ (bottom) for
    components between $z=0.5-1.0$ and $W\ge 10$ m\AA.  The three
    models with winds and the turbulently broadened vzw model are
    generated from a path length of $\Delta z=35$ over a redshift
    range observable by the COS NUV channel.  Note that these
    linewidths are {\it not} corrected for the instrumental broadening
    by the COS LSF, which has $b \approx 24\kms$.}
\label{fig:bparam}
\end{figure}

Component linewidths ($b$-parameters) are also fit with \autovp, but
are almost always dominated by the COS LSF (the instrumental $b\sim 24
\kms$); therefore, linewidths are often much wider than they are in
STIS data.  We show the linewidth histograms for the simulations with
winds in Figure \ref{fig:bparam} along with the vzw-turb model, which
uses the turbulent broadening model motivated by OD09 to match the
observed $b$-parameters of STIS $\OVI$ absorbers, but now applied to
all ions.  The histograms apply to all $\OVI$, $\CIV$, and $\SiIV$
absorbers with $W\ge 30$ m\AA~between $z=0-0.5$, and $\NeVIII$
absorbers with $W\ge 10$ m\AA~at $z=0.5-1.0$.

Without turbulent broadening, the linewidths of the three wind models
are nearly identical.  The COS LSF makes it difficult to determine
temperatures from thermal broadening in our case since we show in the
next section that the majority of our absorbers are near $T\sim 10^4$ K,
corresponding to linewidths much smaller than the LSF.  We do not show
the nw model owing to its low number statistics.

Adding turbulence using the heuristic prescription of OD09
dramatically alters the $b$-parameter histograms for not just $\OVI$,
to which this prescription was calibrated, but even more so for $\CIV$
and $\SiIV$.  {\sc AutoVP} fits a maximum $b = 50 \kms$, which means
that the majority of $\CIV$ and $\SiIV$ absorbers in the 45-50 $\kms$ bin
may even be wider.  Unlike thermal broadening, turbulent broadening is
independent of the atomic species, and the broader linewidths for
$\CIV$ and $\SiIV$ are a result of the OD09 turbulence model, which
adds wider $b_{\rm turb}$ to higher density gas.  We argue in
\S\ref{sec:turb} that such wide absorbers likely represent a
break-down of the turbulence model, because while there are few
published observed linewidths of these species, thinner line profiles
are apparent in the data \citep[e.g.][]{coo10}.  $\NeVIII$ absorbers
are not as sensitive to the turbulence model, because they arise from
lower densities where $b_{\rm turb}$ is small.

\subsection{Components \& Systems} \label{sec:compsys}

\begin{figure}
\subfigure{\setlength{\epsfxsize}{0.45\textwidth}\epsfbox{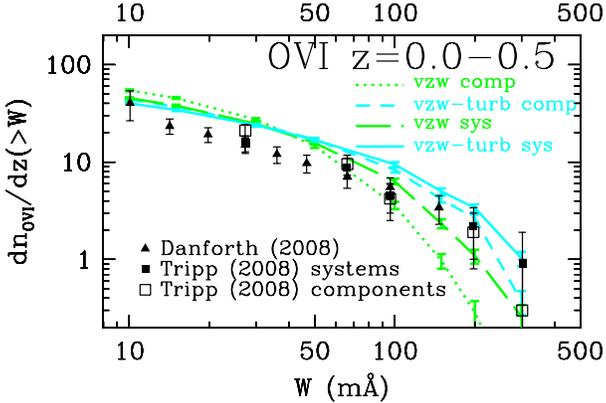}}
\caption[]{Cumulative equivalent width distributions of $\OVI$
  components and systems quantified in our simulations for the vzw and
  vzw-turb models.  Observations of \citet{tri08} have both system
  (filled square) and component (open square) distributions.  Data
  from \citet{dan08} (triangles) groups components based on the
  associated $\HI$ identification.}
\label{fig:EWcompsys}
\end{figure}

The choice to group components into systems or treat them individually
is an important consideration when comparing simulations to
observations.  K. Cooksey (private communication) advises us that it
is more proper to compare grouped systems to the data of
\citet{coo10,coo11}, because they group components into systems if
they lie within a few tens of $\kms$.  \citet{dan08} groups metal-line
components together into systems if their associated $\lya$ lines lie
within $50 \kms$ of each other (C. Danforth, private communication).
\citet{tri08} publish both component and system EWDs, which we show in
Figure \ref{fig:EWcompsys}.

The vzw model indicates how important it is to compare the appropriate
component and system EWDs.  The vzw system EWD would agree with the
component EWD of \citet{tri08}, but the appropriate comparison, the
vzw component EWD, under-predicts the strongest observed $\OVI$
components.  The shortfall of strong $W$ components was part of the
motivation for OD09 to attempt the vzw-turb model (the other
motivation being mismatched linewidths).  The vzw-turb makes less of a
difference between the two types of EWDs, because turbulent broadening
increases the component $W$s so that the strongest component of a
system becomes more dominant relative to weaker components; thus the
strongest component more likely sets the system $W$ resulting in less
differences between the two methods.  The trend of this model's
component and system EWDs matches that of \citet{tri08}, although we
appear to consistently over-estimate component and system frequencies
at $\le 100$ m\AA.

The comparison to \citet{dan08} is more difficult because their search
relies on the identification of the associated $\HI$ absorption, but
their measurements appear consistent with \citet{tri08} systems.  We
also tested the choice of grouping components within 250 $\kms$
instead of 100 $\kms$, but this makes almost no difference and gives
statistically identical results.  If components lie close to each
other, they are almost always within 100 $\kms$ of another component.

\section{Metal-Line Absorber Physical Conditions} \label{sec:absphyscond}

We now link the observed metal-line absorber characteristics with
their physical conditions just as we did for $\lya$ absorbers in \S5
of D10.  We concentrate on ion mass density-weighted physical
parameters, mainly hydrogen density and temperature, using the same
method described in OD09.  This weighting procedure sums the ion mass
density, and hence absorption, of every SPH particle contributing at
the line centre of an identified absorber, accounting for peculiar
velocities and temperature broadening.  This procedure is similar to
the optical depth-weighting of \citet{tep11} with the only difference
being that they sum physical quantities from each pixel across a line
profile whereas we use the value at the line centroid.  \citet{tep11}
find that the difference between these two methods is small, so the
methods should be comparable.  We consider all absorbers where the
strongest line transition has $W \ge 10 $ m\AA~in our 70 S/N=30 COS
sight lines.

\subsection{Absorbers in Phase Space} \label{sec:absphase}

\begin{figure*}
  \subfigure{\setlength{\epsfxsize}{0.45\textwidth}\epsfbox{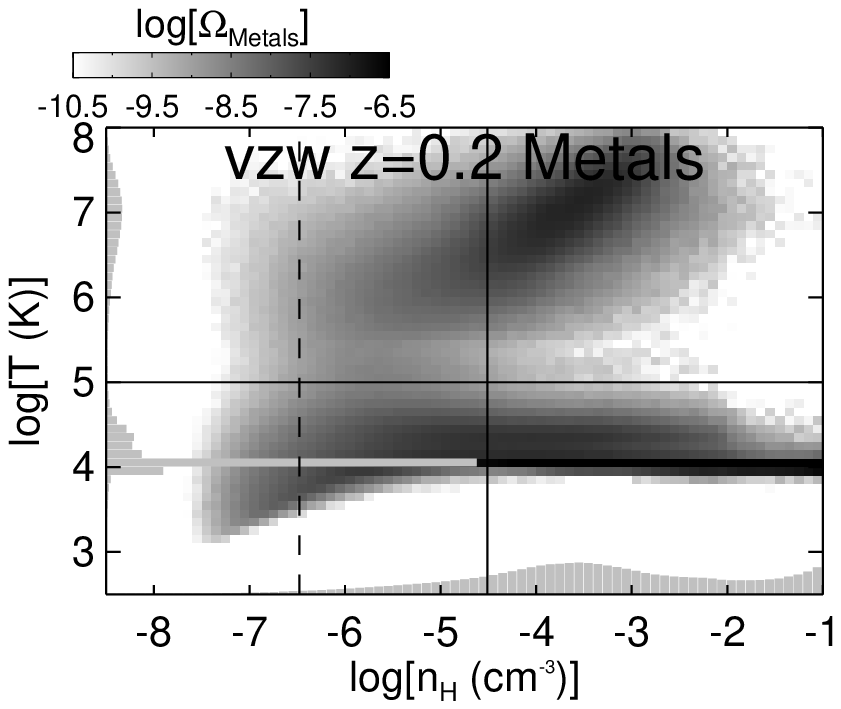}}
  \subfigure{\setlength{\epsfxsize}{0.45\textwidth}\epsfbox{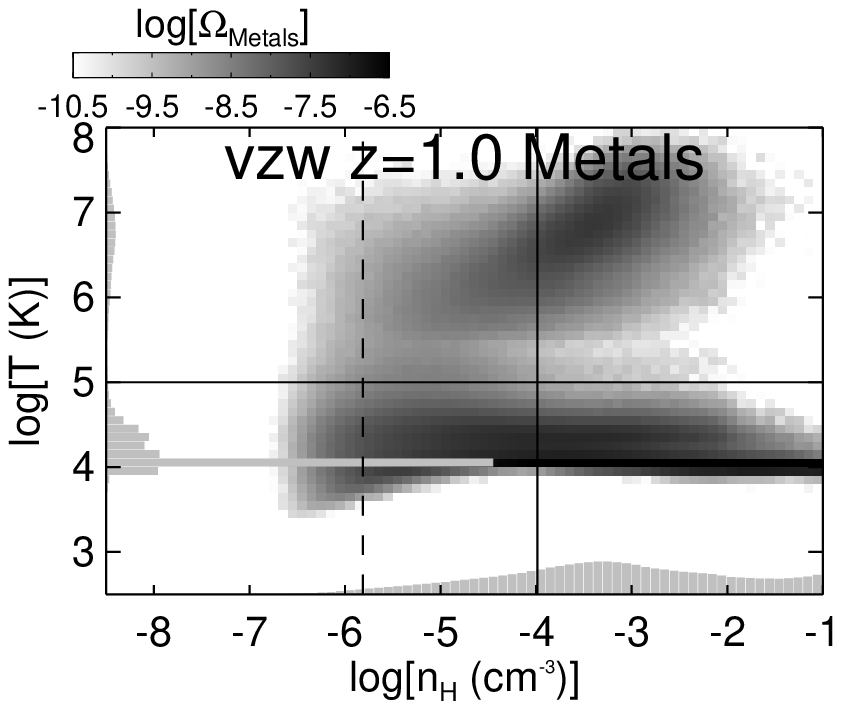}}
  \subfigure{\setlength{\epsfxsize}{0.45\textwidth}\epsfbox{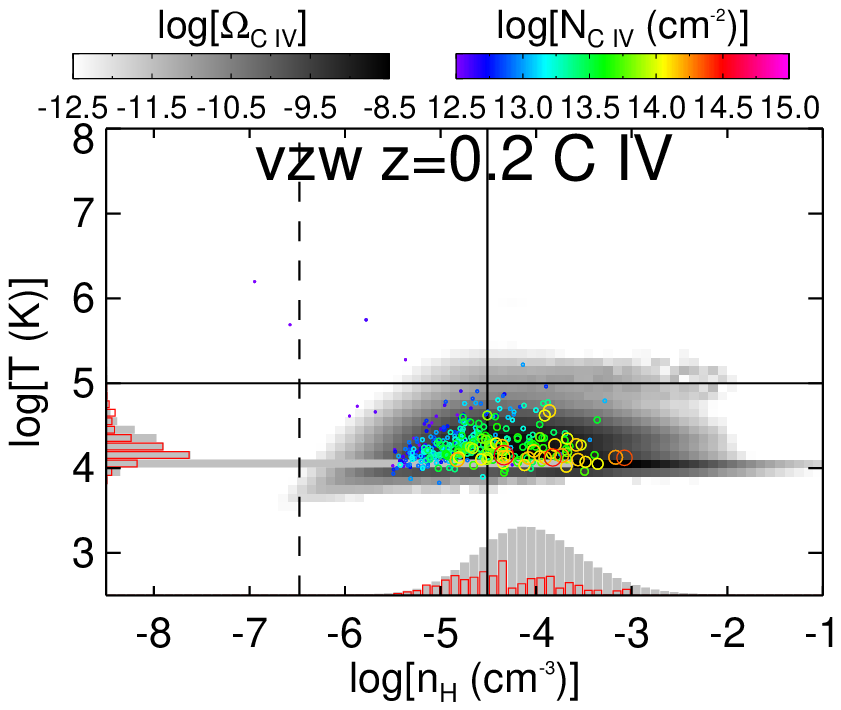}}
  \subfigure{\setlength{\epsfxsize}{0.45\textwidth}\epsfbox{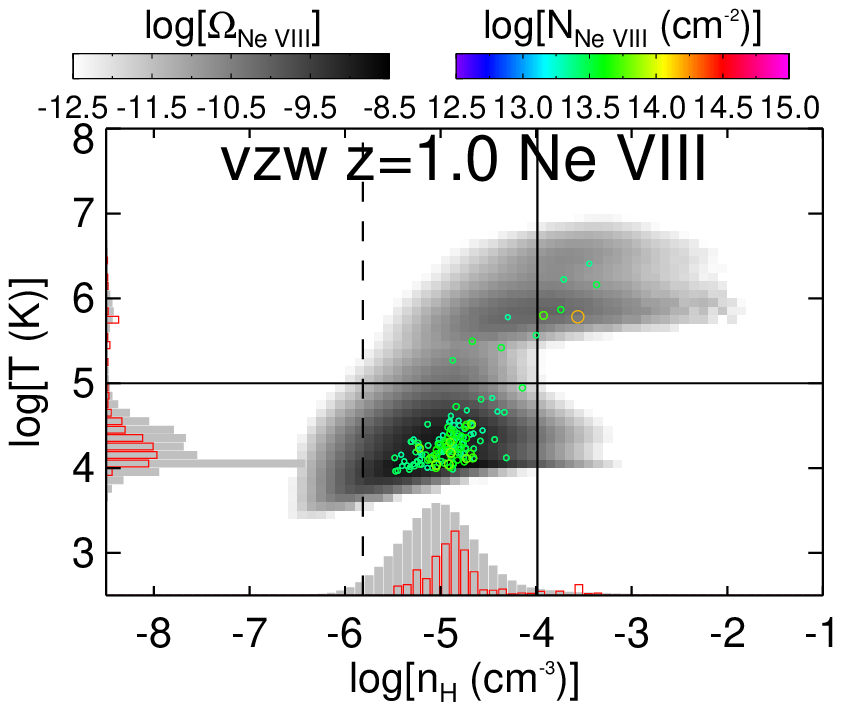}}
  \subfigure{\setlength{\epsfxsize}{0.45\textwidth}\epsfbox{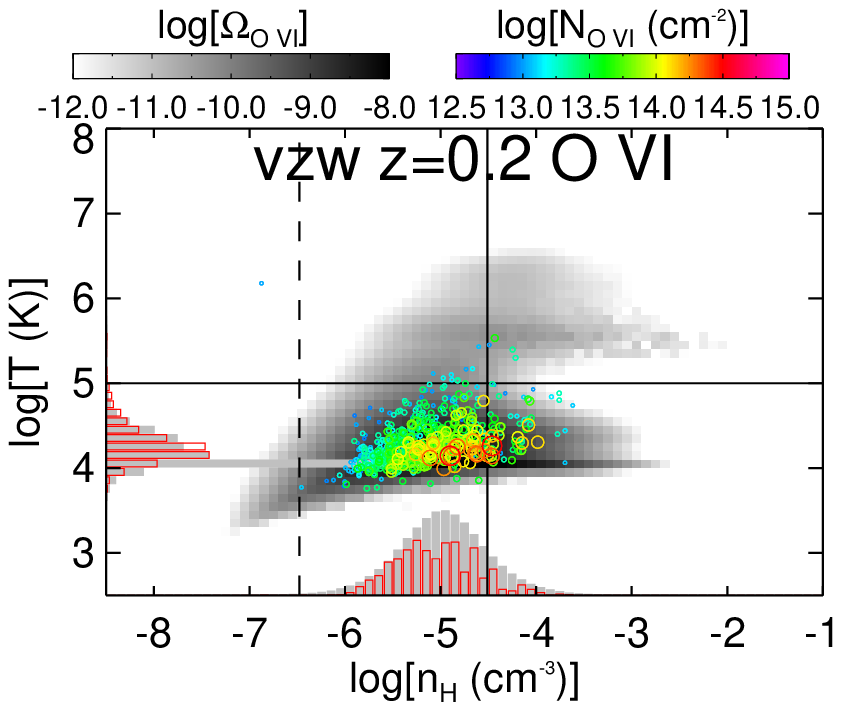}}
  \subfigure{\setlength{\epsfxsize}{0.45\textwidth}\epsfbox{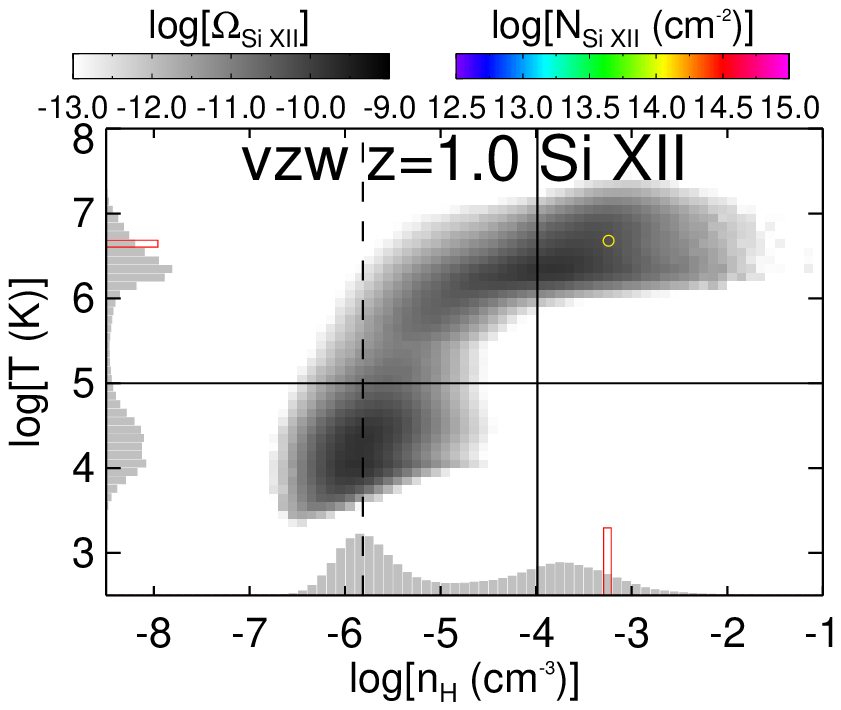}}
  \caption{$\Omega$ phase space diagrams at two redshifts (left-
    $z=0.2$, right- $z=1.0$) for all metals (top) and two common ion
    species observable with COS at each redshift (centre and bottom)
    from the r48n384vzw model.  Grey shading indicates the fractional
    $\Omega$ summed from all SPH particles binned in $0.1\times0.1$
    dex pixels, and grey histograms show the distributions in density
    and temperature for the metals and each ion species.  Overlaid
    coloured circles are simulated COS absorbers over $\Delta z=0.2$
    for 70 sight lines where S/N=30.  Colour and symbol size scale
    with absorber column density.  Red histograms count the summed
    $\Omega$s of absorbers as functions of density and temperature,
    calibrated to the same scale as the grey histograms using Equation
    \ref{equ:omega} and are directly comparable to test how well
    simulated observations recover the total ion density contained in
    the simulations.  Solid lines correspond to the division of the
    four phases as in Figure \ref{fig:Zphasespaces}, and the dashed
    vertical lines indicates mean overdensity.}
\label{fig:Zionphase}
\end{figure*}

Figure \ref{fig:Zionphase} displays total metals (top panels) and
selected ions (middle and bottom panels) in density-temperature phase
space at $z=0.2$ (left) and $z=1.0$ (right).  The grey pixelated
shading is the fractional $\Omega$ summed from SPH particles binned in
$0.1\times 0.1$ dex pixels.  The temperature bimodality of metals
described in \S\ref{sec:rhotcond} is obvious at both redshifts, with
the metals pushing to lower densities and higher temperatures with
time.

We show the phase space occupied by $\CIV$ and $\OVI$ absorption at
$z=0.2$ as the grey shading in the middle left and bottom left panels,
respectively.  These ions trace gas in a limited range in phase space,
particularly weighted towards the cooler gas.  Overlaid using coloured
open circles are the ion mass density-weighted hydrogen densities and
temperatures of simulated absorbers over a pathlength of $\Delta z=
14$ (70 sight lines over $z=0.1-0.3$).  Colour and circle size both
scale with the absorber column density as indicated by the legend
above each panel.

Our identified absorbers overlap with the darkest shading, which
provides a consistency check that our ion weighting accurately
captures the physical conditions of the absorbers.  The two sets of
histograms along the $x$-axis (for density) and $y$-axis (for
temperature) directly compare the SPH-summed quantities (solid grey
histograms) and the absorber-summed quantities (open red histograms);
the amplitudes of these histograms can be directly compared.  For
example, the mismatch of $\CIV$ density histograms in denser halo gas
indicates that summing the halo $\CIV$ absorbers (red) underestimates
the true $\Omega_{\CIV}$ arising from this phase (grey).  The
temperature histograms show that most of this missing absorption is
between $10^{4.0-4.1}$ K.  The same is true for $\SiIV$ probing even
higher densities (not shown).  The main cause for this is that
saturated absorbers tracing halo gas have their column densities
underestimated without turbulent broadening, which is apparent in the
differences between the vzw and vzw-turb CDDs in Figure \ref{fig:CDD}.
The mismatch is less severe but still noticeable for $\OVI$.

At $z=0.2$, both $\CIV$ and $\OVI$ predominantly trace the cooler
phases of gas at temperatures closer to $10^4$ K than $10^5$ K.  $\NV$
(not shown) traces the same temperatures, but at intermediate
overdensities between $\CIV$ and $\OVI$.  For $\OVI$, this is in
marked contrast to some previous work \citep{cen11, smi11, tep11}, all
of whom find most of their $\OVI$ at $T>10^5$ K at the same redshifts.
We believe it is the interplay of our momentum-conserved winds with
the cooling rates in the $T\sim 10^5$~K regime that results in our
differing conclusion.  We will show in \S\ref{sec:Zcool} that using
the photo-ionised metal-line cooling rates of \citet{wie09a} does not
fundamentally change the phase space traced by $\OVI$ absorbers in our
vzw model.  Hence, we reiterate the conclusion by \citet{opp09a} that,
for a physically well-constrained outflow model, $\OVI$ primarily
traces diffuse, photo-ionised metals.

We show two higher ionisation species, $\NeVIII$ and $\SiXII$, at
$z=1.0$, because this is the redshift where COS is able to observe
these lines.  The $\SiXII$ doublet at 499.3,521.1\AA~falls into the
COS sensitivity range at $z>1.1$, but we show simulated absorbers and
phase space centred at $z=1.0$ because these characteristics are not
strongly evolving.  $\SiXII$ has a larger fraction of mass arising
from collisional ionisation at densities corresponding mostly to halo
gas.

Surprisingly, most simulated $\NeVIII$ absorbers are predicted to be
photo-ionised, tracing overdensities of around $\delta \sim 10$.  However,
the strongest simulated absorber is at $N_{\NeVIII}=10^{14.2} \cms$
tracing $T=10^{5.8}$ K gas, and is likely more representative of the
three $\NeVIII$ absorbers \citep{sav05, nar09, nar11} already observed,
since it has a similar column density and hence probably a similar environment.
\citet{mul09} find a 0.25 $L^*$ galaxy at an impact parameter of 73
$\hkpc$ at nearly the same redshift as the $z=0.207$ $\NeVIII$ absorber
along the HE0226-4110 sight line, suggesting that this absorption
originates in collisionally ionised halo gas.  They further suggest
that the origin is a conductive interface between cool clouds moving
through a hot medium, based on the alignment of $\NeVIII$ with lower
ionisation species, something that would not be captured in our simulations.
\citet{nar11} note the existence of a 0.08 $L^*$
galaxy at 110 kpc at $\delta v = 180 \kms$ from the $\NeVIII$ absorber
at $z=0.495$ in the PKS0405-123 sight line, and further argue this is
likely tracing $T\sim 10^{5.7}$ K gas either in a hot halo or nearby WHIM.

We emphasise that our model predicts numerous weaker $\NeVIII$ absorbers
tracing the diffuse photo-ionised component, as a separate and probably
as yet undetected population.  These absorbers are mainly below
$N_{\NeVIII}=10^{13.7} \cms$ (cf. $10^{13.85}$, $10^{13.98}$, $10^{13.96}
\cms$ for the discovered systems mentioned above) and should arise
outside haloes.  We cannot distinguish any other obvious observational
characteristics between the two types of $\NeVIII$ absorbers, as the
simulated $b$-parameters at COS resolution do not straightforwardly translate to
temperature.  Alignment with $\HI$ absorbers tracing overdensities of
$\sim 10$, i.e. $N_{\HI} \sim 10^{13.5} \cms$ at $z=0.75$, is common for
the photo-ionised $\NeVIII$ absorbers.  Upcoming deeper COS observations
might be able to detect this population.

The true frequency of hot $\NeVIII$ absorbers may exceed our
predictions given that our simulations do not adequately resolve
conductive interfaces within haloes.  \citet{nar09} estimates a higher
frequency below $z<0.5$, although their estimate is based on one
absorber and is highly uncertain.  Additionally, we predict a mass
density of $\Omega_{\NeVIII}$ at $z=1.0$ almost $3$ times higher than
at $z=0.2$, which is suggestive of an even greater frequency of such
absorbers at high redshift; the increase is primarily from the
increased neon traced in the photo-ionised component.  Given their
distinct phase spaces (hot halo and dense WHIM vs. diffuse), the
potential for absorber bimodality is greater than for $\OVI$, but
identifying it may require direct observation of an absorber's
environment.  We will return to this discussion in a future paper
discussing the galaxy-absorber connection.

\subsubsection{Photo-ionised Absorption}

\begin{figure}
  \subfigure{\setlength{\epsfxsize}{0.45\textwidth}\epsfbox{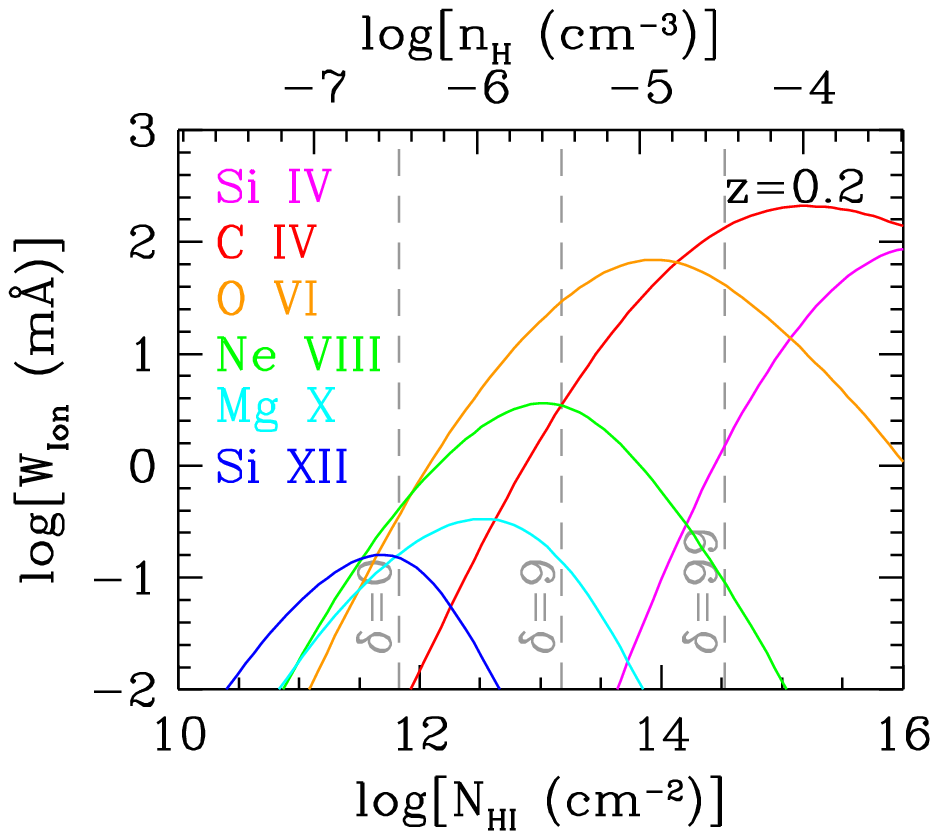}}
  \subfigure{\setlength{\epsfxsize}{0.45\textwidth}\epsfbox{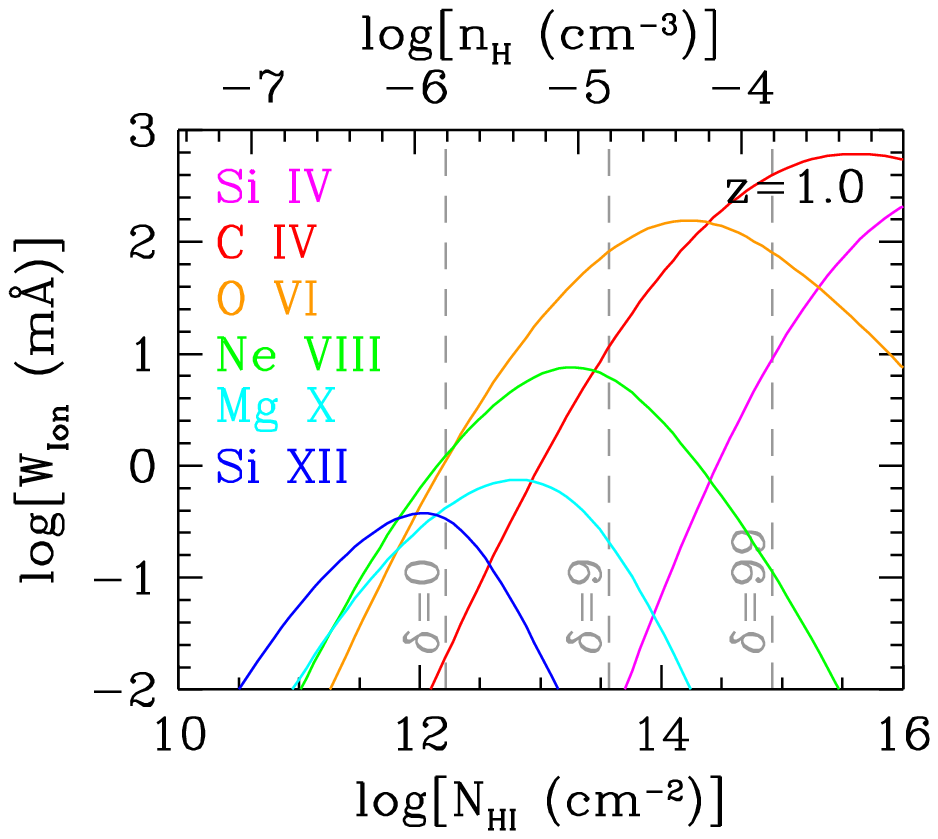}}
  \caption{Line Observability Indexes (LOXs) at $z=0.2$ and 1.0
    showing the photo-ionised equivalent widths for a variety of metal
    species as a function of $\HI$ column density assuming the
    density-$N_{\HI}$ relation from Equation 3 of D10, $Z=0.1
    \Zsolar$, $T=10^{4.2}$ K for the metal species, and a
    \citet{haa01} background renormalised to match the vzw $\lya$
    forest flux decrement (D10).  }
\label{fig:LOX}
\end{figure}

It may initially be surprising that high ionisation species such as
$\NeVIII$, $\MgX$, and $\SiXII$ can be excited via photo-ionisation
given their high ionisation potentials, but the existence of such
absorbers is straightforwardly traceable to the shape of the
\citet{haa01} background.  We demonstrate this in Figure
\ref{fig:LOX}, where we show the Line Observability Index \citep[LOX,
][]{hel98}-- the predicted equivalent width of metals lines associated
with $\HI$ if the IGM is uniformly painted with a metallicity of
$Z=0.1 \Zsolar$ and with a uniform temperature of $T=10^{4.2}$ K.
Although these assumptions are overly simplistic, they are
representative and serve to illustrate the main point.

At $z=1.0$, $\NeVIII$ peaks near $n_{\rm H}\sim 10^{-5.0} \cmc$ with
equivalent widths as high as 8 m\AA.  Our simulated photo-ionised
$\NeVIII$ absorbers are stronger than this owing to the inhomogeneous
enrichment, which results in higher metallicities for the detected absorbers,
and furthermore arise at slightly higher densities owing to the positive
metallicity-density gradient present in the vzw models \citep{opp06}.
Photo-ionised $\SiXII$ should exist around the mean density (assuming
the metallicity is sufficiently high), but it should not exceed even 1
m\AA, well below the detectability threshold of COS.  Photo-ionised $\MgX$
exists at higher overdensities but is still very weak.  A detection of
$\MgX$ or particularly $\SiXII$, while likely rare and exceedingly
difficult, would provide the most unambiguous UV tracer of hot halo
gas at $T\sim 10^{6-7}$ K.

\subsubsection{$\OVI$ in Other Wind Models}

\begin{figure}
  \subfigure{\setlength{\epsfxsize}{0.45\textwidth}\epsfbox{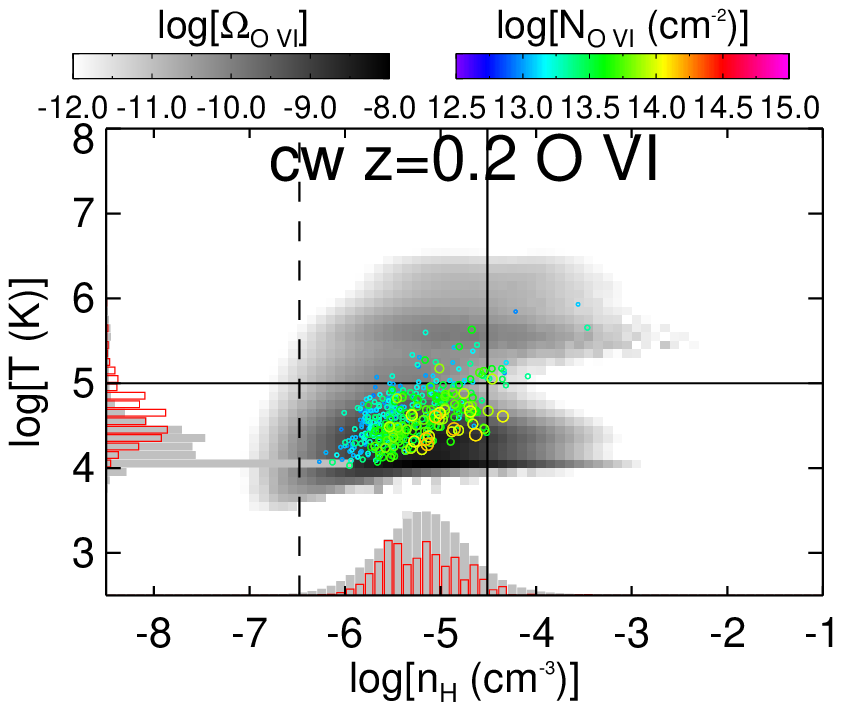}}
  \subfigure{\setlength{\epsfxsize}{0.45\textwidth}\epsfbox{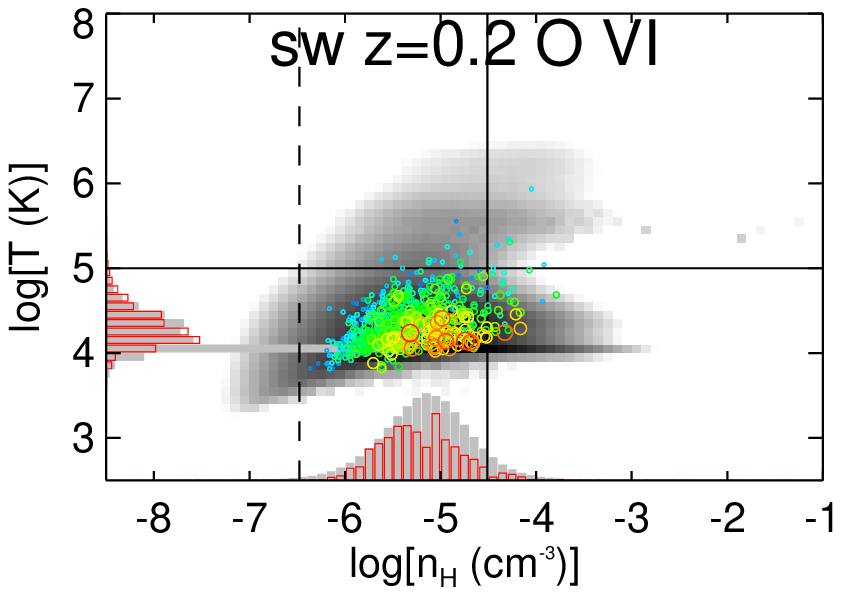}}
  \subfigure{\setlength{\epsfxsize}{0.45\textwidth}\epsfbox{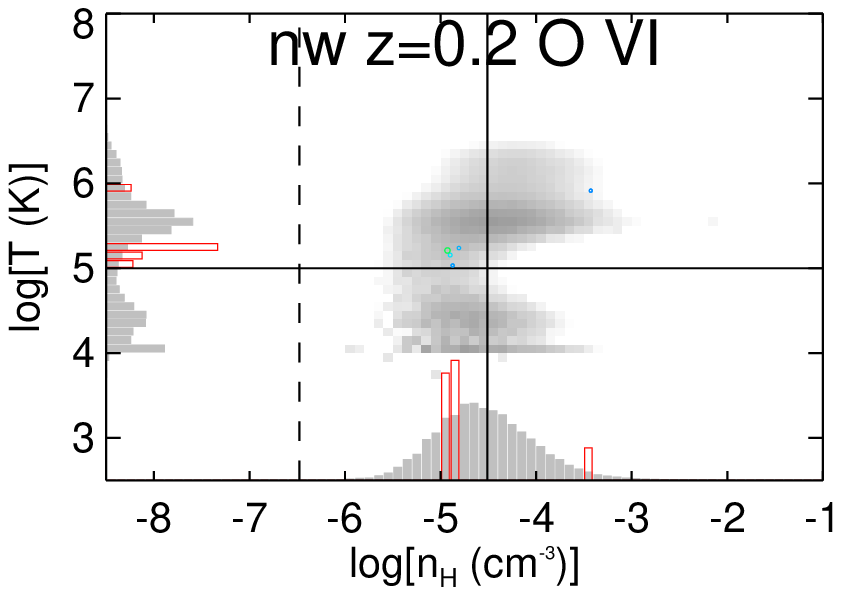}}
  \caption{$\OVI$ $\Omega$ phase space diagrams as described in Figure
    \ref{fig:Zionphase} for other wind models: constant winds (cw,
    $\vw=680 \kms$, $\eta=2$), slow winds (sw, $\vw=340 \kms$,
    $\eta=2$), and no winds (nw) for $z=0.2$ $\OVI$.}
\label{fig:Zmodelionphase}
\end{figure}

The phase space traced by $\OVI$ at $z=0.2$ is shown for the constant,
slow, and no-wind models in Figure \ref{fig:Zmodelionphase}.  While
the cw model appears qualitatively similar to vzw with most absorbers
being diffuse and photo-ionised, there are significantly fewer
absorbers at $N_{\OVI}>10^{14.3} \cms$.  Stronger outflows create
fewer strong absorbers, because metals preferentially reach lower
overdensities where they cannot rapidly cool.  Even though the IGM
is generally hotter in this model relative to vzw, absorbers still
remain predominantly photo-ionised.  Despite the metal content of
the cw WHIM being $\sim 6.4$ times higher than in the vzw model, the limited
phase space traced by the $\OVI$ does not probe this enrichment, which
can only be directly detected in species with X-ray transitions
(e.g. $\OVII$ and $\OVIII$).

Slow winds produce $\OVI$ statistics nearly indistinguishable from the
vzw model as reflected in the phase space plot.  This same simulation
produces a similar $z\sim 0$ galactic stellar mass function to the vzw
model below $M^*$ \citep{opp10}, thus the match with the $\OVI$ statistics
is consistent with sub-$M^*$ galaxies producing the $\OVI$ absorbers
(OD09).  The no-wind simulation shows that winds are required to
enrich the diffuse IGM; no absorbers exist in this phase.

Overall, the conclusion that the $\OVI$ detected by COS (at least under
the S/N assumptions here) will be predominantly photo-ionised is
robust to variations in the wind model and the cooling rates (as we will show
in \S\ref{sec:Zcool}).  Even $\NeVIII$, at columns just below what is
currently accessible, becomes dominated by photo-ionised absorbers.
$\SiXII$ at COS sensitivities probes hot gas, but not at low WHIM
densities, and anyway these lines should be rare.  A key conclusion from this
work is that the bulk of the diffuse collisionally ionised WHIM gas,
i.e. the so-called missing baryons, is only traceable through high ionisation
X-ray transitions.
We will explore such X-ray absorption in a forthcoming paper.

\subsection{Cosmic Densities of Ions} \label{sec:omegacorr}

The central theme in our exploration of the physical conditions of
metal absorbers is the complex link between the enrichment patterns of
metals and their observational tracers.  Simulations show that even
the most commonly observed low-$z$ metal absorption line, $\OVI$,
traces only a small fraction of the metals and baryons outside of galaxies
\citep[OD09; ][]{cen11, tep11}.  Furthermore, even when one uses Equation
\ref{equ:omega} to sum the observed absorbers to find the cosmic density
of a particular ion, $\Omega_{\rm abs}$, this does not necessarily
yield the true physical mass density of that ion.

We compare the mass-weighted cosmic ion density summed via the
contribution of every SPH particle outside of galaxies, $\Omega_{\rm
  SPH}$ (the grey shading in Figure \ref{fig:Zionphase}) to
$\Omega_{\rm abs}$ (Equation \ref{equ:omega}, the volume-weighted
observational probe represented by the coloured circles for individual
absorbers).  Ideally the two methods should yield the same answer;
i.e. $\Omega_{\rm abs}$ should recover the true $\Omega_{\rm SPH}$.
We list the comparisons in Table \ref{table:omegacomp} for a number of
different ions and wind models along with the percentage of the
$\Omega_{\rm SPH}$ recovered ($f_{\rm rec}$) in our simulated
observational sample corresponding to S/N=30 COS observations over a
pathlength of $\Delta z = 35$ and any identified line below $10^{15}
\cms$.  This pathlength appears to adequately sample the densities
from which the species we discuss arise.  

\begin{table} 
\caption{Cosmic Densities of Ions}
\begin{tabular}{lccccc}
\hline
Ion &
Redshift &
Model &
$\Omega_{\rm abs}^{a}$ &
$\Omega_{\rm SPH}^{b}$ &
$f_{\rm rec}^{c}$
\\
\hline
\multicolumn {6}{c}{}\\
$\OVI$   & $0-0.5$ & vzw    &  36.7  &  57.6  & 64 \\ 
"   & "              & cw     &  20.0  &  30.7  & 65 \\
"   & "              & sw     &  41.6  &  65.8  & 63 \\
"   & "              & nw     &  0.11  &  0.30  & 37 \\
"   & "              & vzw-turb     &  45.9  &  57.6  & 80 \\
$\CIV$   & $0-0.5$ & vzw    &  11.5  &  34.0  & 34 \\
"   & "              & cw     &   5.0  &  12.2  & 41 \\
"   & "              & sw     &   8.5  &  24.2  & 35 \\
"   & "              & vzw-turb     &  24.8  &  34.0  & 73 \\
$\SiIV$  & $0-0.5$ & vzw    &  2.3  &  10.3  & 22 \\
"   & "              & cw     &  0.9  &  2.6  & 35 \\
"   & "              & sw     &  1.3  &  5.2  & 25 \\
"   & "              & vzw-turb     &  6.9  &  10.3  & 67 \\
$\NeVIII$& $0.5-1.0$& vzw &  4.5  &  10.7  & 42 \\
"   & "              & cw     &  4.0  &  8.8  & 45 \\
"   & "              & sw     &  7.6  &  15.5  & 49 \\
"   & "              & vzw-turb     &  3.4  &  10.7  & 32 \\
\hline
\end{tabular}
\\
\parbox{15cm}{
$^a$ Summed ion cosmic density by SPH particle, in units of $10^{-8}$.\\
$^b$ Summed ion cosmic density by simulated absorbers, in units of $10^{-8}$.\\
$^c$ Recovery percentage, i.e. $\Omega_{\rm abs}/\Omega_{\rm SPH}\times 100$.
}
\label{table:omegacomp}
\end{table}

Between 22 and 80\% of $\Omega_{\rm SPH}$ is recovered, depending on
the species and the wind model.  Lower ionisation species often have
lower $f_{\rm rec}$, which at least double for $\Omega_{\CIV}$ and
$\Omega_{\SiIV}$ when turbulence is added to the vzw model.  This is
because these species (i) arise from higher densities that are more
stochastically sampled in our sight lines, and (ii) depend heavily on
the derived column densities for the strongest absorbers, from where
most of the cosmic density of an ion arises.  $\SiIV$ is the most
stochastic absorber we follow, and arises from the highest densities
($n_{\rm H}=10^{-4}-10^{-2} \cmc$).

Conversely, $\Omega_{\OVI}$ and $\Omega_{\NeVIII}$ are higher
ionisation species that trace lower densities with less stochasticity.
This explains the lower dispersion of recovery fractions between the
models.  Recovery of $\Omega_{\NeVIII}$ (32-49\% between $z=0.5-1.0$)
suffers because of sensitivity issues; the majority of $\NeVIII$ is
not well-sampled at diffuse overdensities because it is too weak to be
detected.  $\SiXII$ is an even more extreme example with vzw having
$f_{\rm rec}=13\%$; only a portion of the collisionally ionised
$\SiXII$ halo component is recovered.  $\OVI$ traces a statistically
well-sampled phase space with the most sensitivity to provide the most
complete census of $\Omega_{\rm SPH}$, but still manages to recover at
most $80\%$ of the total.

Another practical issue results from saturated absorbers that have
uncertain column densities.  {\sc AutoVP} typically estimates lower
column densities than the true value, and $\Omega_{\rm abs}$ becomes
under-estimated.  Adding turbulent broadening raises the column
density where lines become saturated and, therefore, $\Omega_{\rm
abs}$ rises and also becomes more accurate as we discuss in
\S\ref{sec:turb}.  This effect can be quite dramatic for species with
high stochasticity such as $\CIV$ and $\SiIV$ at $z<0.5$, where
applying turbulent broadening more than doubles $f_{\rm rec}$.

In summary, limited sensitivity and stochasticity can lead to
inaccuracies in the determination of the cosmic ion abundance.
Sensitivity issues always lead to an underestimate, while
stochasticity leads to more uncertainty in an ion's cosmic density.
Additionally, saturated absorbers can hide significant amounts of true
absorption, leading to under-estimates of cosmic ion densities,
especially for low ionisation species.  Hence while cosmic ion
densities are a useful characterisation of the overall metal evolution
in the IGM, detailed comparisons between models and data are more
robustly accomplished using absorber statistics.

\section{Model Variations} \label{sec:modelvar}

\begin{figure}
  \subfigure{\setlength{\epsfxsize}{0.49\textwidth}\epsfbox{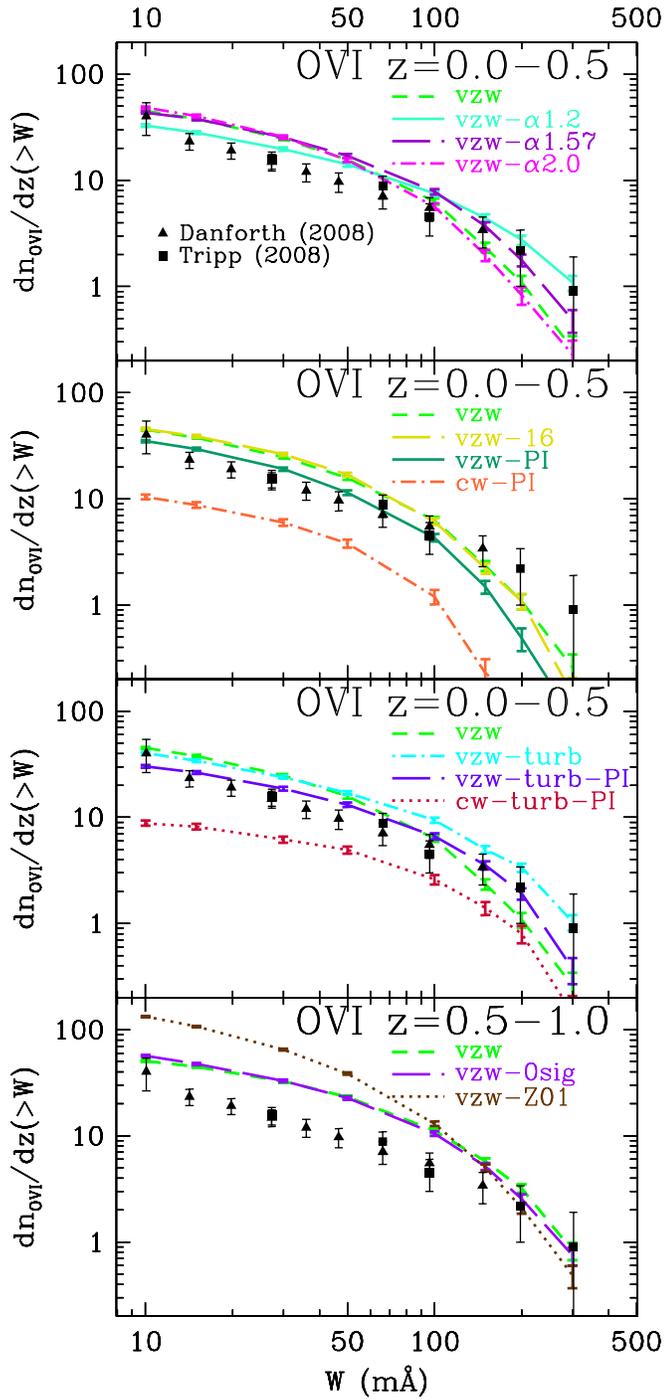}}
\caption{Equivalent width cumulative distribution plots for $\OVI$
  systems at $\langle z \rangle=0.25$ for different model variations
  including the shape of the ionisation background (top panel, \S6.1),
  different types of metal-line cooling (2nd, \S6.2), turbulent
  broadening (3rd, \S6.3), and the metal inhomogeneity (4th, \S6.4).
  Models are described in the text.}
\label{fig:EW_OVI}
\end{figure}

\begin{figure}
  \subfigure{\setlength{\epsfxsize}{0.465\textwidth}\epsfbox{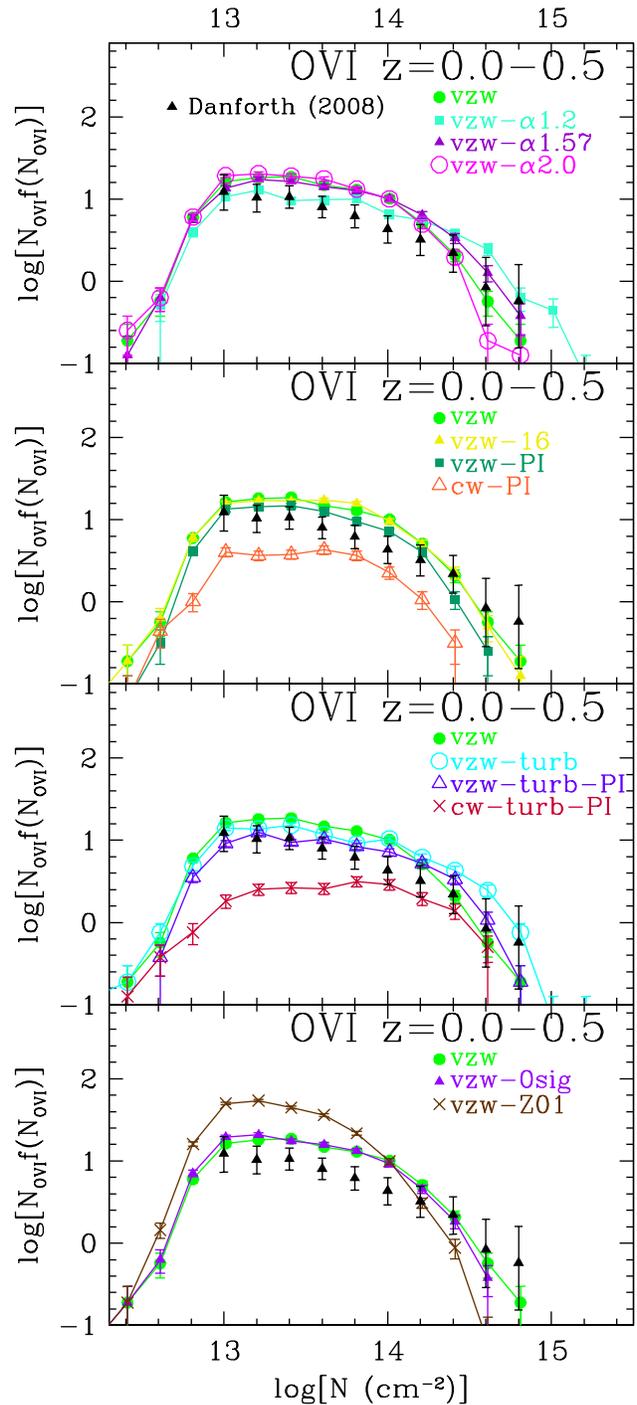}}
\caption{Column density distribution plots for $\OVI$ systems at
  $\langle z \rangle=0.25$ analogous to Figure \ref{fig:EW_OVI}.}
\label{fig:CDD_OVI}
\end{figure}

A fundamental difference between the $\lya$ and metal-line forests is
the latter's high sensitivity to outflow models.  In this section we
explore such sensitivities of the model predictions to variations in
some important physical and chemical assumptions.  We will mainly
focus on $\OVI$, but will also consider other species.  OD09 already
explored a range of model variations for $\OVI$, finding that this ion
consistently traced the diffuse, $T<10^5$ K IGM.  Given that other
groups find different results for $\OVI$
\citep[e.g.][]{cen11,smi11,tep11}, we concentrate our study on varying
four aspects of the modelling: the ionisation background, metal-line
cooling, turbulent broadening, and metal inhomogeneity.  We present
the EWDs and the CDDs for the variations that we explore in Figures
\ref{fig:EW_OVI} and \ref{fig:CDD_OVI}, respectively, and provide
cosmic $\OVI$ ion densities at $z<0.5$ in Table \ref{table:omega_OVI}.

\subsection{Ionisation Background} \label{sec:ionbkgdvar}

Our conclusion that photo-ionisation is responsible for the majority
of the absorption by high-ionisation species is sensitive to the shape
and intensity of the ionisation background at far-UV and soft X-ray
energies.  No direct observational constraints exist between the
observed slope of extreme UV (EUV) spectra of quasars \citep{tel02,
sco04} and the X-ray background \citep{wor05}.  The ionisation
potential required to photo-ionise $\OVI$ at 8.4 Rydbergs lies in this
unconstrained gap (J. M. Shull, private communication).

\begin{table} 
\caption{Cosmic Densities of $\OVI$ between $z=0-0.5$}
\begin{tabular}{lccc}
\hline
Model &
$\Omega_{\rm abs}^{a}$ &
$\Omega_{\rm SPH}^{b}$ &
$f_{\rm rec}^{c}$
\\
\hline

\multicolumn {4}{c}{}\\
vzw              &  36.7  &  57.6 & 64 \\ 
vzw-$\alpha2.0$  &  36.2  &  53.4 & 68 \\
vzw-$\alpha1.57$ &  42.5  &  75.3 & 56 \\
vzw-$\alpha1.2$  &  41.5  & 110.5 & 38 \\
vzw-16           &  37.3  &  56.6 & 66 \\
vzw-PI           &  26.2  &  39.6 & 66 \\
cw-16            &  23.9  &  35.6 & 67 \\
cw-PI            &  7.7   &  11.5 & 67 \\
vzw-turb         &  45.9  &  57.6 & 80 \\
vzw-PI-turb      &  32.5  &  39.6 & 82 \\
cw-PI-turb       &  11.5  &  11.5 & 100 \\
vzw-0sig         &  34.4  & 57.6  & 60 \\
vzw-Z01          &  51.5  & 71.4  & 72 \\
\hline
\end{tabular}
\\
\parbox{15cm}{
$^a$ Summing ion cosmic density by SPH particle, in units of $10^{-8}$.\\
$^b$ Summing ion cosmic density by simulated absorbers, in units of $10^{-8}$.\\
$^c$ Recovery percentage.  Ratio of $\Omega_{\rm abs}/\Omega_{\rm SPH}\times 100$.
}
\label{table:omega_OVI}
\end{table}

We use the CUBA package \citep{haa01} to generate three alternative
ionisation backgrounds varying the slope of the EUV portion of the
quasar spectrum.  Figure \ref{fig:ionbkgd} displays these three other
backgrounds, which are referred to as HM2005 backgrounds (after Haardt
\& Madau 2005) with $\alpha=1.2$, 1.57, and 2.0, where the spectrum
above 1 Ryd has an intensity $F_{\nu}\propto\nu^{-\alpha}$.  The HM2005
background is included as an input into the latest package of CLOUDY
and uses the $\alpha=1.57$ power law based on \citet{tel02}, but the
range of slopes explored here is reasonable given the observed range
and uncertainty in quasar spectral shapes (F. Haardt, private
communication).  The fiducial HM2001 spectrum uses $\alpha=1.8$ and is
most similar to the HM2005 $\alpha=2.0$ spectrum.

\begin{figure}
  \subfigure{\setlength{\epsfxsize}{0.45\textwidth}\epsfbox{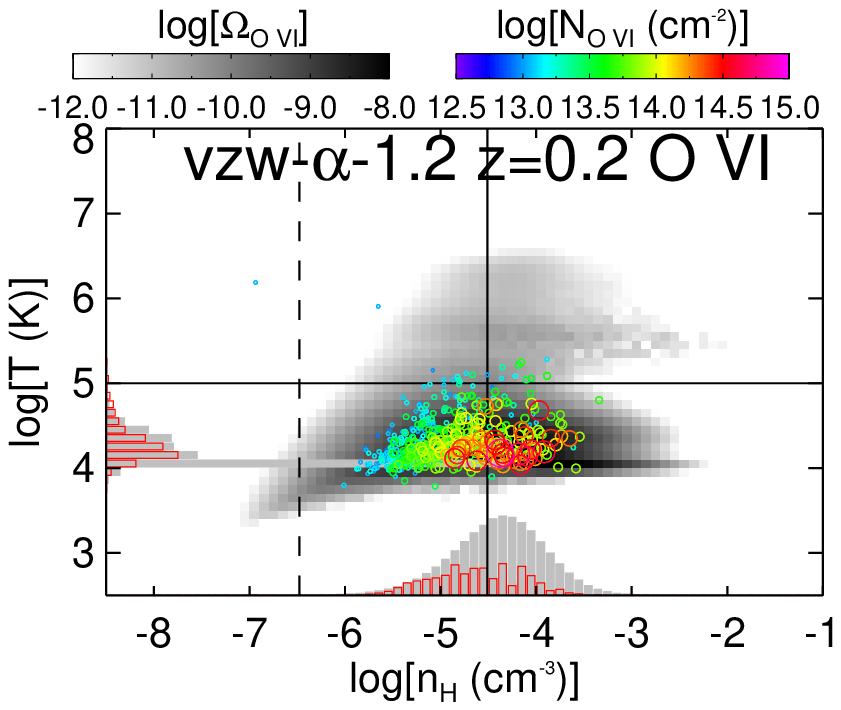}}
  \subfigure{\setlength{\epsfxsize}{0.45\textwidth}\epsfbox{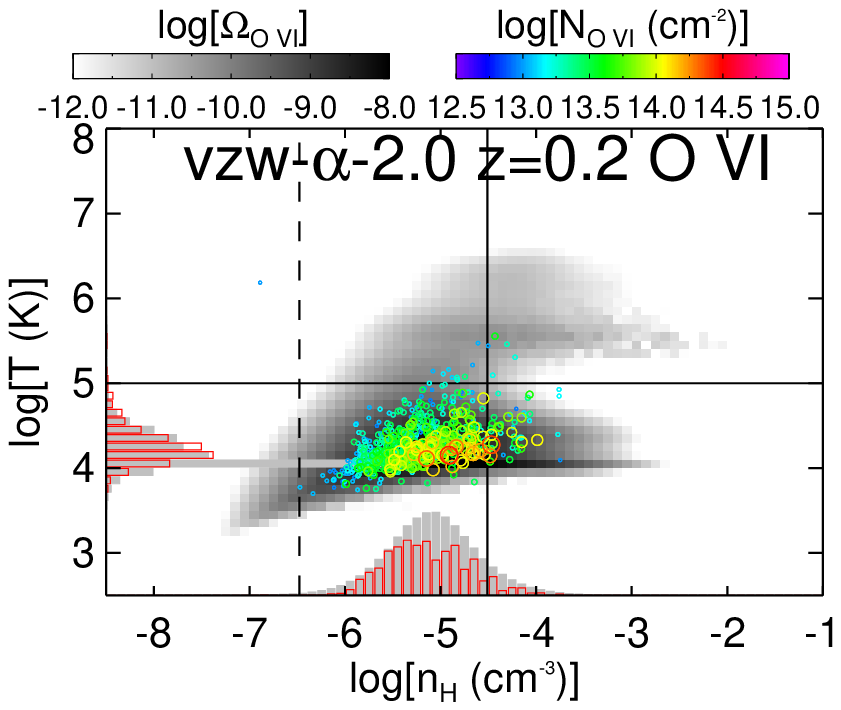}}
\caption{$\Omega$ phase space diagrams as described in Figure
  \ref{fig:Zionphase} for $z=0.2$ $\OVI$ for the two extremes in our assumed
  ionisation background.  The top panel shows the hardest
  background ($\alpha=1.2$ for the EUV power law of quasars) and the
  softest background ($\alpha=2.0$).  }
\label{fig:phaseionbkgd}
\end{figure}

The hardest background, $\alpha=1.2$, has 16 times more intensity at
the $\OVI$ ionisation potential relative to the softest background,
$\alpha=2.0$, while leaving the background at the Lyman limit and,
therefore, the $\lya$ forest nearly unchanged.  Figure
\ref{fig:phaseionbkgd} shows the phase space distribution of $\OVI$
absorbers under these two extreme assumptions, and demonstrates that
the peak density for $\Omega_{\OVI}$ shifts from $10^{-5.1}$ to
$10^{-4.3} \cms$ when moving from the softest to the hardest
backgrounds (cf. grey $x$-axis histograms).  However, typical $\OVI$
absorbers ($N_{\OVI}=10^{13.5-14.5} \cms$) do not reflect such a
dramatic shift.  Recovery fractions of $\Omega_{\OVI}$ also decline
from 56\% to 32\% (Table \ref{table:omega_OVI}), mainly because $\OVI$
absorbers from haloes have underestimated column densities due to
saturation.

The top panel of Figure \ref{fig:CDD_OVI} displays the $\OVI$ CDD for
the four backgrounds.  As the EUV intensity is turned up, photo-ionised
$\OVI$ moves to higher overdensities creating more strong absorbers and
fewer weak absorbers; a power law fit to the CDD would give a shallower
slope.  The $\alpha=1.2$ background model appears consistent with the
EWD in the top panel of Figure \ref{fig:EW_OVI} without the need for
turbulent broadening, but the linewidths are still underestimated for
strong absorbers compared to the observed $b$-parameters (not shown).
When $\alpha=1.2$, strong photo-ionised $\OVI$ absorbers populate haloes
while the $\alpha=2.0$ background predicts barely any photo-ionised
halo absorbers.  If photo-ionised $\OVI$ is clearly identified within
haloes, it suggests that a hard meta-galactic ionisation background is
penetrating halo gas.

Collisionally ionised $\OVI$ is a poor tracer of WHIM gas in all
cases, because even the weakest ionising background efficiently
photo-ionises oxygen in its collisionally ionised temperature range,
peaking at $T=10^{5.45}$ K, to higher ionisation states.  This is why
$\OVI$ fails to trace the copious amounts of metals pushed by cw winds
into the low-density $10^{5-6}$ K WHIM; this oxygen ionises to $\OVII$
and beyond.  In summary, reasonable variations in the hardness of the
EUV background are unlikely to alter our fundamental conclusion that
most $\OVI$ absorbers detectable with COS arise in photo-ionised gas.

\subsection{Metal-Line Cooling} \label{sec:Zcool}

Even if the ionisation background photo-ionises most $\OVI$ at $T>10^5$
K to higher states outside haloes, there still exists a potential for
some $\OVI$ to trace WHIM temperatures.  \citet{cen11}, \citet{smi11},
and \citet{tep11} all predict a significant component of collisionally
ionised $\OVI$ at overdensities as low as $\sim 10$.  One possible
explanation is the efficiency of metal-line cooling in the $\sim 10^5$~K
temperature regime.  To explore this, we now compare two implementations
of metal-line cooling-- (i) the assumption of collisionally ionised
equilibrium (CIE), and (ii) cooling in the presence of the \citet{haa01}
photo-ionising background (PI).

Our main simulations employ the first implementation, namely that
\citet{sut93} CIE metal cooling rates are added to hydrogen and
helium rates calculated in the presence of the \citet{haa01}
background.  However, in the presence of a photo-ionising background,
the number of bound electrons is reduced, leading to less efficient
cooling.  \citet{tep11} in their \S5.3 demonstrate that photo-ionised
cooling times using \citet{wie09a} rates in OWLS (OverWhelmingly
Large Simulations), rates that include the effects of the UV background on
the metal cooling, are significantly longer for $0.1 \Zsolar$ and
$\Zsolar$ gas at $n_{\rm H} = 10^{-5} \cms$.  This leads to a higher
equilibrium temperature where cooling balances photo-heating.

Photo-ionised rates are clearly the more physically correct cooling
rates, therefore we compare the results of using our normal metal
cooling implementation (adding CIE rates) to those using photo-ionised metal
cooling rates based on the \citet{wie09a} solar abundance cooling
tables.  To do so, we evolve four $16 \hmpc$ simulation volumes at the
same resolution as our $48 \hmpc$ simulation volumes: vzw and cw winds
with CIE cooling (vzw-16, cw-16) and PI (vzw-PI, cw-PI) cooling.

We present the EWDs and CDDs for all but the cw-16 model along
with the 48 $\hmpc$ vzw simulation in the second panels of Figures
\ref{fig:EW_OVI} and \ref{fig:CDD_OVI}, respectively.  The vzw-16
simulation is statistically identical to the larger vzw simulation
for all $\OVI$ observations.  This comes as no surprise because OD09
demonstrated that $\OVI$ is mostly associated with low-mass, $<0.3 M^*$,
galaxies, which are well sampled in both volumes.  This justifies using
the smaller volume to explore these physical variations.

\begin{figure*}
  \subfigure{\setlength{\epsfxsize}{0.45\textwidth}\epsfbox{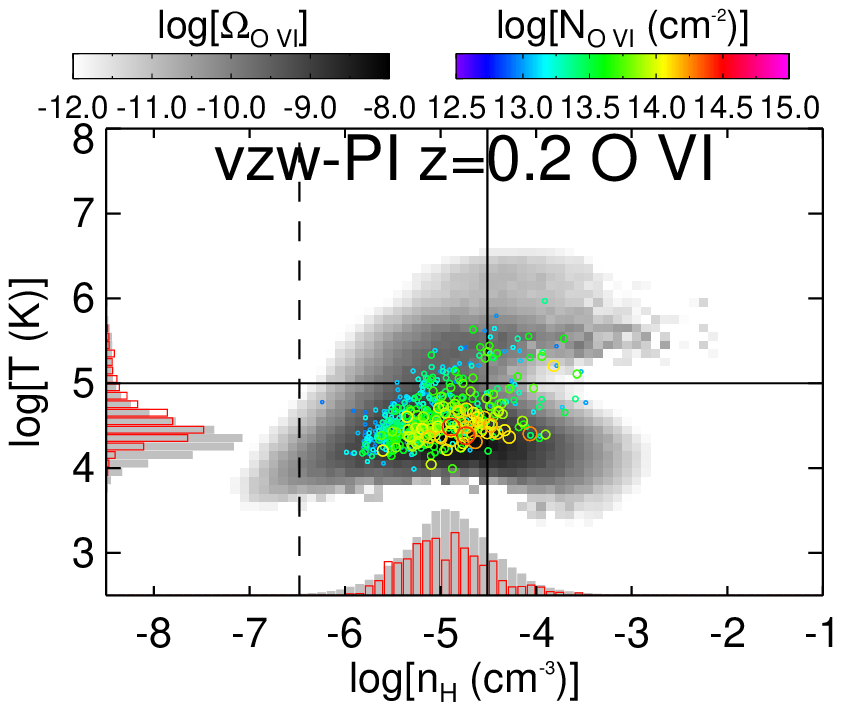}}
  \subfigure{\setlength{\epsfxsize}{0.45\textwidth}\epsfbox{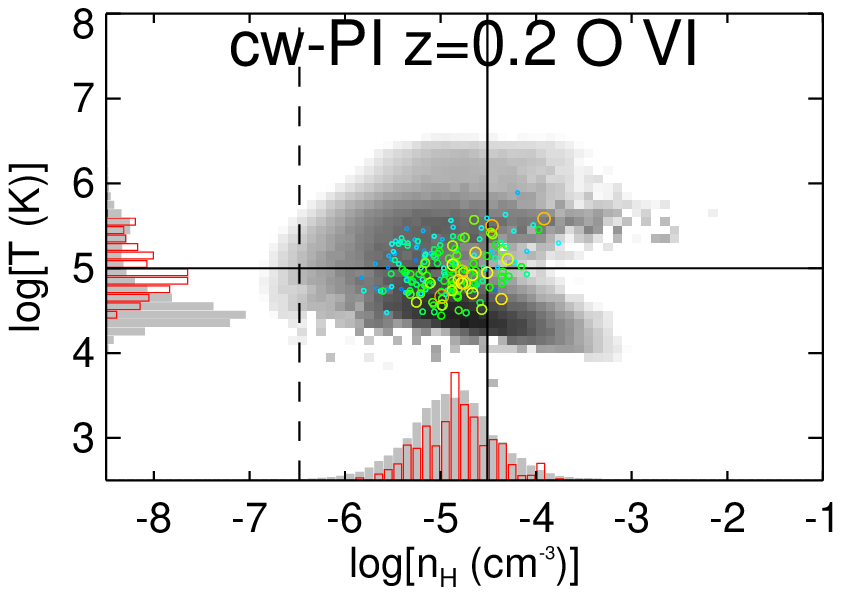}}
  \subfigure{\setlength{\epsfxsize}{0.45\textwidth}\epsfbox{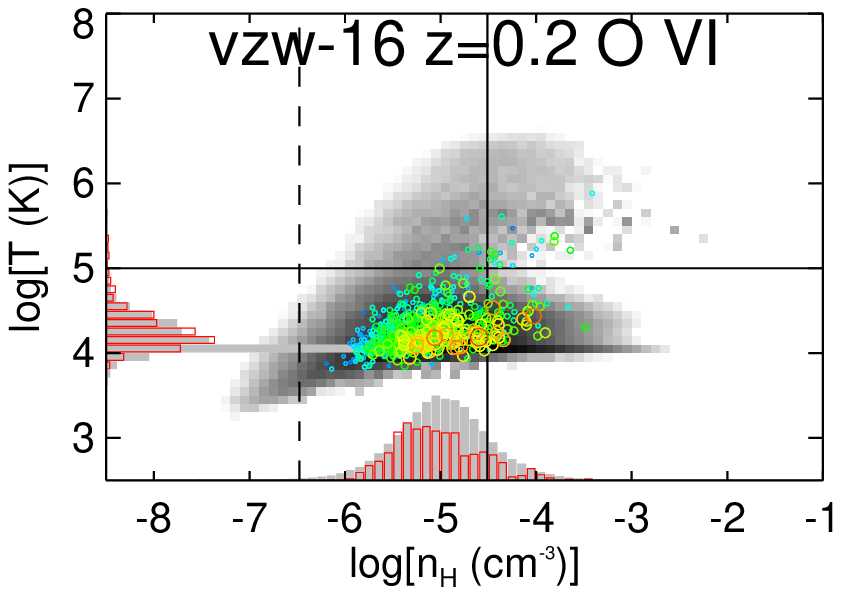}}
  \subfigure{\setlength{\epsfxsize}{0.45\textwidth}\epsfbox{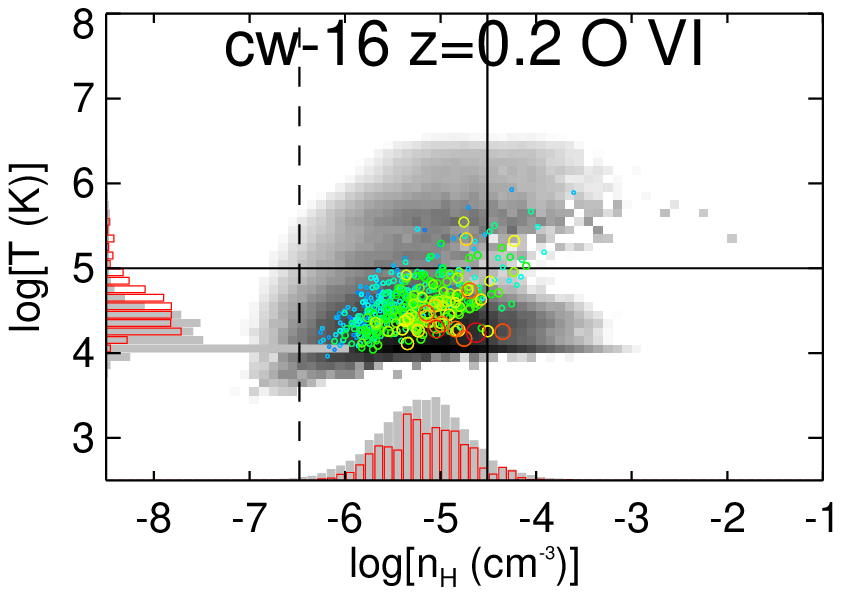}}
\caption{$\Omega$ phase space diagrams as described in Figure
  \ref{fig:Zionphase} for $\OVI$ at $z=0.2$ for simulations with
  photo-ionisational-dependent metal-line cooling (top panels) and the
  assumption of collisional ionisation for metal-line cooling (bottom
  panels).  The left panels show the vzw models and the right panels
  show the cw models.  We show the $16 \hmpc$ volumes in all four cases,
  but the $\OVI$ observables are statistically identical to the $48
  \hmpc$ volumes for collisional ionisation cooling.  }
\label{fig:Ophase_cool}
\end{figure*}

Comparing the vzw-16 and vzw-PI simulation, there exists 70\% as much
$\OVI$ absorption in the latter, specifically $\Omega_{\OVI}$,
$3.26\times 10^{-7}$ versus $2.35\times 10^{-7}$, with the reduction
evenly distributed in column density (Figure \ref{fig:CDD_OVI}, middle
panel).  Examination of the $\OVI$ phase space for vzw-PI in Figure
\ref{fig:Ophase_cool} (upper left panel) reveals that most of the
$\OVI$ absorption still arises from the diffuse $\OVI$ phase.
However, there are subtle differences compared to the standard vzw
model (bottom right panel).  The locus of cool metals at $\delta\ga
10$ (i.e. $n_{\rm H} \ga 10^{-5.5} {\rm cm}^{-3}$) is no longer flat
at $\sim 10^4$ K, below which CIE metal-line cooling no longer
contributes, but declines from $\sim 10^{4.5}$ K to below $10^4$ K for
increasing overdensity as seen also in OWLS \citep[cf.][their Figure
1]{wie10}.  $\OVI$ absorbers, therefore, trace somewhat hotter
temperatures, but are still within the diffuse phase.  Somewhat more
$\OVI$ exists in the WHIM phase owing to this less efficient cooling,
but still only a handful of identified absorbers actually trace this
phase.

Examining the two different cooling models in the cw simulations
reveals much larger differences (Figure \ref{fig:Ophase_cool}, right
panels); $\Omega_{\OVI}$ declines by a factor of almost three using PI
cooling, from $2.08\times 10^{-7}$ to $7.1\times 10^{-8}$.  The main
difference is that cw winds push metals to lower overdensities where
photo-ionisation plays a larger role in retarding cooling.  Less
oxygen cools below $10^{4.5}$ K where photo-ionised $\OVI$ ionisation
fractions are the largest.

As an aside, we note that galaxy growth remains relatively unaffected
by this change.  The amount of stars formed at $z=0$ is reduced by
only 8\% using PI cooling relative to CIE cooling.  This mostly arises
because the reduced metal-line cooling at lower densities hinders the
recycling of wind material back into galaxies~\citep{opp10}.  A small
reduction in overall star formation was also noted by \citet{smi11}
when making the same comparison in their grid-based simulation
(cf. the red and black curves in their Figure~1).

In summary, including photo-ionised cooling rates does not alter our
fundamental conclusions about the phases $\OVI$ and other absorbers
trace in our favoured vzw model -- $\OVI$ still traces mostly diffuse,
photo-ionised gas.  We also demonstrate that using a different wind
model (cw), similar to that used by \citet{tep11}, produces a much
larger change in $\OVI$ absorber properties when using photo-ionised
metal cooling rates, because this model pushes many more metals into
very diffuse gas where the photo-ionised cooling rates have a larger
impact.  We note that the vzw model does a much better job than the cw
model in fitting the observed $z=0$ galactic stellar mass function
below $5\times 10^{10} \msolar$ \citep{opp10}, from which most $\OVI$
absorption arises (OD09).  Hence, we favour the vzw model, although
our tests do illustrate that the interpretation of observed $\OVI$
still could depend on the details of the input physics.

\subsection{Turbulent Broadening} \label{sec:turb}

In OD09 we introduced a heuristic prescription to add sub-grid turbulent
broadening to the $b$-parameters of $\OVI$, after the conclusion of the
simulation, using {\tt specexbin}.  This added turbulent broadening is
a function of the physical density of the gas.  The prescription adds
an additional turbulent $b$-parameter in quadrature where
\begin{equation} \label{eqn:bturb1}
b_{\rm turb} = \sqrt{1405 \logrm[n_{\rm H}]^2 + 15674 \logrm[n_{\rm H}] + 43610}~\kms ,
\end{equation}
over the range $n_{\rm H}=10^{-5.31}$ to $10^{-4.5} \cmc$ and
\begin{equation} \label{eqn:bturb2}
b_{\rm turb} = 13.93 \log(n_{\rm H}) + 101.8~\kms
\end{equation} over the range $n_{\rm H}=10^{-4.5}$ to $10^{-3.0} \cmc$.
Furthermore, $b_{\rm turb}$ retains its maximum value of $60 \kms$ at
higher densities.  See OD09 \S4.4 for an in-depth discussion of the
physical motivation for this model and its relation to the turbulence
observed on kpc scales in $\CIV$ paired absorbers along lensed quasar
sight lines at $z\sim 2-3$ \citep{rau01}.  OD09 argue that the
turbulence is related to outflows and decays on timescales of the
order of a Hubble time.

\begin{figure}
  \subfigure{\setlength{\epsfxsize}{0.49\textwidth}\epsfbox{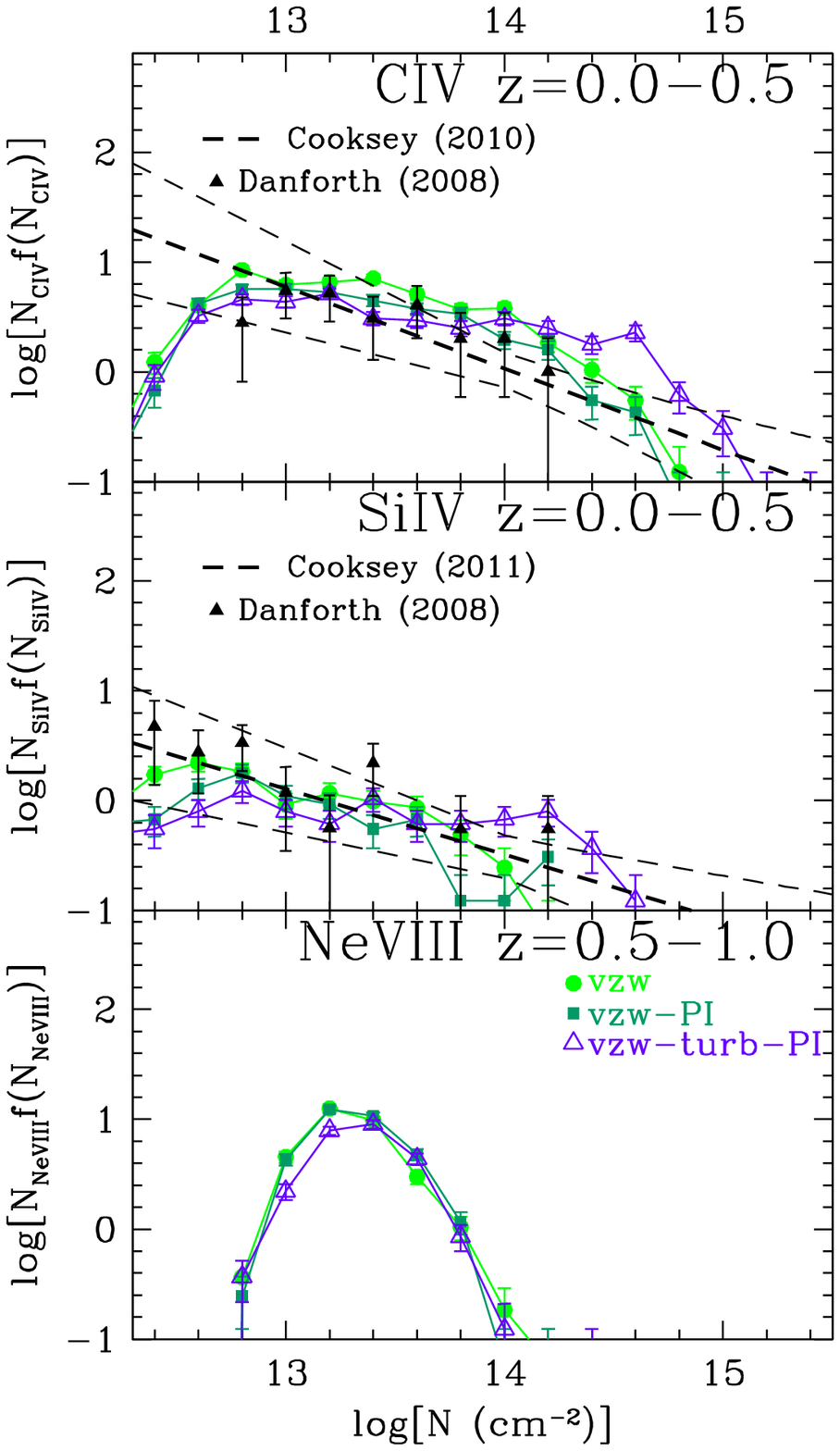}}
\caption{Column density distribution plots for $\CIV$, $\SiIV$, and
$\NeVIII$ for vzw winds with photo-ionisation-dependent cooling rates
and turbulent broadening.  Error bars correspond to Poisson errors.}
\label{fig:CDD_c4_si4_ne8}
\end{figure}

We already showed results for the vzw-turb model in Figures
\ref{fig:EW} through \ref{fig:EWcompsys}.  Here we combine it with the
vzw and cw models with photo-ionised cooling rate (vzw-turb-PI,
cw-turb-PI) shown in the third panels of Figures \ref{fig:EW_OVI} and
\ref{fig:CDD_OVI}.  Photo-ionised metal-line cooling rates reduce vzw
$\OVI$ line frequencies and $\Omega_{\OVI}$ by the same factor,
one-third, either including turbulence or without it.  The vzw-turb-PI
model agrees well with the observed $\OVI$ CDD and EWD, and better
than the vzw-turb model, which over-produces weak $\OVI$ absorbers.
The vzw-turb-PI model is our favoured model for $\OVI$ as it has the
most physically motivated combination of outflows, ionisation
background, physical metal-line cooling, and turbulent broadening to
explain both $\OVI$ and $\lya$ forest observations (D10), as well as
other observations of low-$z$ galaxies \citep{opp10,dav11a,dav11b}.

We show the $\CIV$, $\SiIV$, and $\NeVIII$ CDDs for this model along
with the vzw-PI model in Figure \ref{fig:CDD_c4_si4_ne8}.  As before,
our favoured model continues to overproduce the incidence of the
lower-ionisation species.  We do not believe this discrepancy can be
mitigated by cosmic variance, because even our $16 \hmpc$ box is
sufficient to sample the gas and galaxies responsible for the
absorption of these ions (cf. vzw and vzw-16 show nearly identical
results).  However, the idea of adding turbulent broadening for lower
ionisation species that trace higher densities may not be as valid as
in the case of $\OVI$.  For instance, using lensed quasars probing the
$z\sim 2.3$ IGM, \citet{lop07} observed sub-kpc sub-structure in lower
ionisation species including $\CIV$ and $\SiIV$ that did not exist in
$\OVI$.  At low redshifts, all of the $\SiIV$ and most of the $\CIV$
in our models have densities where our turbulence model predicts
$b_{\rm turb}> 40 \kms$, but the observations of $\CIV$ suggest much
narrower linewidths \citep[e.g.][]{coo10}.  Hence $\OVI$ may be
tracing more diffusely distributed gas that is more affected by
large-scale turbulence injected by outflows, while $\CIV$ and $\SiIV$
could be tracing individual high-density cloudlets.  Regardless,
models both with and without turbulence overpredict the $\CIV$ and
$\SiIV$ absorption.

\begin{figure*}
  \subfigure{\setlength{\epsfxsize}{0.45\textwidth}\epsfbox{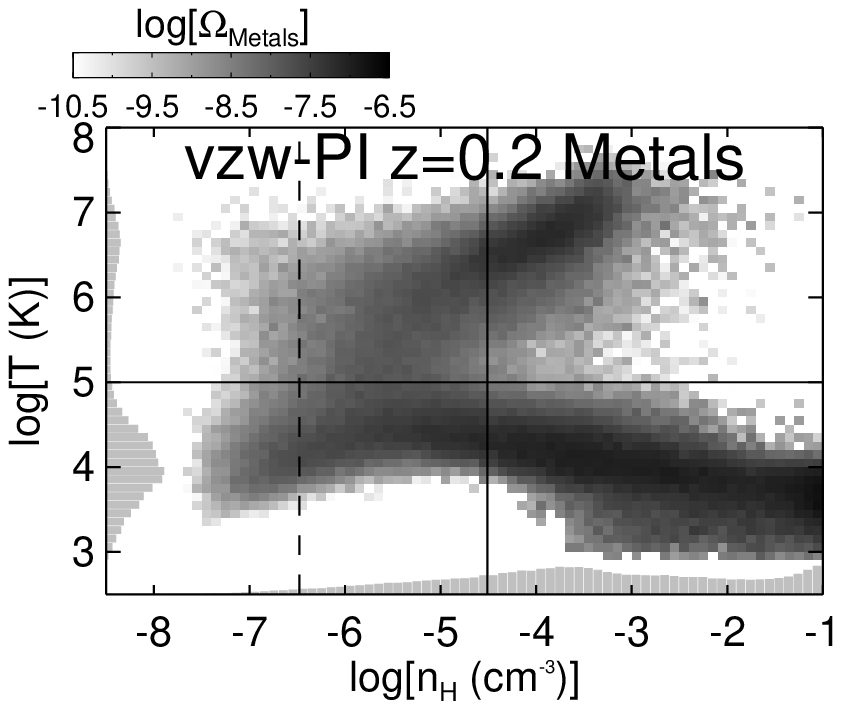}}
  \subfigure{\setlength{\epsfxsize}{0.45\textwidth}\epsfbox{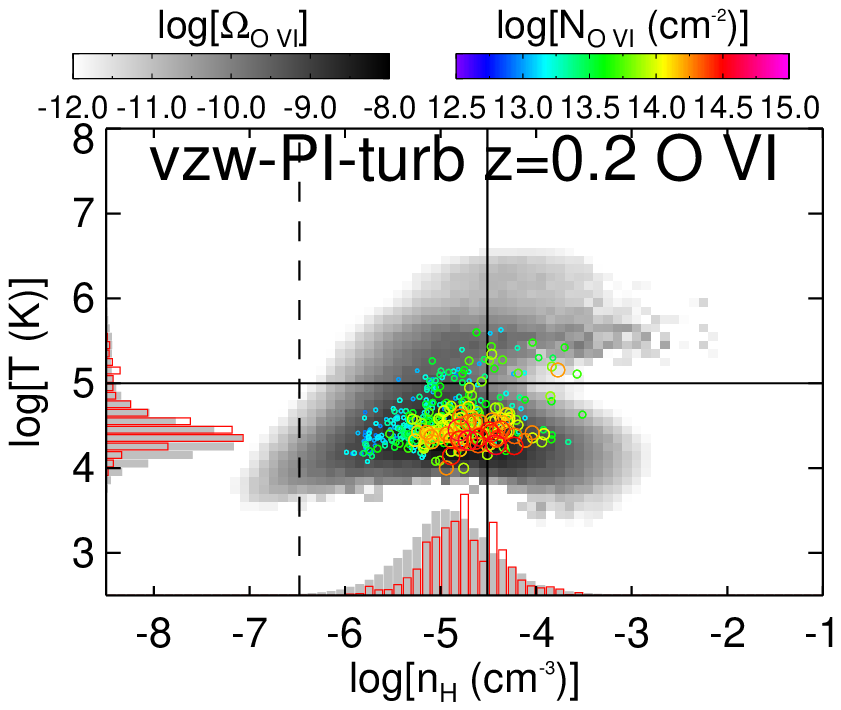}}
  \subfigure{\setlength{\epsfxsize}{0.45\textwidth}\epsfbox{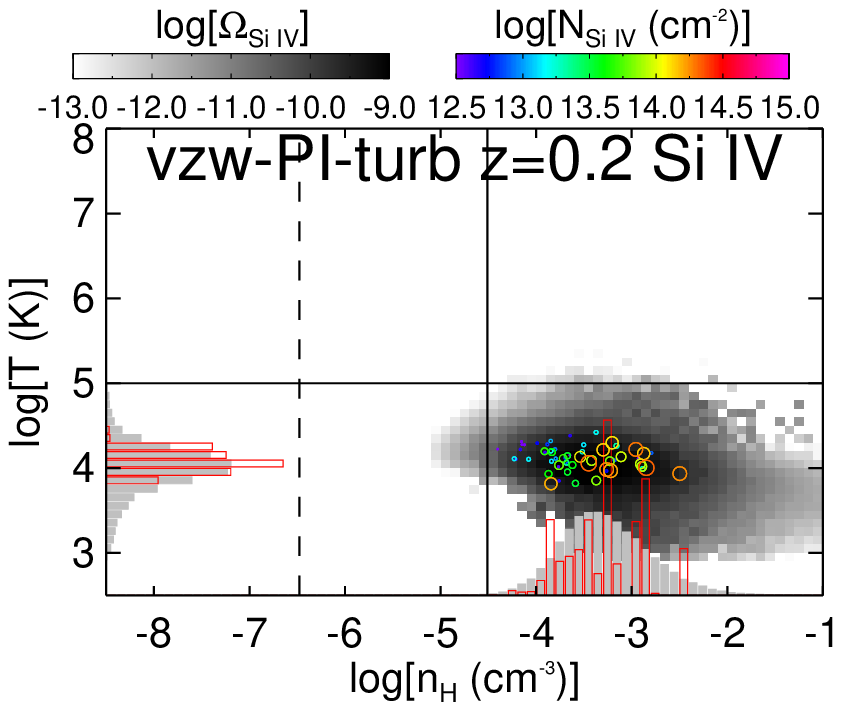}}
  \subfigure{\setlength{\epsfxsize}{0.45\textwidth}\epsfbox{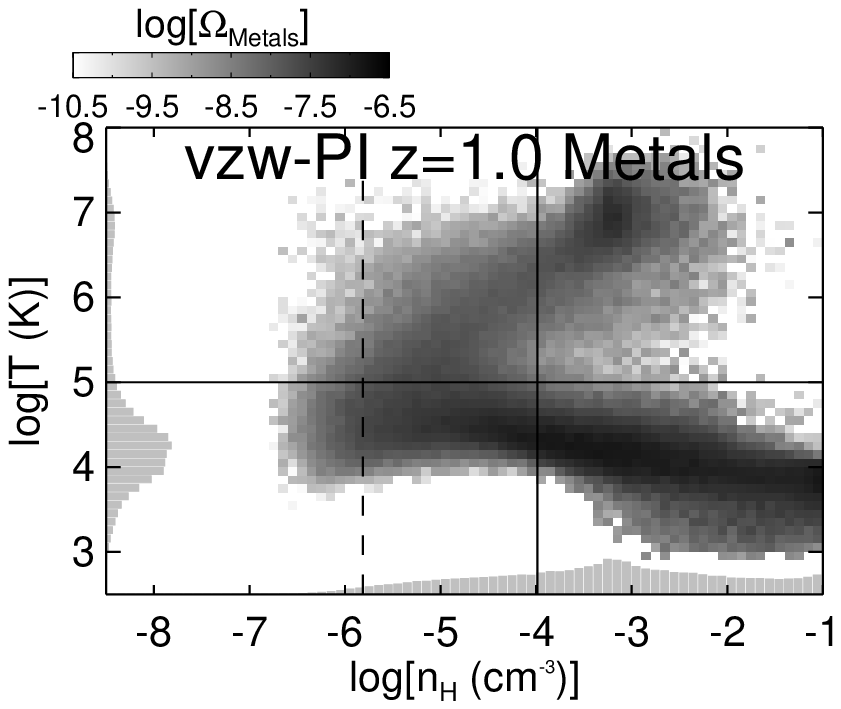}}
  \subfigure{\setlength{\epsfxsize}{0.45\textwidth}\epsfbox{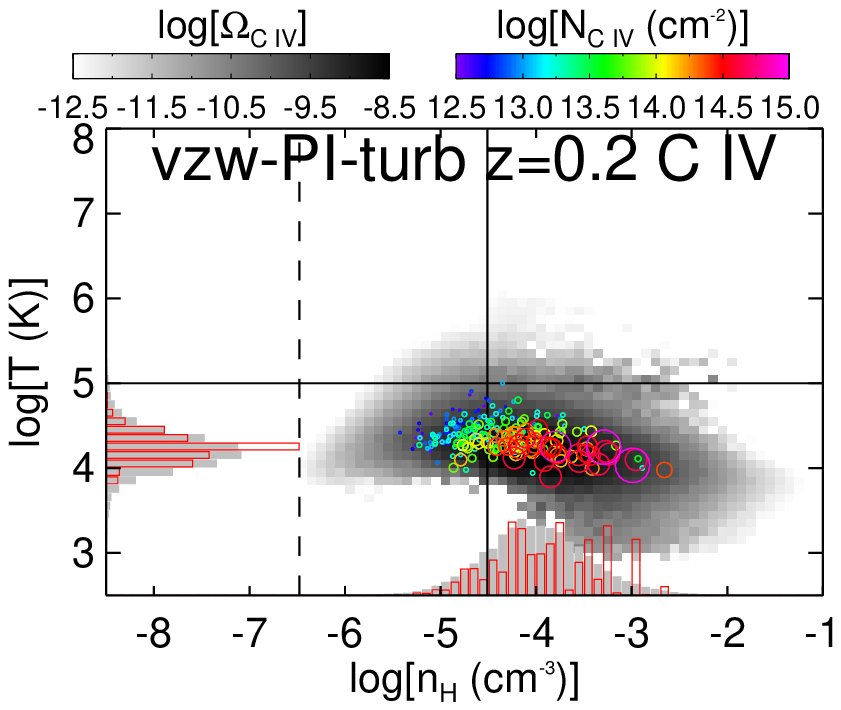}}
  \subfigure{\setlength{\epsfxsize}{0.45\textwidth}\epsfbox{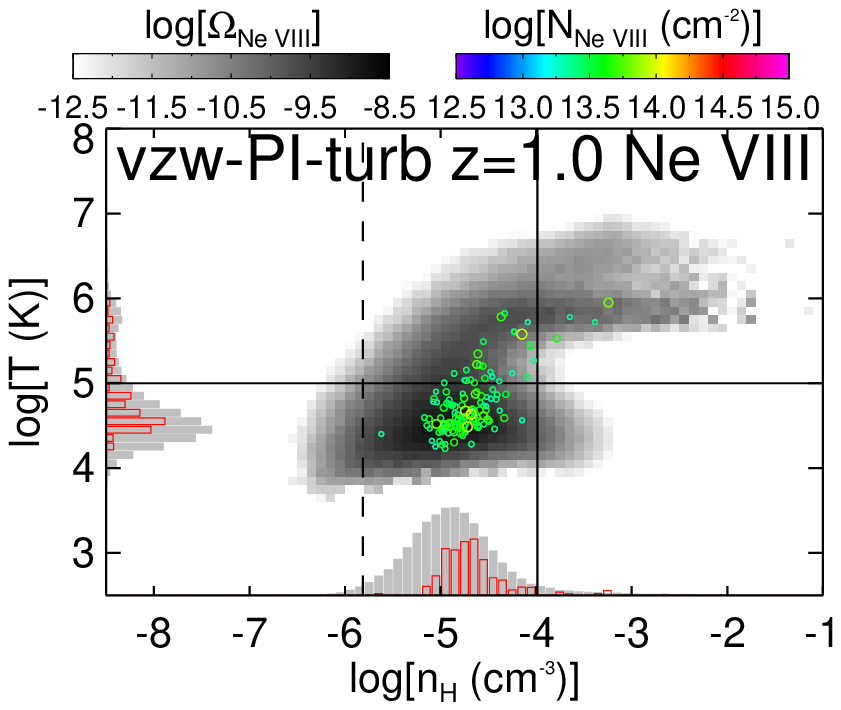}}
  \caption{$\Omega$ phase space diagrams of the vzw-turb-PI model for
    metals, $\SiIV$, $\CIV$, and $\OVI$ at $z=0.2$; plus metals and
    $\NeVIII$ at $z=1$.}
\label{fig:Zionphase_bturb_PI}
\end{figure*}

Figure \ref{fig:Zionphase_bturb_PI} shows the phase-space diagram for
metals and absorbers in the vzw-turb-PI model.  $\SiIV$ and stronger
$\CIV$ absorbers trace much higher densities than $\OVI$.  This
results in many more strong $\CIV$ and $\SiIV$ components than in the
vzw-PI model, because \autovp~finds stronger $N$s for otherwise
non-broadened saturated profiles and weaker absorbers are subsumed
into the stronger ones.  The histograms in Figure
\ref{fig:Zionphase_bturb_PI} show that these absorbers mostly arise in
cool halo gas.  A difficulty with our simulations is that we may not
resolve cooling instabilities in the halo gas~\citep[e.g.][]{mal04}
that could result in much of the carbon being in dense cold cloudlets,
which would favour lower ionisation states than $\CIV$.  This is in
principle easily possible as only 3-5\% of the carbon outside of
galaxies resides in $\CIV$ between $z=0-0.5$ (vs. 13-15\% at $z=2-4$),
and hence a small change in the ionisation rate at $\sim 3-4$ Ryd
could dramatically change the $\CIV$ statistics.  It will be
interesting to compare our predictions of $\CII$ and $\CIII$ to
observations when a sufficient sample becomes available, to see if we
significantly underpredict the amount of carbon in these states
corresponding to our overprediction of $\CIV$.

Another possibility is that the \citet{chi04} Type II carbon yields we
use are too high.  For example, the \citet{woo95} carbon yields are
$\sim 0.2$ dex lower, which could explain some of the differences with
respect to the observations.  The AGB stellar yields we use cannot
explain the over-estimate, because carbon production from AGB stars is
sub-dominant to Type II SNe in our simulations \citep{opp08}.  Yet
another potential explanation is that a substantial fraction of carbon
and silicon are depleted onto dust, while the more volatile oxygen is
not.  \cite{zu11} find that the vzw model can explain the
observational evidence for intergalactic dust \citep{men10} if the IGM
dust-to-gas ratio is about half that in the Milky Way ISM, which about
25\% of intergalactic metals (by mass) in dust.

$\NeVIII$ traces similar overdensities as $\OVI$ and hence should be
affected by the same turbulence, and hence the vzw-turb-PI model
should be the most appropriate.  But as
Figure~\ref{fig:CDD_c4_si4_ne8} shows, the predicted column density
distributions do not vary significantly with PI cooling or turbulence.
In detail, between $z=0.5-1.0$, $\Omega_{\NeVIII}$ slightly increases
with PI cooling, but declines by a factor of 20\% with the addition of
turbulence to $4.0\times 10^{-8}$, because already weak lines are
broadened enough so that \autovp~fails to identify them.  Most of the
$\NeVIII$ absorption still traces photo-ionised gas, but the reduced
cooling raises the bulk of $\NeVIII$ absorbers from $T=10^{4.0-4.4}$ K
to $10^{4.4-4.9}$ K and leads to 22\% compared to 7\% of the $\NeVIII$
absorption arising from $T>10^{5}$ K gas
(Figure~\ref{fig:Zionphase_bturb_PI}).  This percentage could grow
even larger if our simulations adequately resolved interfaces between
cool and hot gas within haloes, which they do not as we discussed in
\S\ref{sec:absphase}.

We show the cw-turb-PI model to demonstrate that turbulent broadening
affects other wind models in much the same way as the vzw model, by
making more wide $\OVI$ absorbers and slightly fewer weak absorbers
(cf. cw-PI model).  This more appropriate cooling rates applied to the
cw model, combined with turbulent broadening to match observed line
widths indicates this model far under-produces $\OVI$, especially
weaker systems, when compared to observed EWDs and CDDs.

\subsection{Metal Inhomogeneity} \label{sec:inhomo}

The degree of metal inhomogeneity in the IGM remains highly uncertain,
but dynamically very important.  The distribution of metals affects
gas cooling, which is critical for galaxy formation because the
temperature of the gas determines the efficiency of gas accretion onto
galaxies \citep{kat03,ker05,dek06,ker09a}.  More metal mixing leads to
more star formation as more efficient metal-line cooling affects more
baryons \citep{wie09b}.  In our simulations, we do not mix metals, and
each wind particle retains its own metals as it propagates out of
galaxies.

\begin{figure*}
  \subfigure{\setlength{\epsfxsize}{0.45\textwidth}\epsfbox{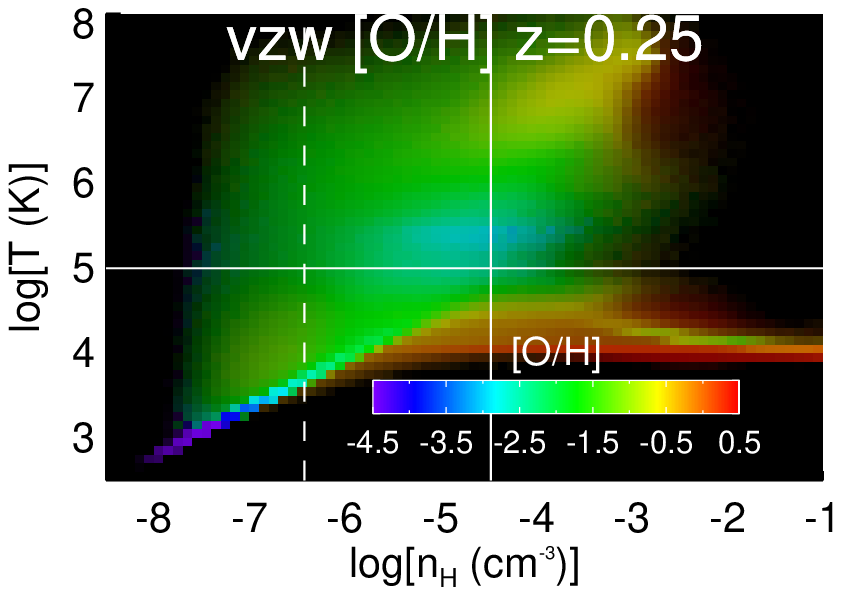}}
  \subfigure{\setlength{\epsfxsize}{0.45\textwidth}\epsfbox{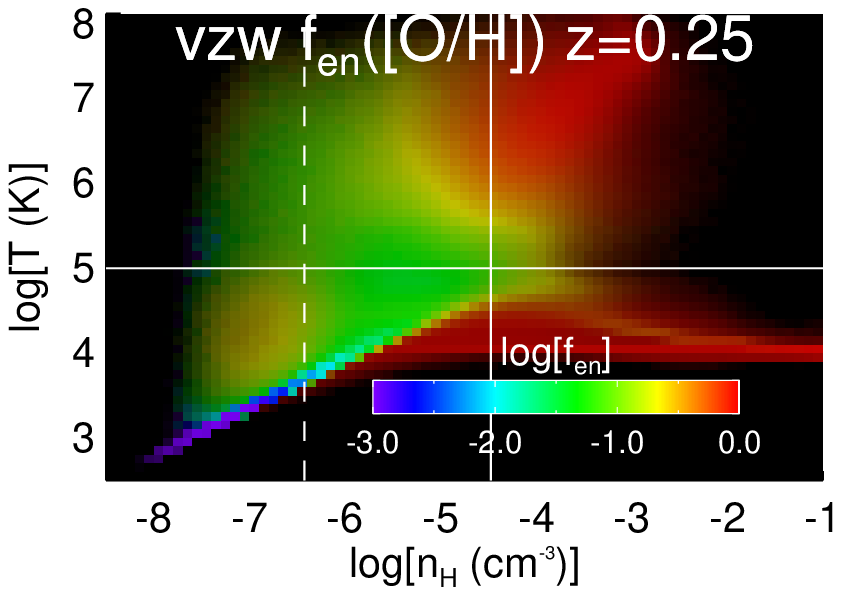}}
  \subfigure{\setlength{\epsfxsize}{0.45\textwidth}\epsfbox{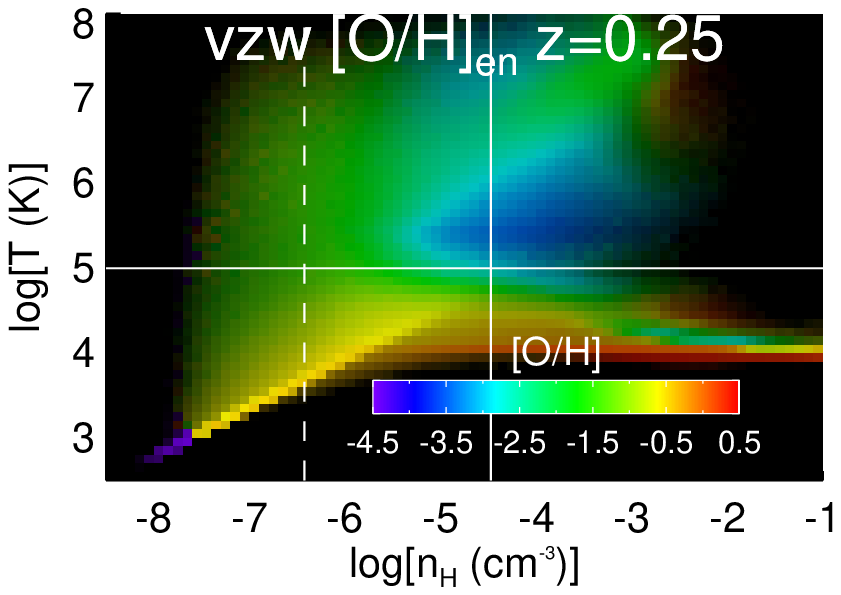}}
  \subfigure{\setlength{\epsfxsize}{0.45\textwidth}\epsfbox{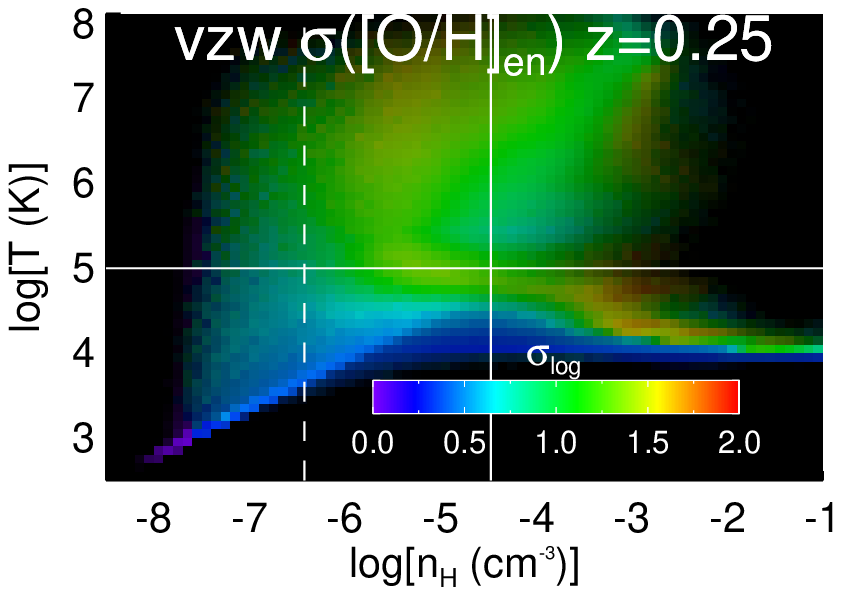}}
\caption[]{Phase space diagrams displaying oxygen metallicity
  characteristics for the r48n348vzw simulation at $z=0.25$.  The
  colour scales correspond to mean metallicity (upper left), the
  fraction of baryons enriched to at least $10^{-6} \Zsolar$ (upper
  right), the median metallicity of the enriched fraction (bottom
  left), and the log-normal metallicity dispersion of the enriched
  fraction (bottom right).}
\label{fig:Ophase_dist}
\end{figure*}

The homogeneity of metals impacts the statistics of absorbers, which
we explore here.  We consider two ad hoc models for distributing metals
throughout the IGM, by replacing particle metallicities in our simulation
with prescriptive metallicities in our vzw model.  In the first (vzw-Z01),
we assume a uniform metallicity of $0.096 \Zsolar$ in the IGM (vzw-Z01),
which is the global $z=0$ IGM metallicity in our vzw model.  Note that
this is similar to $0.1 \Zsolar$ painted onto earlier simulations to
reproduce older observed $\OVI$ absorber frequencies \citep{cen01,
fang01, che03}.  The second model (vzw-0sig) takes the average [O/H]
of the vzw model in each $0.1\times 0.1$ dex pixel of phase space (upper
left panel of Figure \ref{fig:Ophase_dist} at $z=0.25$) and paints this
metallicity onto all SPH particles within that phase space cell, with
zero dispersion.  Both prescriptions preserve the total metal mass, but neither
model is dynamically self-consistent, because metal cooling would alter
the physical state of the gas and feedback would have to enrich the
baryons in the first place.  However, our purpose here is to explore
extreme cases of metal homogeneity to see if absorption line statistics
can distinguish the level of metal inhomogeneity.

Figures \ref{fig:EW_OVI} and \ref{fig:CDD_OVI}, bottom panels, show
the EWDs and the CDDs of these models, compared to our original vzw
model.  The vzw-Z01 model, as found in OD09, produces too many weak
absorbers, too few strong absorbers (in terms of column density), and
a large alignment fraction with $\HI$ (Figure \ref{fig:align_OVI}).
Collisionally ionised $\OVI$ ($T>10^5$ K) accounts for a quarter of
$\Omega_{\OVI}$ and two-thirds of lines above $N_{\OVI}>10^{14} \cms$.
Simply painting metals onto a simulation uniformly overestimates
$\OVI$ and other species in CIE, because the most commonly observed UV
resonance lines ($\CIV$, $\OVI$, and soon $\NeVIII$) are also some of
the most important coolants.

Painting the average metallicity by phase space pixels explores how
the dispersion in metallicity among SPH particles affects the line
statistics.  This model shows negligible differences versus the
original vzw case, showing that the fact that metals are not mixed in
our simulations has no impact on $\OVI$ absorber statistics.  We
reiterate that this test does not dynamically account for differences
in the cooling rates that would result from such metal mixing, and
hence it is not self-consistent; this test isolates the impact of
spatial inhomogeneities in the metals.

Figure \ref{fig:Ophase_dist} provides a broader view of metal
inhomogeneity in our original vzw simulation.  The upper left
shows the oxygen metallicity in phase space, this is similar to
Figure~\ref{fig:Zphasespaces} except here we only consider oxygen and the
general trends are as before.  The upper right panel shows the fraction
of baryons enriched with oxygen, $f_{\rm en}$, using a threshold of
[O/H]=$10^{-6} \Zsolar$.  $f_{\rm en}$ is not very sensitive to this
threshold and is used to discard SPH particles enriched to very low levels
by delayed feedback.  Practically all the ICM gas and most gas, both  hot
and cold, inside of haloes is enriched.  Interestingly, the $f_{\rm en}$
distribution for the nw simulation (not shown) looks nearly identical
to the vzw model above $10^5$ K, indicating that enrichment mechanisms
other than outflows (e.g.  dynamical stripping, delayed feedback) are
getting hot baryons, i.e. hot halo and WHIM gas, above this threshold
metallicity.  In the IGM outside of haloes, most particles remain
pristine, as the typical fractions of enriched particles are $\la 0.1$.
In photo-ionised gas within voids, the fraction of enriched particles
can drop below $10^{-3}$.

The lower left panel shows the average metallicity of only the
enriched particles, [O/H]$_{\rm en}$.  This demonstrates that much
of the enriched portion of the diffuse IGM has [O/H]=$0.1-0.3
\Zsolar$.  Since the outflows carry out metals that are representative
of the ejecting galaxy, this indicates that the dominant galaxies
enriching the IGM tend to have sub-solar metallicities; these
galaxies are generally well below $L^*$~\citep[as found in][]{opp08}.
However, there is a selection effect that highly enriched IGM
particles undergo stronger metal-line cooling, dropping them out
of the warmer phases onto the photo-ionised locus.  Hence enriched
particles in the underdense regions actually have higher metallicities
than the WHIM gas in moderately overdense regions, since these are the
only particles that can cool at such low densities.  Nevertheless,
the $f_{\rm en}$ shows that such particles are quite rare.

The lower right panel shows the log-normal dispersion,
$\sigma$([O/H]$_{\rm en}$) of the metals around the median
log[[O/H]$_{\rm en}$] for the enriched particles only.  Both
\citet{sch03} and \citet{sim04} found that $\CIV$ and $\OVI$
enrichment at $z\sim 2-4$ could be described by a log-normal
distribution, but applied to all baryons.  In general, regions with
dispersions below 0.5$\sigma$ are enriched via outflows consisting of
ISM gas having similar metallicities.  The higher dispersions
correspond to regions enriched via other mechanisms such as dynamical
stripping and delayed feedback from intracluster stars.

\subsection{Alignment Statistics} \label{sec:align}

\begin{figure}
  \subfigure{\setlength{\epsfxsize}{0.49\textwidth}\epsfbox{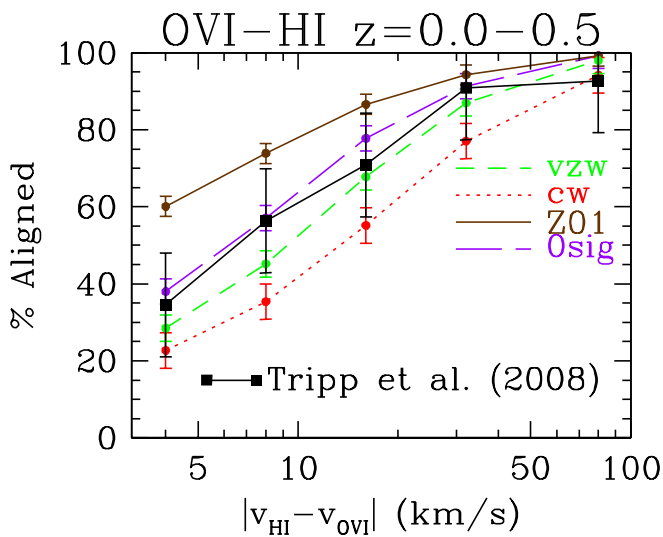}}
  \caption{Alignment fractions for $\OVI$ with $\HI$ components
    between $z=0-0.5$ as a function of inter-species component
    velocity difference.  Both $\OVI$ 1032\AA~and $\HI$ 1216\AA~are
    required to have $W>30$ m\AA.}
\label{fig:align_OVI}
\end{figure}

The association of $\HI$ with $\OVI$ offers a potentially powerful way
to explore metal inhomogeneity.  We show alignment statistics of
$\OVI$ components with nearby $\HI$ components in Figure
\ref{fig:align_OVI}.  We plot the fraction of $\OVI$ absorbers with an
associated $\HI$ absorber at or below the velocity difference
indicated on the $x$-axis, with the criteria that both absorbers have
$W\ge30$ m\AA~and $z<0.5$.  Overlaid are STIS data from \citet{tri08}
using the same selection criterion.  

The data are unable to discriminate between vzw and vzw-0sig, but
disfavour the vzw-Z01 model, which tends to produce more alignment.
OD09 showed that $\OVI$ absorbers trace similar diffuse densities and
temperatures as $\HI$, but that the $\OVI$-traced enriched gas is
inhomogeneous and clumpier than the diffuse baryons traced by the
$\lya$ forest.  The difference in metal inhomogeneity between the vzw
and vzw-0sig models does not show up in the EWD and CDD distributions
in Figures \ref{fig:EW_OVI} and \ref{fig:CDD_OVI}, but the expected
observational sample of the two most common species probed by COS is
expected to provide the ability to statistically distinguish these two
models.  The cw model shows lower alignment owing to faster winds
pushing metals further from the $\lya$ in both physical and velocity
space.  Both turbulence and photo-ionisation-dependent cooling (not
shown) do not alter the statistics from the vzw model.

Alignment statistics versus $\HI$ fundamentally probe the level of
photo-ionisation in metal absorption, because $\HI$ arises almost
exclusively in photo-ionised gas.  Models that produce more
collisionally ionised $\OVI$ will therefore have significantly less
alignment with $\HI$.  Current observations of $\OVI$-$\HI$ alignment
from \citet{tri08} shown in Figure \ref{fig:align_OVI} suggest that
the vzw model produces roughly the correct level of photo-ionised gas
in relation to $\HI$, and therefore strongly favours our conjecture
that much of the $\OVI$ seen with COS is photo-ionised.


\section{Discussion}

Our favoured model includes momentum-conserved winds and uses
photo-ionisation-dependent cooling rates.  The vzw-turb-PI model
matches the $\OVI$ statistics including the $z<0.5$ EWD, CDD,
$\Omega_{\OVI}$, and $b$-parameters when turbulent broadening is added
as argued and theoretically motivated in OD09.  We further advocate
vzw winds for their ability to provide an adequate fit to the $z=0$
galactic stellar mass function below $5\times 10^{10} \msolar$
\citep{opp10}, from which most $\OVI$ absorption arises (OD09), as
well as other properties of the low-redshift galaxy
population~\citep{dav11a,dav11b}; these galaxy properties are not
expected to be greatly impacted by adding IGM turbulence or PI
cooling.  The HM2001 ionisation background multiplied by 1.5 appears
capable of both reproducing the observed $\lya$ forest (D10) and
$\OVI$ that is principally photo-ionised.  The vzw wind model also has
successes fitting higher redshift data, including the $z=2$
mass-metallicity relationship \citep{fin08a}, the $z\ge 6$ galaxy
luminosity function \citep{dav06, fin10}, and observations of IGM
metal enrichment at redshifts $\ge 1.5$ \citep{opp06, opp09b}.  The
most obvious failing of the vzw model is its inability to quench star
formation in $\ga M^*$ galaxies and create the observed red sequence,
although this likely has to do with our simulations lacking a
mass-dependent quenching mechanism such as AGN feedback
\citep{gab10,gab11}.  However, massive galaxies are not responsible
for the vast majority of $\OVI$ enrichment (OD09), because they enrich
a comparatively small volume of the Universe, and so this quenching of
massive galaxies is not expected to significantly impact our IGM
enrichment levels.

Nevertheless, this model also has significant problems that are shared
by the other wind models we consider here.  $\CIV$ and $\SiIV$
observations pose a problem in that our favoured model produces too
many absorbers compared to \citet{coo10, coo11}.  It might be possible
to alter the ionisation background at 3-4 Rydbergs to mitigate this
discrepancy while still satisfying the other observational
constraints.  Updated ionisation backgrounds generated with the new
version of CUBA \citep{haa11} are worth exploring in the future.
Alternatively, our assumed yields could be too high, or carbon and
silicon could be preferentially depleted onto dust \citep{zu11}.
However, the discrepancy may result from limitations present in our
simulations, namely that we fail to resolve cooling instabilities in
halo gas where such lower-ionisation absorbers are expected to arise.
Quantifying $\CII$ and $\CIII$ absorption may help distinguish between
these possibilities.  Our favoured model also under-predicts the
frequency of $\NeVIII$ lines with column densities similar to those
observed so far \citep{nar09}, which again may be a result of
insufficient resolution within the halo gas, in the sense that we fail
to resolve metal-bearing interfaces between cold clumps embedded in
hot, tenuous halo gas.  Relatedly, OD09 (\S5.4) found a lower
frequency of complex, collisionally ionised $\OVI$ absorbers, which
may also arise in shock interfaces within hot halo gas, as implied by
some observations.

In contrast, absorbers arising in more diffuse gas should be more
robustly modelled in our simulations.  Hence, we are confident that,
when probed to sufficiently low column densities, most high ionisation
lines of $\OVI$ and $\NeVIII$ will mostly trace diffuse photo-ionised gas.
Since there have been a number of recent publications exploring $\OVI$
and other metal-line statistics in cosmological simulations to $z=0$,
we now compare and contrast our results with these works.

\subsection{Feedback Methods in Other Simulations}

Our discussion of alternative feedback methods begins with
\citet{cen11}, who generate statistics for $\OVI$ and $\CIV$ from
$z=6\rightarrow 0$ in a 50 $\hmpc$ $2048^3$-cell uniform mesh
simulation.  Their winds are implemented by distributing gas and
energy onto the 27 nearest cells \citep{cen05}, which instantaneously
distributes metals over a 73 comoving $\hkpc$ cube.  Clearly this
model does not self-consistently track the dynamics of enriching this
cube, but it nevertheless provides a heuristic way to distribute
metals into the IGM.  Their favoured wind efficiency is $\epsilon_{\rm
GSW}=7\times10^{-6}$, where $\epsilon_{\rm GSW}$ represents the
fractional rest mass energy of all stars formed put into thermal
feedback.  Their total wind energy is then $E_{\rm wind}= 1.2\times
10^{49}$ erg/$\msolar$, which \citet{cen11} argue is roughly
comparable to the total SNe energy in a Chabrier initial mass function
(IMF).

\citet{smi11} similarly use fixed-mesh Eulerian simulations and
parametrise star formation and feedback the same way.  They can
reproduce much of the observed $z<0.4$ $\OVI$ CDD with a wind energy
efficiency of $\epsilon_{\rm GSW}=10^{-5}$.  We calculate their total
$E_{\rm wind}= 1.8\times 10^{49}$ erg/$\msolar$, which is nearly twice
the SNe energy available in a Chabrier IMF if all stars $\ge 10
\msolar$ explode as $10^{51}$ erg SNe, but about equivalent if all
stars $\ge 6 \msolar$ explode as SNe.  \citet{smi11} advocate the need
of distributing feedback over 27 cells (as in \citealt{cen11}) to fit
the observed star formation rate density history, whereas limiting the
feedback to one cell makes it much less efficient at curtailing star
formation.  This means that in their case, metals are instantaneously
distributed over a 99 to 294 $\hkpc$ cube depending on their
simulations.

Our simulations, in contrast, attempt to directly mimic observations
of galaxies that indicate wind speeds of hundreds of km/s emanating
from galaxies~\citep[e.g.][]{ste10}, and scaling with the circular
velocity~\citep[e.g.][]{mar05a}.  We are inevitably limited by
resolution and numerics in that we cannot fully resolve how winds
escape from galaxies, so we are forced to employ parametrisations for
these winds.  The adopted parametrisations (including hydrodynamic
decoupling) are intended to represent observed winds at the scales at
which they are observed.  It is interesting that our winds typically
travel $\sim 100$ physical kpc \citep{opp08}, which is similar to the
scales over which \citet{cen11} and \citet{smi11} assume outflows
distribute gas.  The difference is that some of our winds can travel
much farther or not as far, depending on the self-consistently tracked
interaction between the gas and the ambient medium; we do not directly
parametrise how far winds and metals travel.  Moreover, in these fixed
mesh simulations the distance metals travel is explicitly governed by
the spatial resolution, while in our case the wind speeds are based on
the galaxy mass, which is a well-converged quantity.  For instance,
the lower-resolution mesh simulations of \citet{cen06b} produce a
significant $\OVI$ signal without any mechanism for distributing gas
to nearby cells, in contrast to our no-wind case that produces
essentially no IGM metal absorbers.  The direct sensitivity of fixed
mesh simulations to ad hoc numerical parameters makes it difficult to
interpret their conclusions about the physics driving IGM enrichment.
Adaptive mesh refinement (AMR) simulations could potentially provide a
less resolution-sensitive implementation of winds.

Finally, we point out that our feedback model has quite modest energy
requirements relative to the above mesh simulations.  Both
\citet{smi11} and \citet{cen11} strain the limits of available
supernova energy.  Our cw model and the feedback prescription in
\citet{tep11} ($\vw=600 \kms$, $\eta=2$) require substantial fractions
of SN energy to be converted directly to kinetic winds-- $E_{\rm
wind}/E_{\rm SN}= 0.75-0.97$ assuming the (conservative) $10 \msolar$
lower mass limit for SNe\footnote{The \citet{tep11} model converts to
$E_{\rm wind}/E_{\rm SN}=0.40$ using the 6 $\msolar$ lower mass limit
for SNe, and the cw model gives an efficiency of $0.52$ using this
limit; however we scale all models to the $10 \msolar$ mass limit in
our comparison.}.  Comparatively, the vzw model has $E_{\rm
wind}/E_{\rm SN}= 0.56$ at $z=1$ and $0.45$ at $z=0$, which is meant
as an illustrative comparison since this wind prescription simulates
energy generated during the lifetimes of O and B stars.\footnote{The
$E_{\rm wind}/E_{\rm SN}=0.72$ r48n384vzw value published in Table 1
of \citet{opp10} is a miscalculation, overestimating the efficiency by
50\%.}  Because vzw feedback energy scales with $\sigma$, the lower
mass galaxies that are primarily responsible for the enrichment of the
diffuse IGM have even less of an energy requirement.  For instance, a
$M_*= 10^{10} \msolar$ galaxy has $E_{\rm wind}/E_{\rm SN}=0.30$ at
$z=1$ and 0.21 at $z=0$.  Note that \citet{str09} in their \S5.4
misinterpret the vzw model as requiring more energy than in the SN
energy budget; this is true only for central galaxies in haloes more
massive than $10^{13.5} \msolar$ at $z=1$, or for between 2-4\% of all
wind particles launched at any redshift.  These galaxies are not
responsible for enriching the IGM, and in any case should not have
significant ongoing star formation today~\citep{gab10}.

\subsection{Metal Mixing} \label{sec:mixing}

Absent in our simulations is any attempt to simulate the mixing
of metals.  For the purposes of cooling and generating mock spectra,
we use particle metallicities, which is in contrast to \citet{tep11},
who smooth metallicities over SPH particle smoothing lengths when
calculating metal-line cooling.  Their metal smoothing was introduced in
\citet{wie09b}, who calculate that smoothed metallicities are $\sim 2/9$th
the individual particle metallicities when all surrounding particles
in their kernel are pristine.  Given that our metal-enriched IGM wind
particles are often surrounded by pristine gas, the cooling times of our
metals could be as much as $\sim 4.5$ times shorter using our individual
particle metallicities if metal-line cooling dominates total cooling.
We suggest this could be the source of the greatest differences with the
results of \citet{tep11}, who find the majority of $\OVI$ above $10^5$ K.
Our cw-PI model is most similar to their ``REFERENCE'' model with notable
other differences including a slightly higher wind speed ($\vw = 680$
vs. $600 \kms$) and that we decouple our winds; both of these should make
it easier for winds to escape galaxies in our simulations.  Nonetheless,
both simulations have similar $\Omega_{\OVI}$ and similar overdensities
bearing $\OVI$ gas, but most of our $\OVI$ gas resides around $10^{4.5}$
K, whereas theirs is at $10^{4.8-5.5}$ K with significantly more $\OVI$
at temperatures $>10^{5.1}$ K.

Fixed mesh simulations with distributed feedback also can lead to much
smoother metallicities, resulting in longer cooling times and hotter
metal-bearing gas.  We argue that the lack of metal smoothing is a
fundamental difference causing a significant photo-ionised component
of $\OVI$ in our simulations versus others \citep{cen11, smi11, tep11}.
Both \citet{tep11} and \citet{smi11} make a strong case that using CIE
metal-line cooling rates over-estimates cooling resulting in too much
photo-ionised $\OVI$ in OD09.  This is no doubt true, especially given the
fact that in our cw simulations, photo-ionised metal-line cooling rates
reduce $\Omega_{\OVI}$ by a factor of three.  But as we demonstrated in
\S\ref{sec:Zcool}, for our favoured vzw case the difference is much less.
Still, metal mixing can be an important determinant for the IGM phase
of $\OVI$ absorbers.

Treatments of metal diffusion have been implemented into other
simulations \citep{gre09, she10}, but it remains unclear to what
extent metals mix in the IGM.  Interpreting observations of metal
absorbers at higher redshift suggest that metals could remain in
concentrated, small cloudlets with metallicities similar to galactic
ISM gas \citep{sim06a, sch07}.  This scenario would favour our lack of
mixing and the fact that winds carry ISM gas directly into haloes and
the IGM.  In summary, the necessity and specific implementation of
metal mixing remains uncertain when considering the low-$z$ IGM.  It
appears that metal mixing is as determinative as implementations of
metal-line cooling or feedback for interpreting IGM metal-line
observations.

\section{Conclusions} 

We examine the absorption and physical conditions of metals residing
in the $\sim$90\% of baryons outside of galaxies between $z=0$ and $2$.
We employ $48 \hmpc$, $2\times 384^3$-particle cosmological simulations,
the same simulations used by \citet{opp10} to explore the $z=0$ galactic
stellar mass function and by \citet{dav10} to explore the $\lya$
forest.  Our work builds on the previous results of \citet{opp09a},
finding fundamentally the same conclusions for $\OVI$, and also makes
predictions for $\SiIV$, $\CIV$, $\NeVIII$, $\MgX$, and $\SiXII$
specifically tailored to the expected instrumental characteristics of
the {\it Cosmic Origins Spectrograph} (COS).  We subdivide the gas phases
between the true IGM ($\delta<120$ at $z=0$) and gas within dark matter
haloes ($120<\delta<10^{5.7}$ at $z=0$), and between hot ($T\ge 10^5$
K, WHIM and hot halo) and cool ($T<10^5$ K, diffuse and condensed) phases.

Metal-line absorption is fundamentally different than its primordial
big brother, the $\lya$ forest, owing to its much greater sensitivity
to the processes of galaxy formation and evolution.  While galactic
outflows suppress star formation and therefore metal production, they
concurrently distribute metals into haloes and the IGM.  Our simulation
without winds (nw) produces 2.1 times more metals by $z=0$ than the
momentum-conserved wind (vzw) model, but the vzw simulation has 3.5 times
more gaseous metals outside of galaxies, and {\it 380 times more metals
in the diffuse IGM phase} from which most IGM metal absorbers arise.
The resulting enrichment patterns are quite sensitive to the details
of the enriching outflows.  The combination of wind velocities and
environment determines to how low a density the winds reach, e.g., whether
or not metals leave their dark matter haloes.  Low velocities from dwarf
galaxies easily reach the IGM, while $>1000 \kms$ winds from 
$>M_*$ galaxies often fail to
escape their haloes because of the greater ambient density of
the surrounding gas \citep{opp10}.  The wind velocity also determines
the temperature to which the gas shock heats as the wind decelerates.
Fast winds that shock gas to $\ga 10^6$ K are more likely to join the
WHIM or hot halo gas.  The density and temperature where the metals are
deposited in turn impacts their cooling rate, which can return enriched
gas to photo-ionisation temperatures.  Tracking all these processes
dynamically within growing large-scale structure presents a complex and
challenging problem that impacts our understanding of a wide range of
phenomena related to the cosmic evolution of baryons.

Our main results are as follows:\\

$\bullet$ Metals progressively are found in more overdense regions
from $z=2\rightarrow 0$.  By $z=0$ in our favoured momentum-conserved
wind (vzw) model, only 4\% of metals are outside of haloes (i.e. at
densities below the threshold defined by Equation \ref{eqn:deltath}),
compared with $20\%$ at $z=2$.  The constant wind (cw) model has many
more metals outside of haloes, but they mostly reside in the
difficult-to-detect WHIM phase.  The majority of metals today reside
within galaxies, in contrast to the bulk of baryons that reside outside
of haloes (\S3).

$\bullet$ For all the wind models, the mean metallicity of most of the
baryonic phases is very slowly rising since $z\sim 2$.  The exception
is the condensed phase, which shows a more rapid rise as winds
progressively enrich halo gas closer to galaxies at later epochs.  By
$z=0$, the IGM metallicities in both the cool (diffuse) and the hot
(WHIM) phases are $\sim 1/50\Zsolar$ in the vzw model, and
$\bar{Z}=0.096\Zsolar$ in the 90\% of baryons outside galaxies (\S3).

$\bullet$ The IGM is enriched in an ``outside-in'' fashion where lower
overdensities are enriched at higher redshift.  Higher ionisation metal
species generally trace older metals at lower overdensities owing to these
species being primarily photo-ionised.  The $\vw\propto \sigma$ relation
for vzw outflows results in a continuously rising metallicity-overdensity
gradient and a gradual age-overdensity anti-correlation, which contrasts
with the constant wind simulations (\S3.3).

$\bullet$ Constant (cw), momentum-conserved (vzw), and slow (sw) wind
models all predict $z<0.5$ $\OVI$ observational statistics (equivalent
width distributions, column density distribution, cosmic ion densities)
that agree within a factor of two with observations by STIS and FUSE.
By contrast, the no-wind model produces an $\OVI$ incidence 100 times
lower than the other feedback models and observations, which strongly
emphasises the need for galactic superwinds to enrich the IGM to the
observed levels.  All wind models over-produce the $\CIV$ and $\SiIV$
incidence relative to available observations (\S4).

$\bullet$ $\OVI$, $\CIV$, and $\SiIV$ absorbers predominantly (by
number) arise in photo-ionised, cool gas.  Photo-ionisation results in
higher ionisation lines tracing lower gas densities: $\SiIV$ and
$\CIV$ arise primarily from halo gas at $z<0.5$, while $\OVI$ arises
mainly in the diffuse IGM (\S4,5.1).  We predict a population of
$\NeVIII$ absorbers with $N_{\NeVIII}\la 10^{13.6} \cms$ tracing
photo-ionised diffuse $T<10^5$ K gas at overdensities of $\sim 10$.
Stronger $\NeVIII$ absorbers (like stronger $\OVI$ systems) are more
likely to be associated with $T=10^{5.5-6.0}$ K gas and may share more
in common with the absorbers already detected (\S4,5.1).  These
results do not change when using metal-line cooling rates including
the presence of a photo-ionising background (\S6.2,6.3).

$\bullet$ $\MgX$ and $\SiXII$ absorbers are very rare, with frequencies
of $dn/dz<< 1$ for COS S/N=30 spectra in all of our models.  These
absorbers likely trace halo gas at $T> 10^6$ K (\S4,5.1).  These
species and $\NeVIII$ all have a significant photo-ionised component,
which is consistent with the shape of the \citet{haa01} ionisation
background; however $\MgX$ and $\SiXII$ are too weak in this state to
be detected using COS (\S5.1.1).  No UV tracer we examine consistently
and predominantly traces WHIM gas, i.e. shock-heated gas outside of
galaxy haloes.

$\bullet$ Using metal-line cooling rates accounting for
photo-ionisation by the ionisation background \citep{wie09a} reduces
the incidence of $\OVI$ absorbers at all column densities by one-third
in the vzw model, which still provides an acceptable match with the
observations.  The difference is more dramatic for cw winds since it
pushes metals to lower overdensities; $\OVI$ statistics are reduced by
two-thirds and arise from gas closer to $T=10^5$ K.  \citet{tep11} and
\citet{smi11} correctly argue that assuming collisional ionisation
equilibrium for metal cooling rates (with no photo-ionisation) leads
to over-efficient metal-line cooling; however we find that the
moderate vzw winds leave metals at higher overdensities where
metal-line cooling is still efficient even in the presence of
photo-ionisation and results in most of the $\OVI$ absorbers tracing
$T<10^{5}$ K gas (\S6.2).

$\bullet$ We reiterate the result of OD09 that heuristic turbulent
broadening is required to reproduce the observed $\OVI$ linewidths,
but we suggest that this prescription may not be applicable to $\CIV$
and $\SiIV$ primarily arising from haloes.  $\OVI$ arises from longer
columns with potentially many cloudlets with random/turbulent velocities
at sub-particle scales, while lower ionisation species are more likely
to trace individual higher density cloudlets (\S6.3).

$\bullet$ Absorber alignment statistics of $\HI$ with $\OVI$ provide an
independent constraint on the level of metal inhomogeneity in the IGM,
which we quantify in \S6.4.  We disfavour a uniform metallicity of $Z=0.1
\Zsolar$ from comparisons with existing data, and future COS observations
can potentially quantify the fraction of enriched gas and the dispersion
of metals in the IGM (\S6.5).

$\bullet$ The metal-line statistics to be probed by COS rarely vary
by a factor of more than two between the different feedback models,
despite enrichment patterns varying by much greater factors (e.g., 6.4
times more metals in the $z=0$ WHIM of cw vs. vzw).  The relatively
restricted density-temperature phase space traced by $\OVI$, $\CIV$,
$\SiIV$, and $\NeVIII$ are enriched similarly in all wind models relative
to the total distribution of metals outside galaxies (\S3.2,5.1.2).

$\bullet$ The observed cosmic ion density $\Omega$ summed from
absorbers typically underestimates the total $\Omega$ of an ion.
The stochastic effects of high overdensities being probed rarely in a
volume-averaged measurement plus a large fraction of an ion's cosmic
density being contained in saturated lines make accurate determinations
of $\Omega_{\CIV}$ and $\Omega_{\SiIV}$ more difficult.  We estimate
recovery fractions of 56-69\% for $\Omega_{\OVI}$ and 30-39\% for
$\Omega_{\NeVIII}$ in S/N=30 COS observations.  These fractions may be
sensitive to the quality of the spectra and the fitting method used to
estimate the column densities (\S5.2).

$\bullet$ Exploring alternative ionisation backgrounds given the
observed range of quasar emission slopes beyond the Lyman limit, we
find that the slope of the $\OVI$ column density distribution becomes
shallower using a harder background.  The ionisation background
strength varies by a factor of 16 at the $\OVI$ photo-ionisation edge
(8.4 Rydberg), but even for the weakest field $\OVI$ remains primarily
photo-ionised.  Slopes of $\alpha=1.57$ and harder predict more
photo-ionised $\OVI$ arising from halo gas (\S6.1).\\

While metal lines observed by COS are more sensitive to the processes
regulating galaxy evolution, they are also more uncertain in their
modelling and interpretation compared to the $\lya$ forest.  Our parameter
exploration finds good fits to existing $\OVI$ statistics, but appears
to over-produce $\CIV$ and possibly under-produce $\NeVIII$ arising in
hot haloes.  Like OD09, we favour a momentum-conserved wind model with
turbulent broadening, although we note that metal-line cooling in the
presence of the photo-ionising background reduces $\OVI$ by a third and
makes the average absorber hotter (30,000 vs. 15,000 K); however our
$\OVI$ results are still consistent with the current data.  We further
advocate this model for its ability to reproduce the galactic stellar mass
function below $M=5\times10^{10} \msolar$ \citep{opp10}, which are the
same galaxies that are primarily responsible for IGM enrichment (OD09).

We double down on our prediction of a primarily photo-ionised origin for
most UV absorbers (OD09), but realise the risky nature of our gamble.
Our model variations show just how sensitive metal-line statistics
are to the shape of the ionisation background at poorly constrained
energies, the stiff behaviour of metal-line cooling, and a turbulence
model that is surely over-simplified.  How metals mix in haloes and the
IGM is perhaps the most under-appreciated uncertainty for simulating
such observations.  Our scheme of launching ISM gas with no mixing may
be the largest difference with other works with smoother metallicities
leading to hotter temperatures \citep{cen11, smi11, tep11}; however,
there is observational evidence and theoretical motivation that small,
highly-enriched cloudlets remain un-mixed in the IGM \citep{sim06a,
sch07}.  The other numerical uncertainty is the accuracy of the basic
\gad~method of implementing galactic winds by one-at-a-time particle
ejection, also used in many other SPH codes.  We have considered several
variations of the scaling of outflow rate and ejection speed with galaxy
mass, but we have not altered this basic scheme.  

The work presented here represents a broad brush approach to
simulating metal line absorbers along random sight lines sampling the
volume-weighted Universe.  With COS, new statistics can be generated
including correlation functions and line of sight variance statistics,
probing the stochasticity of various ion species and perhaps the phase
of gas that they trace.  Relating the distribution of metal absorbers
to the surrounding large-scale structure, particularly the low-mass
galaxies responsible for enriching the IGM, will provide independent
constraints on the nature of outflows.  Studying the galaxy-absorber
connection uniquely links the physics regulating the growth of
galaxies to the dominant reservoir of primordial and enriched baryons
in the IGM, which we will explore in the next paper in this series.

We expect confronting the galaxy-absorber connection in particular
with these and future simulations will provide unique tests that can
both better distinguish wind models from each other, and push the
limits of our scheme.  Our claim that photo-ionisation can explain the
majority of the high ion species explored here will face a significant
test as we focus on the circumgalactic environment within several
hundred kpc of galaxies.  Given that interface physics and chemistry
is not resolved in our simulation while it appears fundamental for
absorption in Galactic high-velocity clouds, suggests we are
underestimating this phase of metal-absorbing gas.  Conversely, this
phase may represent a small minority of the universal quantity of
diffuse metal-enriched absorbing gas, which is the view we advocate
here.

\section*{Acknowledgements}

We thank Kristian Finlator, Joop Schaye, Mike Shull, Thorsten
Tepper-Garcia, Chris Thom, Jason Tumlinson, and Rob Wiersma for useful
discussions and contributions.  We especially acknowledge Kathy
Cooksey for detailed discussions and her effort in re-formulating her
data specifically for our paper, Francesco Haardt for providing and
explaining his CUBA code, Charles Danforth for providing updated forms
of his data, and Mark Fardal for his diligence in debugging and
testing code.  We also are grateful for the hospitality at the
University of Cape Town and the South African Astronomical Observatory
where much of this work was completed.  Partial support for this work
came from NASA ATP grant NNX10AJ95G, HST grants HST-GO-11598 and
HST-GO-12248, NASA ADP grant NNX08AJ44G, and NSF grant AST-133514.
The simulations used here were run on University of Arizona's SGI
cluster, ICE.

\end{document}